\documentclass[3p]{elsarticle}

\usepackage[colorlinks=true]{hyperref}
\usepackage[english]{babel}
\usepackage[utf8]{inputenc}

\usepackage{amsmath}
\usepackage{amsfonts}
\usepackage{mathrsfs}
\usepackage[figbotcap]{subfigure}
\usepackage{enumitem}
\usepackage{MnSymbol}%
\usepackage{wasysym}%
\usepackage{capt-of}
\usepackage[toc,page]{appendix}
\usepackage{verbatimbox}			
\usepackage[normalem]{ulem} 		
\usepackage{accents}				
\usepackage{empheq}
\usepackage{arydshln} 				
\usepackage{pdflscape} 				

\usepackage{todonotes}

\usepackage{etoolbox}

\makeatletter
\newlength\appendixwidth
\preto\appendix{\addtocontents{toc}{\protect\patchl@section}}
\newcommand{\patchl@section}{%
  \settowidth{\appendixwidth}{\textbf{Appendix }}%
  \addtolength{\appendixwidth}{1.5em}%
  \patchcmd{\l@section}{1.5em}{\appendixwidth}{}{\ddt}%
}
\makeatother

\newcommand*{\vcenteredhbox}[1]{\begingroup
\setbox0=\hbox{#1}\parbox{\wd0}{\box0}\endgroup}

\usetikzlibrary{plotmarks}

    \newcommand{\drawline}[3]{%
        \begin{tikzpicture}[inner sep=0pt, baseline=(base)]%
        \draw[#1,#3,line width=1.5pt](0,0) -- (10mm,0);%
        \node[thick, mark size=3pt,color=#1] at (5mm,0){%
            \pgfuseplotmark{#2}%
        };
        \node (base) at (0,-.5ex) {};
        \end{tikzpicture}%
    }

\newcommand{\norm}[1]{\left\lVert#1\right\rVert}
\newcommand{\mat}[1]{\uline{\mathbf{#1}}}
\newcommand\acclrvec[1]{\accentset{\,\leftrightarrow}{#1}}	
\newcommand{\numflux}[1]{\hat{\mathbf{#1}} } 
\newcommand{\blocktensor}[1]{\acclrvec{{\mathbf #1}}\ }		
\newcommand{\Nabla} {\vec{\nabla}}
\newcommand{\NDOF} {\mathrm{NDOF}}
\def\Re {\mathrm{Re}_{\infty}}
\def\Ma {\mathrm{Ma}_{\infty}}
\newcommand\uuuline{\bgroup\markoverwith%
   {%
     \rule[-0.5ex]{2pt}{0.5pt}%
     \llap{\rule[-0.8ex]{2pt}{0.5pt}}%
     \llap{\rule[-1.1ex]{2pt}{0.5pt}}%
   }%
   \ULon}
\newcommand{\thirdtensor}[1]{\uuline{\mathbf{#1}}}
\newcommand{\fourthtensor}[1]{\uuuline{\mathbf{#1}}}

\newcommand{\bunderline}[1]{\underline{#1\mkern-4mu}\mkern4mu }
\def\ncons{n_{\mathrm{cons}}}
\def\d{\mathrm{d}}
\newcommand{\bigpartialderiv}[2]{ \frac{\partial {#1}}{\partial {#2} } }

\journal{XXX}

\biboptions{sort&compress}

\begin{document}
\hypersetup{
urlcolor=black
}

\begin{frontmatter}
\title{A Statically Condensed Discontinuous Galerkin Spectral Element Method on Gauss-Lobatto Nodes for the Compressible Navier-Stokes Equations}

\author[DMAE,CCS]{Andrés M. Rueda-Ramírez \corref{mycorrespondingauthor}}
\cortext[mycorrespondingauthor]{Corresponding authors:}
\ead{am.rueda@upm.es}

\author[DMAE,CCS]{Esteban Ferrer}

\author[FSU]{David A. Kopriva}

\author[DMAE,CCS]{Gonzalo Rubio}

\author[DMAE,CCS]{Eusebio Valero}

\address[DMAE]{ETSIAE-UPM (School of Aeronautics - Universidad Politécnica de Madrid)	, Plaza de Cardenal Cisneros 3, 28040 Madrid, Spain.}

\address[CCS]{Center for Computational Simulation, Universidad Politécnica de Madrid, Campus de Montegancedo, Boadilla del Monte, 28660 Madrid, Spain.}

\address[FSU]{Department of Mathematics, Florida State University and Computational Science Research Center, San Diego State
University.}

\begin{abstract}
We present a static-condensation method for time-implicit discretizations of the Discontinuous Galerkin Spectral Element Method on Gauss-Lobatto points (GL-DGSEM).
We show that, when solving the compressible Navier-Stokes equations, it is possible to reorganize the linear system that results from the implicit time-integration of the GL-DGSEM as a Schur complement problem, which can be efficiently solved using static condensation. 
The use of static condensation reduces the linear system size and improves the condition number of the system matrix, which translates into shorter computational times when using direct and iterative solvers.

The statically condensed GL-DGSEM presented here can be applied to linear and nonlinear advection-diffusion partial differential equations in conservation form.
To test it we solve the compressible Navier-Stokes equations with direct and Krylov subspace solvers, and we show for a selected problem that using the statically condensed GL-DGSEM leads to speed-ups of up to $200$ when compared to the time-explicit GL-DGSEM, and speed-ups of up to three when compared with the time-implicit GL-DGSEM that solves the global system. 

The GL-DGSEM has gained increasing popularity in recent years because it satisfies the summation-by-parts property, which enables the construction of provably entropy stable schemes, and because it is computationally 
very efficient.
In this paper, we show that the GL-DGSEM has an additional advantage: It can be statically condensed.
\end{abstract}


\end{frontmatter}


\section{Introduction}

High-order Discontinuous Galerkin (DG) methods have become popular to solve linear and nonlinear Partial Differential Equations (PDEs) because of their high accuracy and flexibility \cite{Wang2013High,Ferrer2012,Ferrer2017}.
These methods rely on a variational formulation where the continuity constraint on element interfaces is relaxed, allowing for discontinuities in the numerical solution.
This feature makes DG methods robust for solving advection dominated problems, such as those encountered in fluid dynamics applications.

In spite of the increased popularity of high-order methods, most production-quality CFD codes in use are still low order: e.g. in the aerospace industry \cite{Antoniadis2012,Schwamborn2006,Cambier2019}, in weather prediction \cite{Marras2016}, in astrophysics \cite{Mignone2007,Bryan2014}, etc.
A likely reason is that the long years of development on low-order methods have produced very computational-cost-effective solvers that are difficult to compete against.
In this work, we seek to reduce the cost of computing high-order accurate solutions of the Navier-Stokes equations by choosing a computationally efficient DG method, doing implicit time-integration and using static condensation.

In this paper, we use the Discontinuous Galerkin Spectral Element Method (DGSEM), as it is very efficient computationally \cite{Beck2016}.
The DGSEM is usually called a collocation method since it stores the solution variables at the nodes of the quadrature rule and reconstructs the solution using discretely orthogonal basis functions.
The collocation property gives the DGSEM a computational advantage over other DG methods, where the conservative variables and the fluxes must be evaluated on the quadrature nodes as an additional step.
Moreover, the use of a collocated quadrature and orthogonal bases equips the DGSEM with a diagonal mass matrix and relatively cheap-to-compute operators. 

The standard choices for the position of the nodes of the DGSEM are the Gauss and the Gauss-Lobatto quadrature points.
Several advantages have been identified for the DGSEM on Gauss-Lobatto nodes (the GL-DGSEM) over the one that uses Gauss nodes (the G-DGSEM).
The first advantage is that in the GL-DGSEM the solution is stored where it is needed for the evaluation of the surface integrals, whereas it is not in the G-DGSEM \cite{Kopriva2009implementing}.
This makes the GL-DGSEM computationally cheaper and easier to implement since the solution does not have to be interpolated to the element boundaries, as in the G-DGSEM.
In addition, it has been shown that the GL-DGSEM allows larger time steps than the G-DGSEM because of the spectrum of the spatial operator \cite{Gassner2011}.
Furthermore, the GL-DGSEM can be used to unify certain diffusive numerical fluxes as it has been shown that the Bassi-Rebay 1 scheme is a special case of the symetric Interior Penalty method when Gauss-Lobatto nodes are used \cite{Manzanero2018a}.

The only disadvantage of the Gauss-Lobatto quadrature is that it is not as accurate as the Gauss quadrature: The numerical integration is only exact for polynomials of order up to $2N-1$ when using a Gauss-Lobatto quadrature with $N+1$ nodes, whereas it is exact for polynomials of order up to $2N+1$ when using a Gauss quadrature with the same number of nodes \cite{Kopriva2009implementing}.
An inexact evaluation of the integrals induces aliasing errors (in under-resolved simulations) that can trigger instabilities, which in turn can make the solution blow up, especially at high Reynolds numbers \cite{Manzanero2018}. 
The traditional DGSEM on Gauss nodes also suffers from those instabilities, but they appear \textit{sooner} when Gauss-Lobatto nodes are used \cite{Gassner2011,Manzanero2018}.

The common solution to aliasing-driven instabilities is to increase the number of quadrature points to reduce (or eliminate) the aliasing errors, in an approach that is called polynomial dealiasing or over-integration.
The over-integration strategy increases the computational cost since collocation is no longer possible.
Moreover, an exact numerical integration is not always achievable, no matter the number of quadrature points. 
For some systems of equations, like the compressible Euler or Navier-Stokes, the fluxes are not polynomials, but ratios of polynomials \cite{Gassner2016}.

Recent studies point out that the GL-DGSEM has a numerical superiority over the G-DGSEM to eliminate aliasing-driven instabilities without using over-integration.
A cure to these instabilities has been found with the use of split forms of the governing equations \cite{Kravchenko1997,nordstrom2006conservative,Fisher2013} which can be designed according to the entropy-stability framework developed by Tadmor \cite{tadmor1984skew,tadmor1986minimum,tadmor2003entropy}.
Entropy stability through split forms can be achieved while keeping discrete conservation properties if the discretization scheme satisfies the summation-by-parts simultaneous-approximation-term (SBP-SAT) property \cite{Fisher2013}.
Moreover, Gassner \cite{Gassner2013} showed that the GL-DGSEM satisfies the SBP-SAT property.
As a result, entropy stable split forms can be formulated for the GL-DGSEM, e.g. for the Burgers equation \cite{Gassner2013}, the compressible Navier-Stokes equations \cite{Gassner2016}, the MHD equations \cite{Bohm2018}, the Cahn-Hilliard equation \cite{Manzanero2019}, the incompressible Navier-Stokes equations \cite{Manzanero2019a}, the incompressible Navier-Stokes/Cahn-Hilliard system \cite{Manzanero2019b}, etc.
The split form has a dealiasing effect that makes the GL-DGSEM provably stable (something that is not achievable with simple polynomial dealiasing), while still being computationally cheaper than over-integrated methods and yielding comparable results \cite{Winters2018}. 
In a recent publication, Chan et al. \cite{chan2019efficient} proved that entropy-stable DGSEM discretizations can also be obtained using Gauss nodes.
However, this recent technology requires $12(N+1)^3$ extra flux evaluations, which renders it more expensive than the GL-DGSEM.\\

Traditionally, explicit Runge-Kutta methods have been employed to advance high-order DG methods in time, e.g. \cite{Cockburn2001,Kopriva2002,Manzanero2018b,RuedaRamirez2019a}.
These methods allow one to match in time the high-order spatial accuracy of DG methods and have a very low computational cost per time step.
However, due stability constraints explicit time-integration methods can be very inefficient to solve stiff problems, such as steady-state or turbulent simulations, where they may require a time-step size that is much smaller than the one needed to resolve the physical scales \cite{Wang2007}.
On the other hand, implicit time-integration methods can be designed to be stable for larger time-step sizes, but require solving several systems of linear equations at every time step, which makes them more expensive per time step than explicit methods.

The development of techniques that reduce the computational cost per time step of time-implicit high-order DG methods can contribute to their industrialization.
Important advances have been made in the development of efficient preconditioner techniques \cite{Pazner2018,Birken2013,Versbach2018}, of multigrid methods \cite{Nastase2006,Wang2007,Botti2017}, in the formulation of DG schemes with very sparse Jacobian matrices \cite{Persson2006a,Peraire1997}, in the identification of computationally efficient implicit time-marching schemes \cite{Pazner2017,Bassi2015,Bassi2014}, among others.

Static condensation \cite{guyan1965reduction} is a technique to reduce the size of linear systems.
It has been widely used in the continuous Galerkin (CG) community to solve time-implicit discretizations \cite{fraeijs1965displacement,karniadakis2013spectral,Vos2010}.
However, static condensation is not directly applicable to general DG methods in an efficient manner because of the tight coupling of the degrees of freedom between neighboring elements.

To the authors' knowledge, only two techniques have been developed to efficiently apply static condensation to DG schemes.
The first technique was developed by Sherwin et al. \cite{Sherwin2006}, and uses specially tailored basis functions with which it is possible to statically condense the linear system that arises from the time-implicit DG discretization.
Unfortunately, the new basis functions are neither orthogonal nor tensor-product expansions.
Therefore, some properties are lost, like the existence of diagonal mass matrices \cite{Kopriva2009implementing}, Kroenecker multiplication matrices, or the ability to perform anisotropic $p$-adaptation\footnote{In this work, the polynomial order is denoted by $N$, but the action of locally adapting
the polynomial order is called $p$-adaptation, and discretizations with the same polynomial order in all directions are called $p$-isotropic, as they are commonly called in the literature} with anisotropic truncation error estimates \cite{RuedaRamirez2019a,RuedaRamirez2019}, among others.

The second technique is the hybridizable DG (HDG) method, which was developed by Carrero and Cockburn et al. \cite{Carrero2005,Cockburn2009}.
This method expands the linear system to include the numerical trace of the solution as a new unknown, to statically condense it around this new variable.
This technique has been proved to be computationally efficient \cite{Nguyen2009,Peraire2011}, but it imposes certain constraints on the surface numerical fluxes, such as the need for the elliptic fluxes to be adjoint consistent and compact \cite{Cockburn2009}, and the requirement for the numerical fluxes of nonlinear advection-diffusion equations to have a specific mathematical form \cite{Nguyen2009,Peraire2011}.
As a result, not all the known Riemann solvers can be classified as suitable for HDG.
Further insights into the relations between HDG and GL-DGSEM are provided here in Appendix \ref{App:Static}.\\

In this paper, we show that static condensation can be directly applied to the time-implicit GL-DGSEM in an efficient manner without the strong constraints that other static condensation DG schemes have.
Namely, we keep tensor-product orthogonal bases in a method that can be used with any choice of the numerical flux functions, with the only requirement for the viscous numerical flux that it be compact.
We show, by means of a numerical experiment, that the method developed here provides significant speed-ups when compared to the time-explicit GL-DGSEM and the time-implicit GL-DGSEM that does not use static condensation.
Therefore, this investigation reveals a further advantage of Gauss-Lobatto quadratures over traditional Gauss quadratures in the DGSEM.
The method here exposed is a novel contribution, as to the authors' knowledge, it has not been attempted before.

The paper is organized as follows. 
Section \ref{sec:Background} provides the mathematical background that is needed for the derivations of this work.
We briefly describe the time-implicit DGSEM in Section \ref{sec:ImplicitDGSEM}, and provide an overview of the static-condensation method in Section \ref{sec:BackStaticCond}.
In Section \ref{sec:StaticCondForGL-DGSEM}, we analyze the sparsity patterns of DGSEM methods and demonstrate that static condensation can be efficiently applied when Gauss-Lobatto nodes are selected. 
Next, we detail the implementation of the statically condensed GL-DGSEM that is used in this work. 
Finally, the method is tested for solving the compressible Navier-Stokes equations in Section \ref{sec:Static_Case}.

\section{Mathematical Background} \label{sec:Background}

In this section, we provide the necessary mathematical background to formulate a statically condensed Discontinuous Galerkin Spectral Element Method on Gauss-Lobatto nodes.
Note that we adopt the notation of \cite{Bohm2018,Gassner2018}.
For completeness, we include a description of the notation in Appendix \ref{App:Notation}.

\subsection{Time-Implicit DGSEM} \label{sec:ImplicitDGSEM}

We consider the approximation of systems of conservation laws, 
\begin{equation}\label{eq:NScons}
\partial_t \mathbf{q} + \Nabla \cdot \blocktensor{f} = \mathbf{0}, \ \ \text{in } \Omega,
\end{equation}
subject to appropriate boundary conditions, where $\mathbf{q}$ is the state vector of conserved variables, and $\blocktensor{f}$ is the flux block vector, which depends on $\mathbf{q}$. 
In an advection-diffusion conservation law, such as the Navier-Stokes equations (Appendix \ref{App:NS}), the flux vector can be written as
\begin{equation}
\blocktensor{f} = \blocktensor{f}^a(\mathbf{q}) - \blocktensor{f}^{\nu}(\mathbf{q}, \Nabla \mathbf{q}),
\end{equation}
where $\blocktensor{f}^a$ is the advective flux and $ \blocktensor{f}^{\nu}$ is the diffusive flux.
Because of the dependency of the diffusive flux on $\Nabla \mathbf{q}$, \eqref{eq:NScons} is a second order PDE.
Following Arnold et al. \cite{Arnold2002}, \eqref{eq:NScons} can be rewritten as a first-order system,
\begin{subequations}\label{eq:SystemAdvDiff}
\begin{empheq}[left=\empheqlbrace]{align} 
\partial_t \mathbf{q} 
+ \Nabla \cdot \left( \blocktensor{f}^a (\mathbf{q}) - \blocktensor{f}^{\nu} (\mathbf{q}, \blocktensor{g}) \right)&= \mathbf{0} \ , \ \text{in } \Omega,  \label{eq:OuterEq} \\
\Nabla \mathbf{q}
&=  \blocktensor{g}  , \  \text{in } \Omega. \label{eq:InnerEq}
\end{empheq}
\end{subequations}

To obtain the DGSEM-version of \eqref{eq:SystemAdvDiff}, all variables are approximated by piece-wise Lagrange interpolating polynomials of order $N$ that are continuous in each element: $\mathbf{q} \leftarrow \mathbf{q}^N$, $\blocktensor{f} \leftarrow \blocktensor{f}^N$ and $\blocktensor{g} \leftarrow \blocktensor{g}^N$.
Furthermore, \eqref{eq:OuterEq} and \eqref{eq:InnerEq} are multiplied by an arbitrary polynomial (test function) of order $N$ and numerically integrated by parts inside each element of a mesh with a quadrature rule of order $N$, to obtain
\begin{subequations}\label{eq:DGSEMsystem}
\begin{empheq}[left=\empheqlbrace]{align} 
J_j w_j \partial_t \mathbf{q}^N_j 
- \int_{\Omega^e}^N {\blocktensor{f}}^N \cdot \Nabla {\phi_j} \d \Omega^e 
+ \int_{\partial \Omega^e}^N \numflux{f} {\phi_j} \textrm{d} S^e 
&= \mathbf{0}, \label{eq:DGSEMsystem:1}\\
- \int _{\Omega^e}^N \mathbf{q}^N \Nabla \phi_j \d \Omega^e 
+ \int_{\partial \Omega^e}^N \phi_j \hat{\mathbf{q}} \vec{n} \d S^e
&=
J_j w_j  \blocktensor{g}^N_j \label{eq:DGSEMsystem:2}
\end{empheq}
\end{subequations}
for each degree of freedom of each element.
In \eqref{eq:DGSEMsystem}, $\numflux{f}$ and $\numflux{q}$ are the numerical traces of the flux and the solution, respectively, the functions $\phi_j$ are the so-called basis functions, the $J_j$ are the Jacobians of the geometry transformation the mesh is created with, and the $w_j$ are the weights of the quadrature rule. 
The derivation of \eqref{eq:DGSEMsystem} is given in \cite{Kopriva2009implementing,Gassner2009a}. 
However, for completeness, we include the derivation in our notation in Appendix \ref{App:DGSEM}.

Since \eqref{eq:DGSEMsystem} is valid for every degree of freedom of each element, it is possible to gather the contributions of all degrees of freedom and write the nonlinear system
\begin{equation} \label{eq:DiscretNS}
\mat{M} \bigpartialderiv{\mathbf{Q}^N}{t} + \mathbf{H}^N(\mathbf{Q}^N) = \mathbf{0},
\end{equation}
where $\mat{M}$ is the so-called mass matrix, $\mathbf{Q}^N$ is the vector that contains the solution on all the degrees of freedom of the discretization, and $\mathbf{H}^N(\cdot)$ is a nonlinear operator that contains all the DGSEM operations.

We replace the time derivative in \eqref{eq:DiscretNS} by an implicit time integration scheme,
\begin{equation}
\bigpartialderiv{\mathbf{Q}^N}{t}
\leftarrow
\frac{\delta \mathbf{Q}^N}{\delta t} (\mathbf{Q}^N_{s+1},\mathbf{Q}^N_s, \cdots),
\end{equation}
where the operator $\delta \mathbf{Q}^N/ \delta t$ is a function of the solution on the next time step, $\mathbf{Q}^N_{s+1}$ (the unknown), the current time step, $\mathbf{Q}^N_{s}$, and possibly previous time steps. 
Equation \eqref{eq:DiscretNS} can then be approximated as
\begin{align} 
\mat{M} \frac{\delta \mathbf{Q}^N}{\delta t} (\mathbf{Q}^N_{s+1},\mathbf{Q}^N_s, \cdots) + \mathbf{H}^N(\mathbf{Q}^N_{s+1}) &= \mathbf{0}, \label{eq:ImplicitDisc} \\
\mathbf{R}^N (\mathbf{Q}^N_{s+1}) &= \mathbf{0},  \label{eq:NeqtonEq}
\end{align}
where the nonlinear operator $\mathbf{H}^N$ is evaluated on the unknown solution, $\mathbf{Q}^N_{s+1}$.

The system of nonlinear equations, \eqref{eq:NeqtonEq}, can be solved with Newton's method.
Using a Taylor expansion and neglecting terms with high-order derivatives the problem yields
\begin{equation} \label{eq:NewtonLinSys1}
\mat{A} \Delta \mathbf{Q}^N = \mathbf{B},
\end{equation}
where $\mat{A} = \partial \mathbf{R}^N / \partial \mathbf{Q}^N (\tilde{\mathbf{Q}}^N_{s+1})$ is the Jacobian matrix, $\mathbf{B} = - \mathbf{R} (\tilde{\mathbf{Q}}^N_{s+1}) $ is the right-hand-side (RHS), and $\tilde{\mathbf{Q}}^N_{s+1}$ is an approximation to the unknown solution, ${\mathbf{Q}}^N_{s+1}$.
Equation \eqref{eq:NewtonLinSys1} is a linear system that must be solved multiple times to obtain better approximations of $\tilde{\mathbf{Q}}^N_{s+1} \leftarrow \tilde{\mathbf{Q}}^N_{s+1} + \Delta \mathbf{Q}^N$. 

In following sections we derive the analytical expressions for the Jacobian matrix of a DGSEM discretization.
To do so, we part from the DGSEM discretization, \eqref{eq:DGSEMsystem}, and linearize locally to obtain expressions that depend linearly on $\Delta \mathbf{q}^N$.

\subsubsection{Advective Term Linearization}

To facilitate the derivation of the advective terms of the Jacobian matrix, let us first consider the purely advective equations with the temporal term (e.g. the compressible Euler equations of gas dynamics).
The first step is to linearize the advective flux using a Taylor expansion,
\begin{align}\label{eq:linearizeAdv}
\blocktensor{f}^a(\mathbf{q}) 
&= \blocktensor{f}^a (\mathbf{q}_0) 
+ \bigpartialderiv {\blocktensor{f}^a} {\mathbf{q}} \Delta \mathbf{q} 
+ \mathcal{O} \left(  (\Delta \mathbf{q})^2 \right) \\
&\approx \blocktensor{f}^a (\mathbf{q}_0) 
+ \thirdtensor{J}^a \Delta \mathbf{q}, \label{eq:FluxLinearized}
\end{align}
where $\thirdtensor{J}^a$ is the Jacobian of the advective flux evaluated at $\mathbf{q}_0$. 

A linearized expression for the advective numerical flux of an internal interface can be obtained in the same way, now taking into account that it depends on the solution on both sides of the interface,
\begin{align}\label{eq:AdvNumFluxLinearized}
\numflux{f}^{\ a} (\mathbf{q}^+,\mathbf{q}^-,\vec{n}) \bigg \rvert_{\partial \Omega \setminus \Gamma}
&= \numflux{f}^{\ a} (\mathbf{q}_0^+,\mathbf{q}_0^-,\vec{n})
+ \bigpartialderiv {\numflux{f}^{\ a}} {\mathbf{q}^+} \Delta \mathbf{q}^+
+ \bigpartialderiv {\numflux{f}^{\ a}} {\mathbf{q}^-} \Delta \mathbf{q}^- 
+ \mathcal{O} \left(  \max \left( (\Delta \mathbf{q}^+)^2 , (\Delta \mathbf{q}^-)^2 \right) \right) \nonumber \\
&\approx \numflux{f}^{\ a} (\mathbf{q}_0^+,\mathbf{q}_0^-,\vec{n})
+ \numflux{f}^{\ a}_{\mathbf{q}^+} \Delta \mathbf{q}^+
+ \numflux{f}^{\ a}_{\mathbf{q}^-} \Delta \mathbf{q}^-.
\end{align}
In \eqref{eq:AdvNumFluxLinearized}, $\numflux{f}^{\ a}_{\mathbf{q}^+}$ and $\numflux{f}^{\ a}_{\mathbf{q}^-}$ denote the Jacobians of the advective numerical flux function with respect to $\mathbf{q}^+$ and $\mathbf{q}^-$, respectively, evaluated in $\mathbf{q}^+_0$ and $\mathbf{q}^-_0$. 

When the face of an element belongs to a physical domain boundary, $ \partial \Omega \subseteq \Gamma$, the solution on the outer side of the face may depend on the solution on the inner side because of the boundary condition, $\mathbf{q}^-(\mathbf{q}^+)$. 
Therefore, the numerical flux function depends only on the solution on the inner side of the face, $\numflux{f}^{\ a}(\mathbf{q}^+,\vec{n})$. As a consequence, \eqref{eq:AdvNumFluxLinearized} on a physical domain boundary is actually
\begin{equation} \label{eq:AdvNumFluxLinearized2}
\numflux{f}^{\ a} (\mathbf{q}^+,\vec{n}) \bigg \rvert_{\partial \Omega \cap \Gamma}
\approx \numflux{f}^{\ a} (\mathbf{q}^+_0,\vec{n}) 
+ \left( \numflux{f}^{\ a}_{\mathbf{q}^+} 
+ \numflux{f}^{\ a}_{\mathbf{q}^-} \mathbf{q}^-_{\mathbf{q}^+} \right)
\Delta \mathbf{q}^+,
\end{equation}
where $\mathbf{q}^-_{\mathbf{q}^+} = \partial \mathbf{q}^- / \partial \mathbf{q}^+$ is the Jacobian of the Dirichlet boundary condition. 

Inserting \eqref{eq:FluxLinearized}, \eqref{eq:AdvNumFluxLinearized} and \eqref{eq:AdvNumFluxLinearized2} into \eqref{eq:DGSEMsystem:1}, we obtain
\begin{multline} \label{eq:AdvImpDG}
J_j w_j
\partial_t
\mathbf{q}_j^N 
+ \left( 
- \int_{\Omega^e}^N (\thirdtensor{J}^a \phi)_r \cdot \Nabla {\phi_j} \textrm{d} \Omega^e 
+ \int_{\partial \Omega^e}^N  \numflux{f}^{\ a}_{\mathbf{q}^+}  \phi_r  {\phi_j} \textrm{d} S  
+ \int_{\partial \Omega^e \cap \Gamma }^N \numflux{f}^{\ a}_{\mathbf{q}^-} \mathbf{q}^-_{\mathbf{q}^+} \phi_r   {\phi_j} \textrm{d} S \right) 
\Delta \mathbf{q}_r^{N}  \\
+ \left( 
\int_{\partial \Omega^e \setminus \Gamma }^N \numflux{f}^{\ a}_{\mathbf{q}^-} \phi_r^-   {\phi_j} \textrm{d} S \right)
\Delta  \mathbf{q}_r^{\bunderline{N}} 
=
\hat{\mathbf{b}}^a_j,
\end{multline}
where we are using Einstein notation convention with the index $r$ to simplify the expression. 
Note that $\phi^-_r$ is the shape function that corresponds to the degree of freedom $r$ of an external element ($\neq e$), $\mathbf{q}_r^{\bunderline{N}}$ is the solution state vector on that external degree of freedom $r$, and $\hat{\mathbf{b}}^a_j$ is the advective contribution to the right-hand-side that depends only on known values of the solution ($\mathbf{q}^N_0$).

The first term of \eqref{eq:AdvImpDG} contributes to the diagonal blocks of the Jacobian matrix and to the RHS, in amounts that depend on the discretization of the time derivative, $\partial_t \mathbf{q}_j^N$. 
The second term of \eqref{eq:AdvImpDG} contributes to the diagonal blocks of the Jacobian matrix alone, as it multiplies the variation of the solution of the element.
Finally, the third term of \eqref{eq:AdvImpDG} contributes to the off-diagonal blocks of the matrix, as it multiplies the solution on neighbor elements.

\subsubsection{Diffusive Term Linearization}

We first consider \eqref{eq:DGSEMsystem:1} without the time derivative and without the advective fluxes to facilitate the analysis. 
Rewriting \eqref{eq:DGSEMsystem:1} with the explicit dependencies for a compact scheme (see Appendix \ref{App:DGSEM}) yields
\begin{equation} \label{eq:weakBasisDiffop1}
\int _{\Omega^e}^N \blocktensor{f}^{\nu}(\mathbf{q}^N, \blocktensor{g}^N) \cdot \Nabla \phi_j \d \Omega^e 
- \int_{\partial \Omega^e}^N  \numflux{f}^{\nu} (\mathbf{q}^+, \Nabla \mathbf{q}^+, \mathbf{q}^-, \Nabla \mathbf{q}^-, \vec{n}) \phi_j \d \Omega^e = \mathbf{0},
\end{equation}

As for the advective terms, we start by obtaining a linearized version of the viscous flux, which now depends on $\mathbf{q}$ and $\blocktensor{g}$,
\begin{align}\label{eq:linearizeDiff}
\blocktensor{f}^{\nu}(\mathbf{q},\blocktensor{g}) &= 
\blocktensor{f}^{\nu}(\mathbf{q}_0,\blocktensor{g}_0) 
+ \frac{\partial\blocktensor{f}^{\nu}} {\partial \mathbf{q}} \Delta \mathbf{q} 
+ \frac{\partial\blocktensor{f}^{\nu}} {\partial \blocktensor{g}}  \Delta \blocktensor{g}
+ \mathcal{O}\left( \max \left( (\Delta \mathbf{q})^2, (\Delta \blocktensor{g})^2 \right) \right) \nonumber \\
&\approx \blocktensor{f}^{\nu}(\mathbf{q}_0,\blocktensor{g}_0) 
+ \thirdtensor{J}^{\nu} \Delta \mathbf{q} 
+ \fourthtensor{G} \Delta \blocktensor{g},
\end{align}
and a linearized version of the viscous numerical flux,
\begin{align}\label{eq:linearizeViscNumFlux}
\numflux{f}^{\nu} (\mathbf{q}^+, \Nabla \mathbf{q}^+, \mathbf{q}^-, \Nabla \mathbf{q}^-, \vec{n})
= & \numflux{f}^{\nu} (\mathbf{q}^+_0, \Nabla \mathbf{q}^+_0, \mathbf{q}^-_0, \Nabla \mathbf{q}^-_0, \vec{n}) \nonumber \\
& + \bigpartialderiv {\numflux{f}^{\nu}} {\mathbf{q}^+} \Delta \mathbf{q}^+
+ \bigpartialderiv {\numflux{f}^{\nu}} {\Nabla \mathbf{q}^+} \Delta (\Nabla \mathbf{q}^+) \nonumber\\
&+ \bigpartialderiv {\numflux{f}^{\nu}} {\mathbf{q}^-} \Delta \mathbf{q}^- 
+ \bigpartialderiv {\numflux{f}^{\nu}} {\Nabla \mathbf{q}^-} \Delta (\Nabla \mathbf{q}^-) 
+ \mathcal{O} \left(  \max \left( (\Delta \mathbf{q}^+)^2 , (\Delta \mathbf{q}^-)^2 \right) \right)  \nonumber \\
\approx &
\numflux{f}^{\nu}_0 
+ \numflux{f}^{\nu}_{\mathbf{q}^+} \Delta \mathbf{q}^+
+ \numflux{f}^{\nu}_{\Nabla \mathbf{q}^+} \Delta (\Nabla \mathbf{q}^+)
+ \numflux{f}^{\nu}_{\mathbf{q}^-} \Delta \mathbf{q}^-
+ \numflux{f}^{\nu}_{\Nabla \mathbf{q}^-} \Delta (\Nabla \mathbf{q}^-).
\end{align}

Inserting \eqref{eq:linearizeDiff} and \eqref{eq:linearizeViscNumFlux} into \eqref{eq:weakBasisDiffop1}, and again replacing the functions by the corresponding polynomial expansions, yields
\begin{multline} \label{eq:DiffImpDGouter}
\left( \int^N_{\Omega^e} (\thirdtensor{J}^{\nu} \phi)_r \cdot \Nabla \phi_j \d \Omega^e \right)
\Delta \mathbf{q}^N_r
+ \left( \int^N _{\Omega^e} (\fourthtensor{G} \phi)_m \cdot \Nabla \phi_j \d \Omega^e \right) \cdot 
	\Delta \blocktensor{g}^N_m \\
- \left( 
\int^N_{\partial \Omega^e \setminus \Gamma}  \left(
\numflux{f}^{\nu}_{\mathbf{q}^+} \phi_r
+ \numflux{f}^{\nu}_{\Nabla \mathbf{q}^+} \cdot \Nabla \phi_r
\right) \phi_j \d \Omega^e 
%
+ \int^N_{\partial \Omega^e \cap \Gamma} \left(
  \bigpartialderiv {\numflux{f}^{\nu}_{\Gamma}} {\mathbf{q}^+} \phi_r 
+ \bigpartialderiv {\numflux{f}^{\nu}_{\Gamma}} {\Nabla \mathbf{q}^+} \cdot \Nabla \phi_r  \right) \phi_j \d \Omega^e \right)
\Delta \mathbf{q}^N_r \\
- \left( \int^N_{\partial \Omega^e \setminus \Gamma} \left(
  \numflux{f}^{\nu}_{\mathbf{q}^-} \phi^-_r 
+ \numflux{f}^{\nu}_{\Nabla \mathbf{q}^-} \Nabla \phi^-_r \right) \phi_j \d \Omega^e \right)
\Delta \mathbf{q}^{\bunderline{N}}_r 
=
\hat{\mathbf{b}}^{\nu}_j,
\end{multline}
where Einstein notation is again used for the indexes $r$ and $m$.
The Jacobian of the numerical fluxes on the faces that belong to the physical boundaries, $\partial \Omega^e \cap \Gamma$, can be expressed as
\begin{align}
\bigpartialderiv {\numflux{f}^{\nu}_{\Gamma}} {\mathbf{q}^+} &=
\numflux{f}^{\nu}_{\mathbf{q}^+} 
+ \numflux{f}^{\nu}_{\mathbf{q}^-} \mathbf{q}^-_{\mathbf{q}^+}
+ \numflux{f}^{\nu}_{\Nabla \mathbf{q}^-} (\Nabla \mathbf{q}^-)_{\mathbf{q}^+}, \ \rm{and} \\
\bigpartialderiv {\numflux{f}^{\nu}_{\Gamma}} {\Nabla \mathbf{q}^+} &= 
\numflux{f}^{\nu}_{\Nabla \mathbf{q}^+}
+ \numflux{f}^{\nu}_{\Nabla \mathbf{q}^-} (\Nabla \mathbf{q}^-)_{\Nabla \mathbf{q}^+},
\end{align}
where $\mathbf{q}^-_{\mathbf{q}^+}$ is again the Jacobian of the Dirichlet boundary condition, and $(\Nabla \mathbf{q}^-)_{\mathbf{q}^+}$ and $(\Nabla \mathbf{q}^-)_{\Nabla \mathbf{q}^+}$ are the Jacobians of the Neumann boundary condition.
 
Since we want an expression with linear dependencies on $\Delta \mathbf{q}^N$ to construct $\mat{A}$, we now have to rewrite $\blocktensor{g}_m^N$ in \eqref{eq:DiffImpDGouter} with its dependencies on $\mathbf{q}^N$.
This can be done by performing the same local linearization procedure on \eqref{eq:DGSEMsystem:2}.
Since in all classical viscous numerical fluxes \cite{Arnold2002} $\numflux{q}$ is linear with respect to the solution on both sides of the interface, $\mathbf{q}^+$ and $\mathbf{q}^-$, \eqref{eq:DGSEMsystem:2} can be reduced to
\begin{multline} \label{eq:GradDGImp}
J_m w_m \blocktensor{g}^N_m
= 
\left( - \int^N_{\Omega} \phi_r \Nabla \phi_m \d \Omega 
+ \int^N_{\partial \Omega} \numflux{q}_{\mathbf{q}^+} \phi_r \phi_m \vec{n} \d S 
+ \int^N_{\partial \Omega \cap \Gamma} \numflux{q}_{\mathbf{q}^-} \mathbf{q}^-_{\mathbf{q}^+} \phi_r \phi_m \vec{n} \d S \right)
\mathbf{q}^N_r \\
+ 
\left( \int^N_{\partial \Omega \setminus \Gamma} \numflux{q}_{\mathbf{q}^-} \phi^-_r \phi_m \vec{n} \d S \right) 
\mathbf{q}^{\bunderline{N}}_r.
\end{multline}

\subsection{Static Condensation} \label{sec:BackStaticCond}

The static-condensation method, or Guyan reduction \cite{guyan1965reduction}, is a well-known technique to reduce the size of linear systems that can be written in blocks as
\begin{equation} \label{eq:BlockSystem}
\begin{bmatrix}
\mat{B} & \mat{C} \\
\mat{D} & \mat{E}
\end{bmatrix}
\begin{bmatrix}
\mathbf{X}_1 \\ \mathbf{X}_2
\end{bmatrix}
=
\begin{bmatrix}
\mathbf{F}_1 \\ \mathbf{F}_2
\end{bmatrix},
\end{equation}
where $\mat{B} \in \mathbb{R}^{n_1 \times n_1}$, $\mat{C} \in \mathbb{R}^{n_1 \times n_2}$, $\mat{D} \in \mathbb{R}^{n_2 \times n_1}$, and $\mat{E} \in \mathbb{R}^{n_2 \times n_2}$.

We start by performing block Gauss elimination, which can be summarized as multiplying the system \eqref{eq:BlockSystem} on the left by the matrix
\begin{equation}
\begin{bmatrix}
\mat{I} & -\mat{C}\mat{E}^{-1} \\
\mathbf{0} & \mat{I}
\end{bmatrix},
\end{equation}
to obtain
\begin{equation} \label{eq:StaticCondBlock}
\begin{bmatrix}
\mat{B} - \mat{C} \mat{E}^{-1} \mat{D} & \mathbf{0} \\
\mat{D} & \mat{E}
\end{bmatrix}
\begin{bmatrix}
\mathbf{X}_1 \\ \mathbf{X}_2
\end{bmatrix}
=
\begin{bmatrix}
\mathbf{F}_1 - \mat{C} \mat{E}^{-1} \mathbf{F}_2 \\ 
\mathbf{F}_2
\end{bmatrix}.
\end{equation}

In \eqref{eq:StaticCondBlock}, the system of equations for $\mathbf{X}_1$ is decoupled from the rest of the system with a block of zeros in the upper off-diagonal.
As a result, the original system can be solved in two steps:
\begin{enumerate}
\item Solve the statically-condensed system for $\mathbf{X}_1$,
\begin{equation} \label{eq:SC_step1}
[\mat{B} - \mat{C} \mat{E}^{-1} \mat{D}] \mathbf{X}_1 = \mathbf{F}_1 - \mat{C} \mat{E}^{-1} \mathbf{F}_2,
\end{equation}
where the condensed matrix is also known as the Schur complement of the original global matrix.

\item Compute $\mathbf{X}_2$ as a function of $\mathbf{X}_1$,
\begin{equation} \label{eq:SC_step2}
\mathbf{X}_2 = \mat{E}^{-1} (\mathbf{F}_2 - \mat{D} \mathbf{X}_1).
\end{equation}
\end{enumerate}

This approach is computationally efficient if $n_1$ is small and the matrix $\mat{E}$ is easily invertible. 
In fact, the action of  $\mat{E}^{-1}$ on a vector is needed in \eqref{eq:SC_step1} to construct the statically condensed system matrix ($n_1$ times, i.e. the number of columns of $\mat{D}$) and to construct the statically-condensed RHS (one time), and in \eqref{eq:SC_step2} to recover $\mathbf{X}_2$ (one time).

The static-condensation method has been applied to time-implicit Continuous Galerkin (CG) \cite{fraeijs1965displacement,karniadakis2013spectral,Vos2010} and Discontinuous Galerkin (DG) \cite{Sherwin2006,Carrero2005,Cockburn2009} methods, where the linear system is of the form
\begin{equation} \label{eq:SC_linsys}
\mat{A} \mathbf{Q} = \mathbf{B}.
\end{equation}
Note that the delta symbol is omitted for readability, $\Delta \mathbf{Q} \leftarrow \mathbf{Q}$, which does not necessarily imply that we are dealing with linear fluxes.

In Appendix \ref{App:Static}, we present a brief description of the state-of-the-art implementations of the static-condensation method for CG and DG.
Specifically, we describe the statically condensed method of Sherwin et al. \cite{Sherwin2006} and the HDG method \cite{Carrero2005,Cockburn2009}.

\section{Statically Condensed GL-DGSEM} \label{sec:StaticCondForGL-DGSEM}

We now show that the GL-DGSEM is suitable for static condensation. 
In Section \ref{sec:SparsityDGSEM}, we analyze and compare the Jacobian matrix sparsity patterns of the time-implicit G-DGSEM and GL-DGSEM.
From this analysis, it will follow that the linear systems for the DGSEM on Gauss-Lobatto points can be directly organized as \eqref{eq:StaticCondBlock}, where $\mat{E}$ is a block-diagonal matrix.
Hence, static condensation can be directly and efficiently applied to the GL-DGSEM.
In Section \ref{sec:Implementation}, we analyze the properties of the statically condensed GL-DGSEM system and detail its implementation.

\subsection{Jacobian Sparsity Patterns} \label{sec:SparsityDGSEM}

Following the methodology of Section \ref{sec:ImplicitDGSEM}, we first present the analysis for the advective terms and then for the viscous terms.

\begin{figure}[htbp]
\begin{center}
\subfigure[1D mesh]{
	\label{fig:Static_1Dmesh} 
	\includegraphics[width=0.35\textwidth]{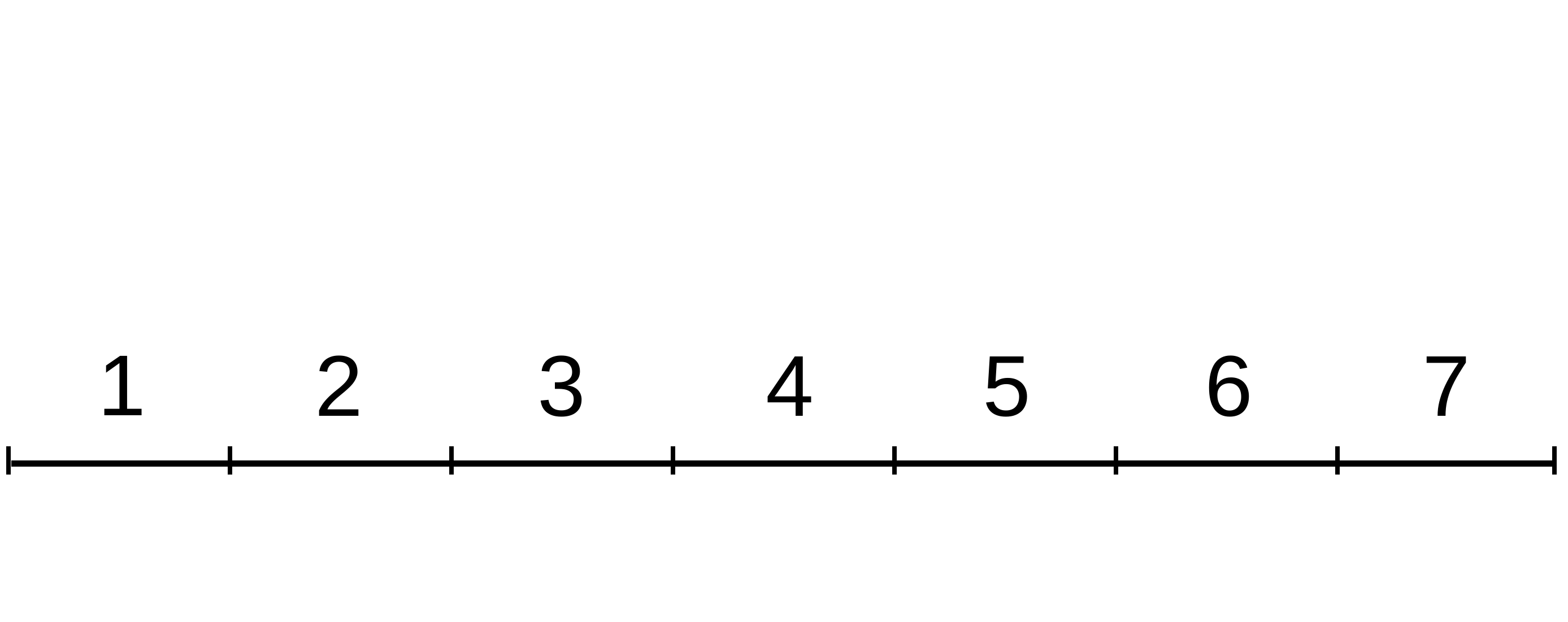}
} \qquad
\subfigure[3D mesh]{
	\label{fig:Static_3Dmesh} 
	\includegraphics[width=0.35\textwidth]{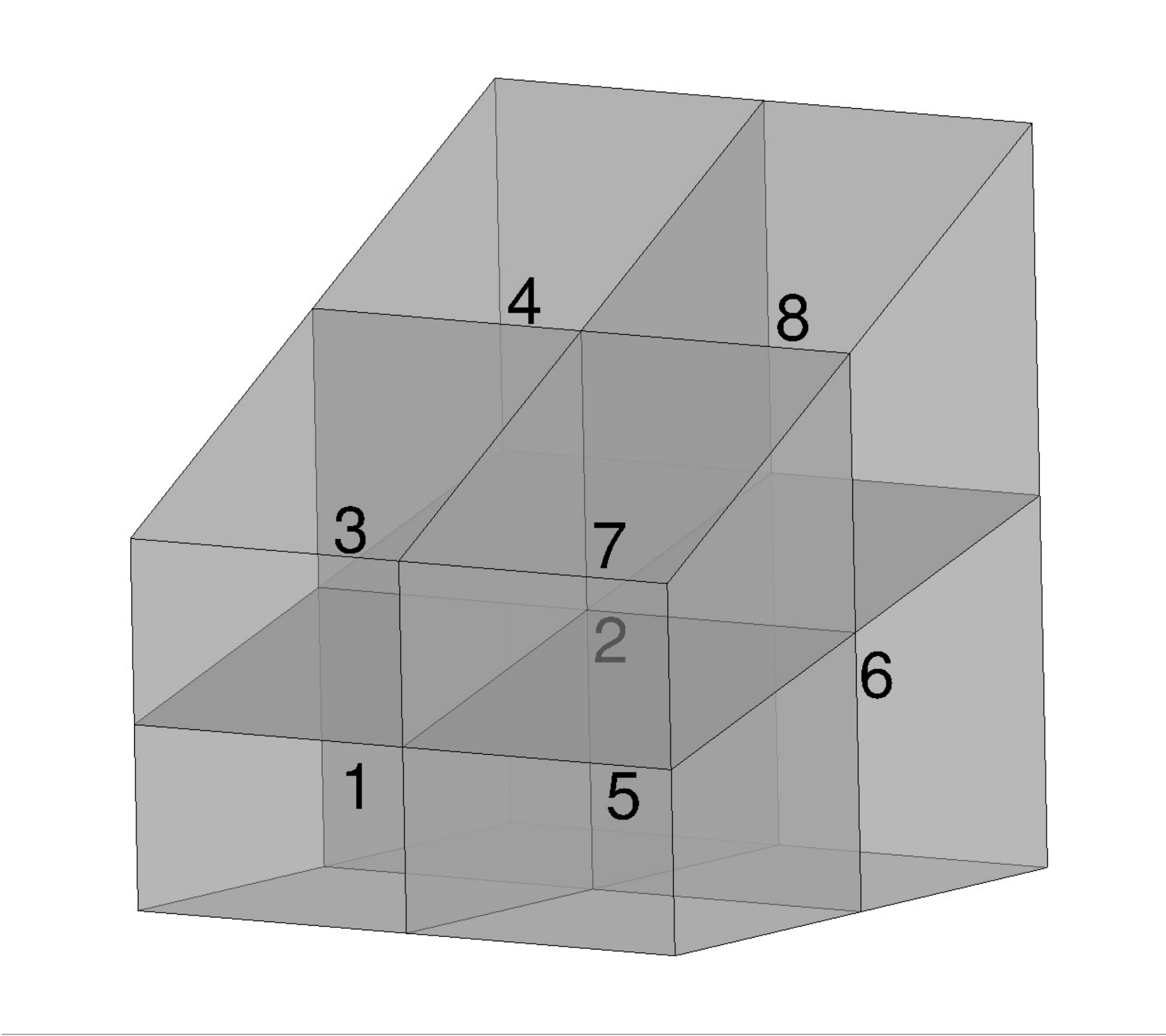}}
\caption{Meshes analyzed for sparsity patterns showing the element numbering.} \label{fig:Static_Meshes}
\end{center}
\end{figure}

The sparsity patterns of a seven-element 1D discretization of order $N=9$ (Figure \ref{fig:Static_1Dmesh}) and an eight-element 3D discretization of order $N=3$ (Figure \ref{fig:Static_3Dmesh}) will be illustrated in this section.

\subsubsection{Advective Terms}

The entries of the Jacobian matrix are presented in \eqref{eq:AdvImpDG} for a time-implicit DGSEM discretization of an advective conservation law. 
From \eqref{eq:AdvImpDG}, it can be inferred that the advective off-diagonal term ($\mathrm{ODT}^{a}_{jr}$) that corresponds to the degree of freedom $j$ of the element $e$, and the degree of freedom $r$ of a certain neighbor element (i.e. the term that multiplies $\Delta \mathbf{q}^{\bunderline{N}}_r$), is
\begin{equation} \label{eq:AdvOffDiag}
\mathrm{ODT}^a_{jr} = \int_{\partial \Omega^e \setminus \Gamma }^N \numflux{f}^{\ a}_{\mathbf{q}^-} \phi_r^-   {\phi_j} \d S^e.
\end{equation}
This term is guaranteed to be zero if the basis functions, $\phi_j$ or $\phi^-_r$, are zero on the element interface.

As explained in Appendix \ref{App:DGSEM}, if one uses Gauss nodes, all basis functions take nonzero values on the element interfaces. 
Therefore, $\phi_j$ and $\phi^+_r$ always contribute to the surface integral (for any $j$ and $r$).
On the other hand, if one uses Gauss-Lobatto nodes, only the basis functions that correspond to interface degrees of freedom take nonzero values on the element interfaces.
Therefore, $\phi_j$ and $\phi^+_r$ only contribute to $\mathrm{ODT}^{a}_{jr}$ if $j$ and $r$ are degrees of freedom that sit on the element boundary.
See Figure \ref{fig:GaussAndGaussLobatto} for details on how the basis functions look on the two node distributions.

\begin{figure}[htbp]
\begin{center}
\subfigure[G-DGSEM]{
	\label{fig:1D_LG_Adv} 
	\includegraphics[width=0.33\textwidth]{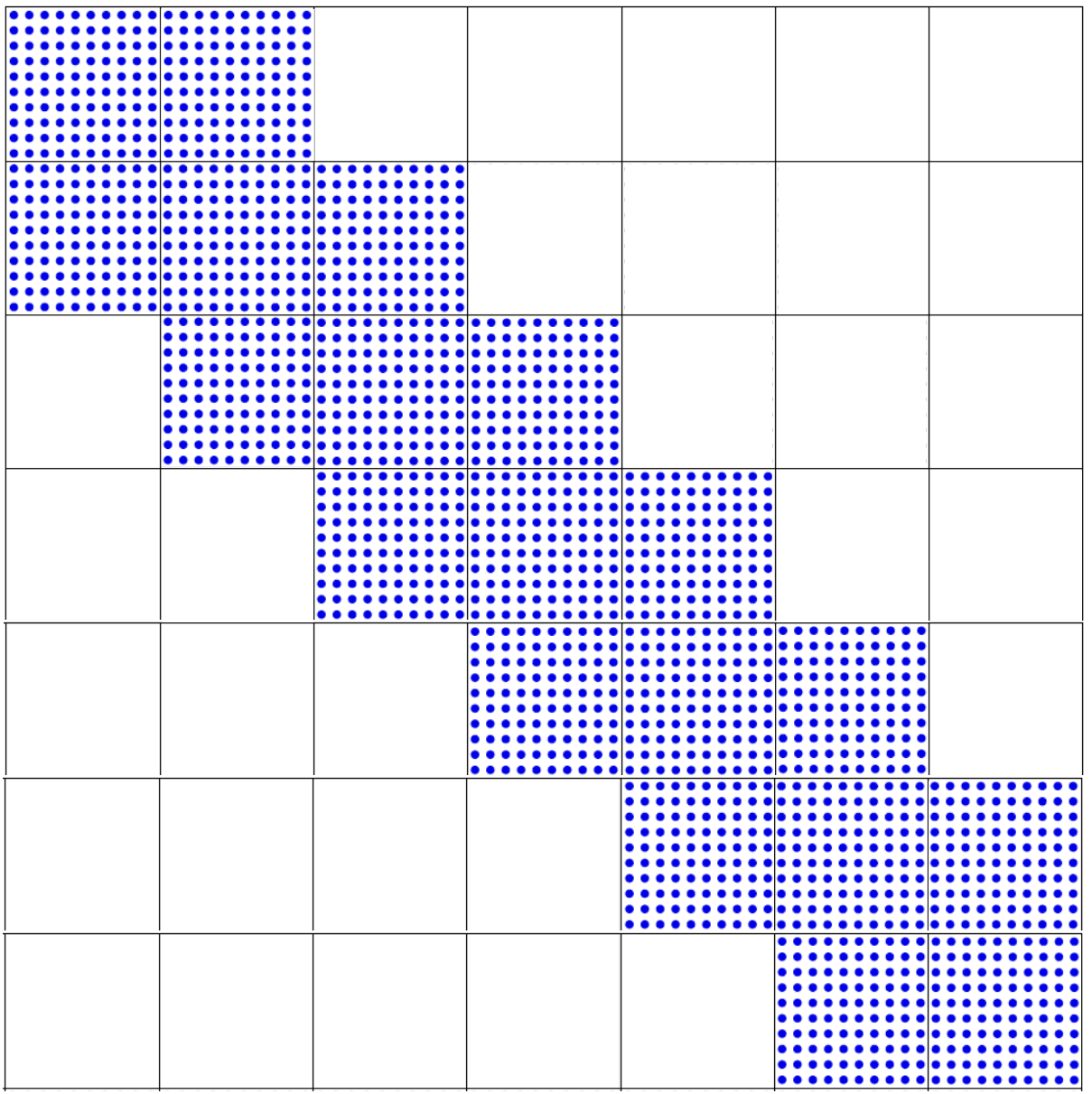}
} \qquad
\subfigure[GL-DGSEM]{
	\label{fig:1D_LGL_Adv} 
	\includegraphics[width=0.33\textwidth]{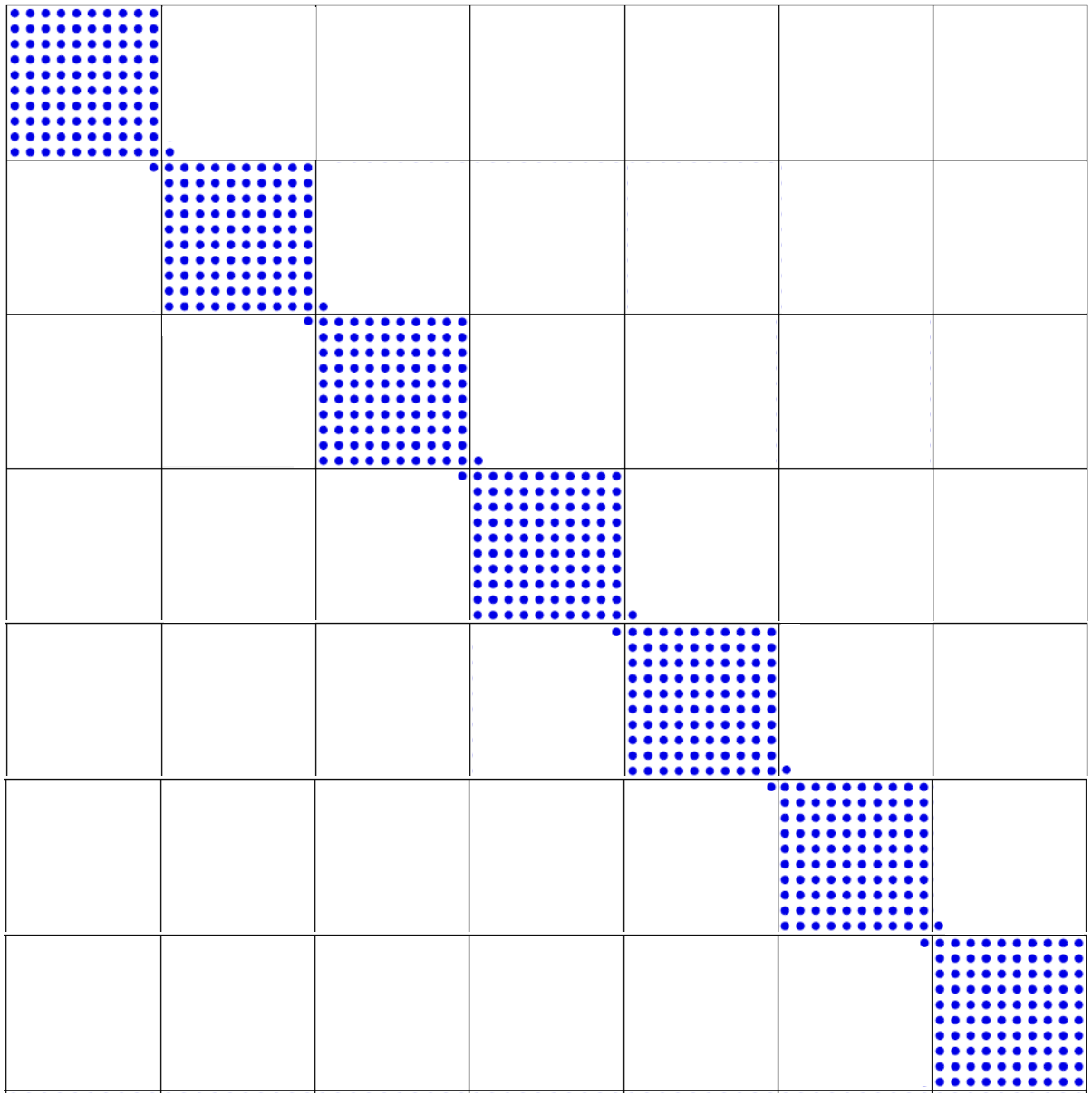}}
\caption{Sparsity patterns of system matrices for 1D scalar advective  equations on the mesh of Figure \ref{fig:Static_1Dmesh}.} \label{fig:1DSparsityAdvective}
\subfigure[G-DGSEM ]{
	\label{fig:3D_LG_Adv} 
	\includegraphics[width=0.35\textwidth]{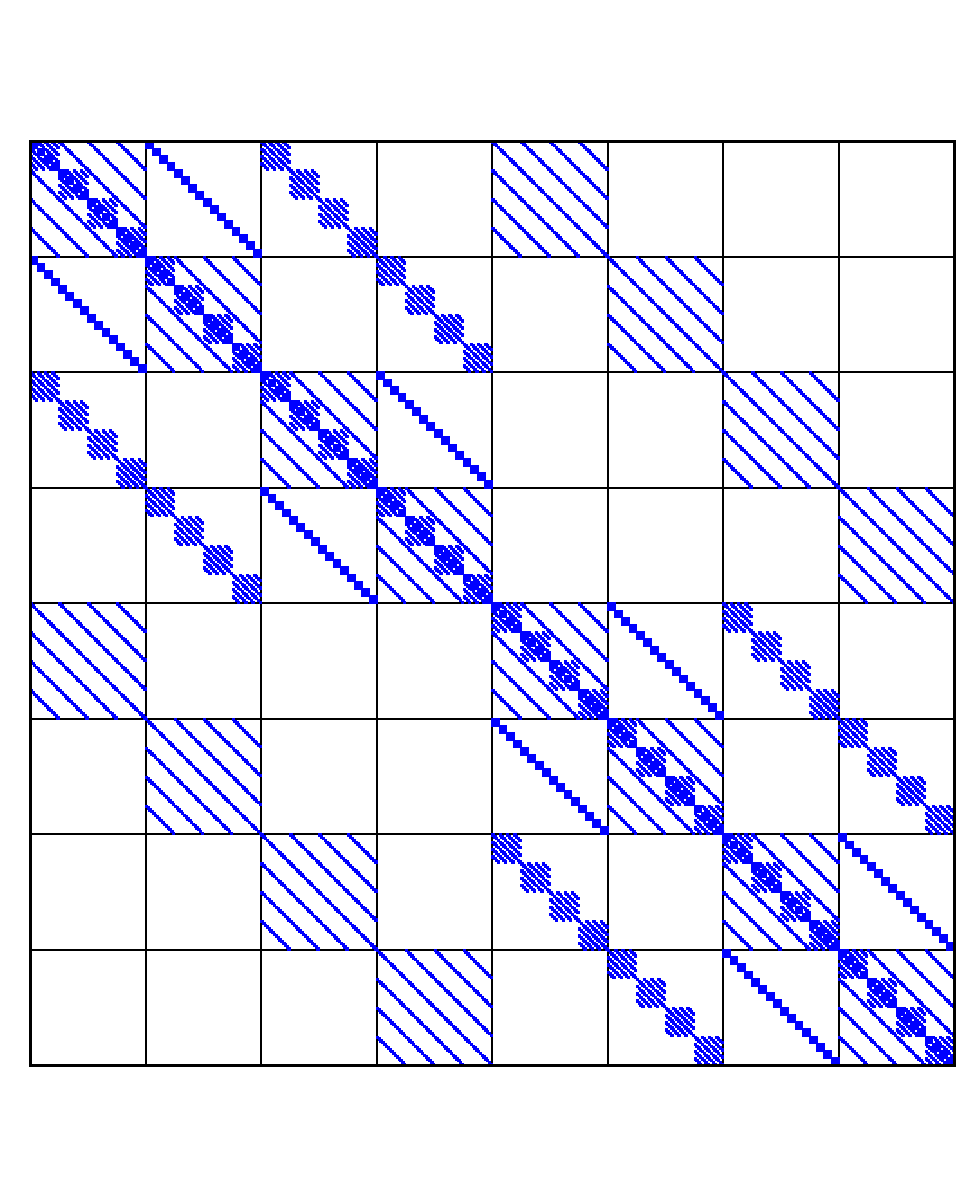}
} \qquad
\subfigure[GL-DGSEM]{
	\label{fig:3D_LGL_Adv} 
	\includegraphics[width=0.35\textwidth]{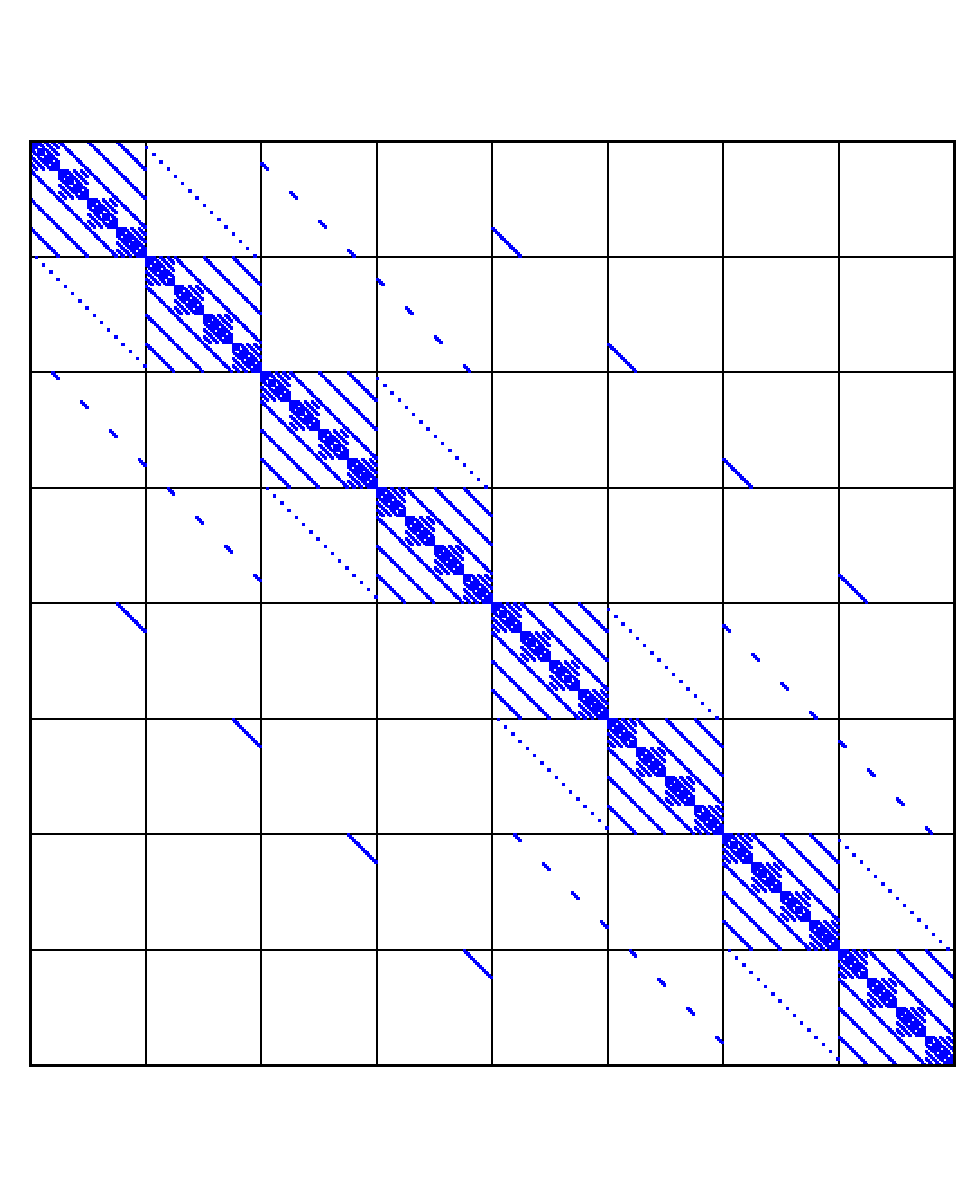}}
\caption{Sparsity patterns of system matrices for 3D scalar advective equations on the mesh of Figure \ref{fig:Static_3Dmesh}.} \label{fig:3DSparsityAdvective}
\end{center}
\end{figure}

The difference between the basis functions on Gauss and Gauss-Lobatto nodes causes different matrix sparsity patterns for the two node distributions, which are illustrated in Figure \ref{fig:1DSparsityAdvective} for the 1D mesh of Figure \ref{fig:Static_1Dmesh} and a scalar ($\ncons=1$) advection equation.
Note that in the 1D DGSEM with Gauss nodes, all the degrees of freedom of a given element are coupled with the degrees of freedom of a neighbor element through entries in the corresponding off-diagonal block.
Contrarily, in the DGSEM with Gauss-Lobatto nodes, only the boundary degrees of freedom are coupled through entries on the off-diagonal block.
As a result, the 1D GL-DGSEM matrix is almost block-diagonal (there is only one entry in each off-diagonal block).
Consequently, it is possible to reorganize the rows and columns that correspond to boundary degrees of freedom, as in Continuous Galerkin methods (Appendix \ref{sec:CGStatic}), to obtain an equivalent linear system,
\begin{equation} \label{eq:StaticDGSEM_QbQiSystem}
\begin{bmatrix}
\mat{A}_{bb} & \mat{A}_{ib} \\
\mat{A}_{bi} & \mat{A}_{ii}
\end{bmatrix}
\begin{bmatrix}
\mathbf{Q}_b \\ \mathbf{Q}_i
\end{bmatrix}
=
\begin{bmatrix}
\mathbf{B}_b \\ \mathbf{B}_i
\end{bmatrix},
\end{equation}
where $\mathbf{Q}_b$ is the solution on the degrees of freedom that sit on the element boundaries (interfaces), and $\mathbf{Q}_i$ is the solution on the inner degrees of freedom. 
Moreover, $\mat{A}_{bb}$ is the boundary-to-boundary matrix, $\mat{A}_{ib}$ is the interior-to-boundary matrix, $\mat{A}_{bi}$ is the boundary-to-interior matrix, and $\mat{A}_{ii}$ is the interior-to-interior matrix.
As in Continuous Galerkin methods, $A_{ii}$ is a block-diagonal matrix, which is inexpensive to invert locally.
Note that system \eqref{eq:StaticDGSEM_QbQiSystem} is equivalent to system  \eqref{eq:BlockSystem}, for
\begin{equation} \label{eq:SC_DGSEMequivalence}
\begin{bmatrix}
\mat{B} & \mat{C} \\
\mat{D} & \mat{E}
\end{bmatrix}
\leftrightarrow
\begin{bmatrix}
\mat{A}_{bb} & \mat{A}_{ib} \\
\mat{A}_{bi} & \mat{A}_{ii}
\end{bmatrix}, \ \
\begin{bmatrix}
\mathbf{X}_1 \\ \mathbf{X}_2
\end{bmatrix}
\leftrightarrow
\begin{bmatrix}
\mathbf{Q}_b \\ \mathbf{Q}_i
\end{bmatrix}, \ \
\begin{bmatrix}
\mathbf{F}_1 \\ \mathbf{F}_2
\end{bmatrix}
\leftrightarrow
\begin{bmatrix}
\mathbf{B}_b \\ \mathbf{B}_i
\end{bmatrix},
\end{equation}

Figure \ref{fig:3DSparsityAdvective} shows that a similar behavior is observed for the 3D mesh of Figure \ref{fig:Static_3Dmesh}.
In this context, the sparsity patterns are also shown for a scalar PDE, where each pixel is an entry of the Jacobian matrix.
In multi-equation PDEs, the sparsity pattern is very similar, but each pixel would contain an $\ncons \times \ncons$ matrix that is not necessarily dense.
In 3D, the diagonal and off-diagonal blocks are no longer dense because of the tensor-product basis expansions.
This can be exploited to reduce the storage requirements, especially for very high orders ($N > 3$).

In the 3D DGSEM with Gauss nodes, all the degrees of freedom of a given element are coupled (through the off-diagonal block) to some degrees of freedom of the neighbor elements in a way that makes it impossible to reorganize the system as \eqref{eq:StaticDGSEM_QbQiSystem}, with $A_{ii}$ as a block-diagonal matrix.
However, when Gauss-Lobatto nodes are used and the matrix is reorganized, $A_{ii}$ is indeed a block-diagonal matrix since only the boundary degrees of freedom are coupled with other boundary degrees of freedom.

\subsubsection{Diffusive Terms}

The Jacobian entries produced by the diffusive terms of the PDE are computed from \eqref{eq:DiffImpDGouter} and \eqref{eq:GradDGImp}.
The diffusive off-diagonal term ($\mathrm{ODT}^{\nu}_{jr}$) that corresponds to the degree of freedom $j$ of the element $e$, and the degree of freedom $r$ of a certain neighbor element (i.e. the term that multiplies $\Delta \mathbf{q}^{\bunderline{N}}_r$), is
\begin{multline} \label{eq:DiffOffDiag}
\mathrm{ODT}^{\nu}_{jr} = 
\sum_{m=1}^{\NDOF^e} 
\left[
	\frac{1}{J_m w_m}
	\left(
		\int_{\Omega^e}^N \fourthtensor{G}{}_m \phi_m \cdot \Nabla \phi_j \d \Omega^e
	\right) \cdot
	\left(
		\int_{\partial \Omega^e \setminus \Gamma}^N \phi^-_r \phi_m \vec{n} \d S^e
	\right)
\right]	\\
- \int_{\partial \Omega^e \setminus \Gamma}^N \left(
  \numflux{f}^{\nu}_{\mathbf{q}^-} \phi^-_r 
+ \numflux{f}^{\nu}_{\Nabla \mathbf{q}^-} \Nabla \phi^-_r \right) \phi_j \d \Omega^e.
\end{multline}

\begin{figure}[htbp]
\begin{center}
\subfigure[G-DGSEM]{
	\label{fig:1D_LG_Diff} 
	\includegraphics[width=0.33\textwidth]{Figs/1D_Sparsity_LG.png}
} \qquad
\subfigure[GL-DGSEM]{
	\label{fig:1D_LGL_Diff} 
	\includegraphics[width=0.33\textwidth]{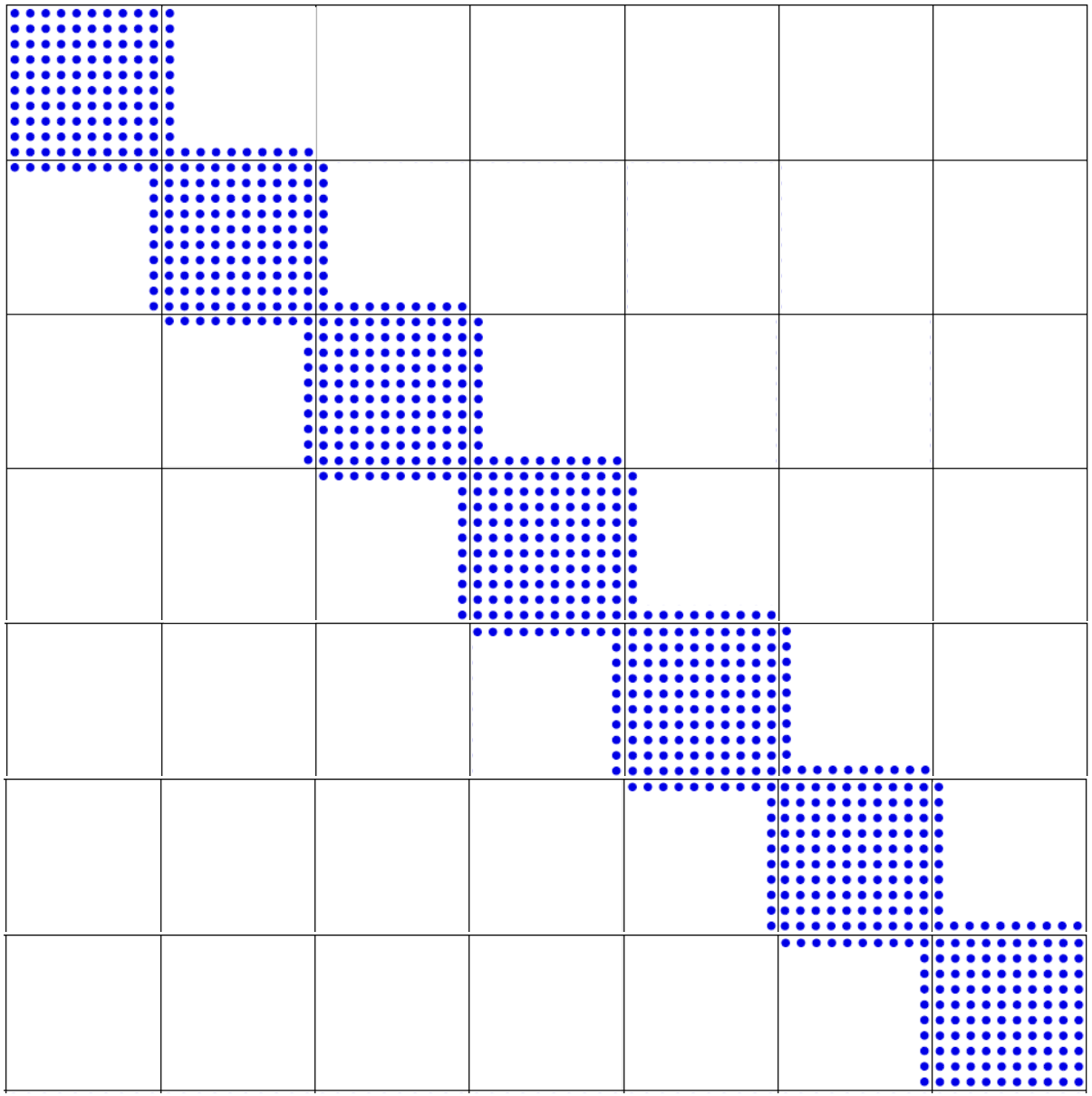}
}
\caption{Sparsity patterns of system matrices for 1D advection-diffusion scalar equations on the mesh of Figure \ref{fig:Static_1Dmesh}.} \label{fig:1DSparsityDiffusive}
\end{center}
\end{figure}

It is evident that \eqref{eq:DiffOffDiag} generates much denser off-diagonal blocks than the ones obtained for the advective terms. 
As a matter of fact, at first sight one could think that static condensation is not applicable to the time-implicit GL-DGSEM when diffusive terms are present. 
However, it is indeed, as can be observed in Figure \ref{fig:1DSparsityDiffusive}, which shows the resulting sparsity patterns for the 1D mesh of Figure \ref{fig:Static_1Dmesh}.

The 1D sparsity pattern for the DGSEM on Gauss nodes is the same as in the advective case, but there are substantial differences when Gauss-Lobatto nodes are used.
To begin, the first term of \eqref{eq:DiffOffDiag} takes nonzero values for any degree of freedom $j$ of the element that is being analyzed, if and only if $r$ corresponds to a boundary degree of freedom of a neighbor element, i.e. when $\phi^-_r \ne 0$. 
Moreover, the second term takes nonzero values for any $r$, if and only if $\phi_j \ne 0$, i.e. for a boundary degree of freedom of the analyzed element. 
As a result, a whole row and a whole column of each off-diagonal block of the 1D system matrix are filled with nonzero values.
These blocks are indeed denser than in the advective case, but the static-condensation method can still be applied with the same row/column reordering.

\begin{figure}[htbp]
\begin{center}
\subfigure[G-DGSEM]{
	\label{fig:3D_LG_Diff} 
	\includegraphics[width=0.35\textwidth]{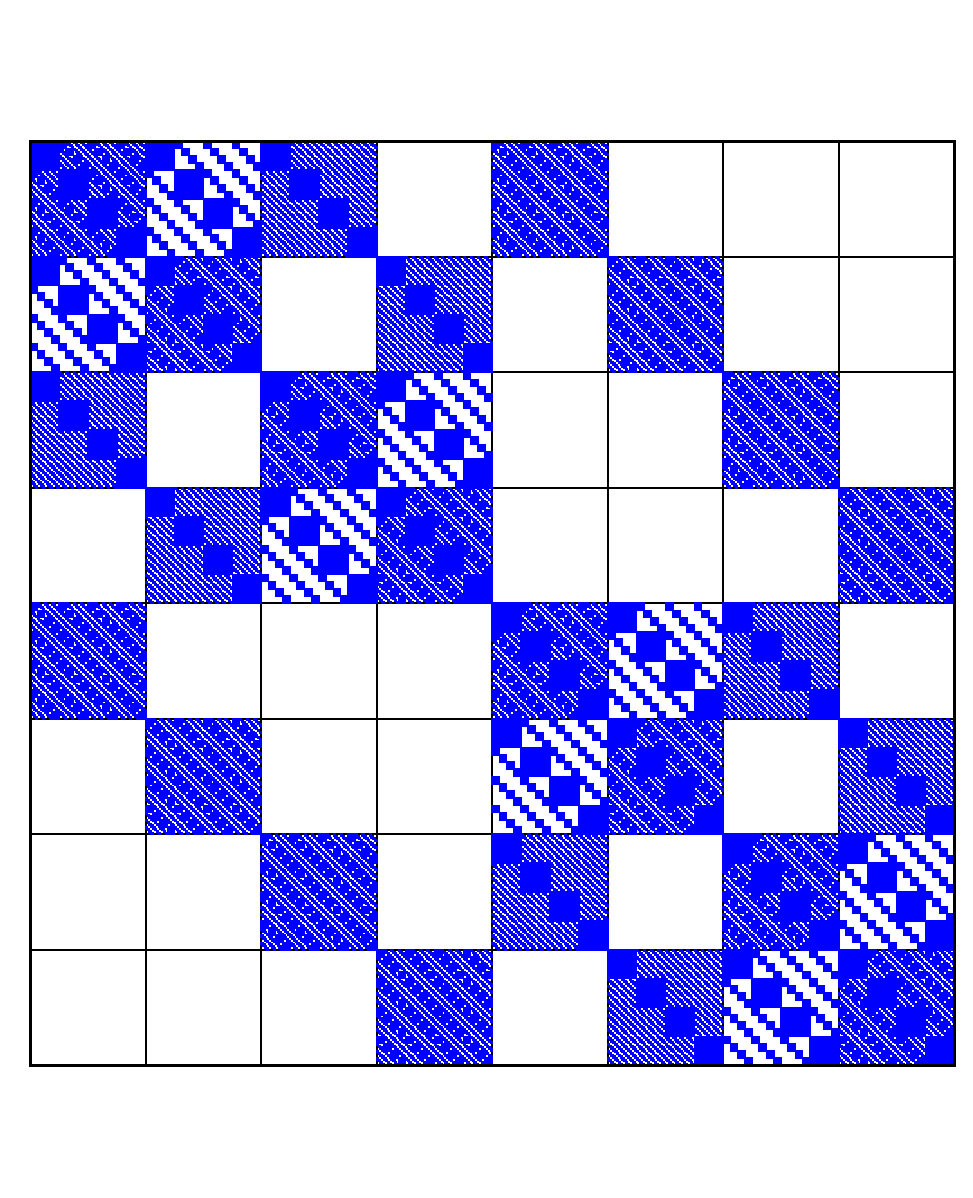}
} \qquad
\subfigure[GL-DGSEM]{
	\label{fig:3D_LGL_Diff} 
	\includegraphics[width=0.35\textwidth]{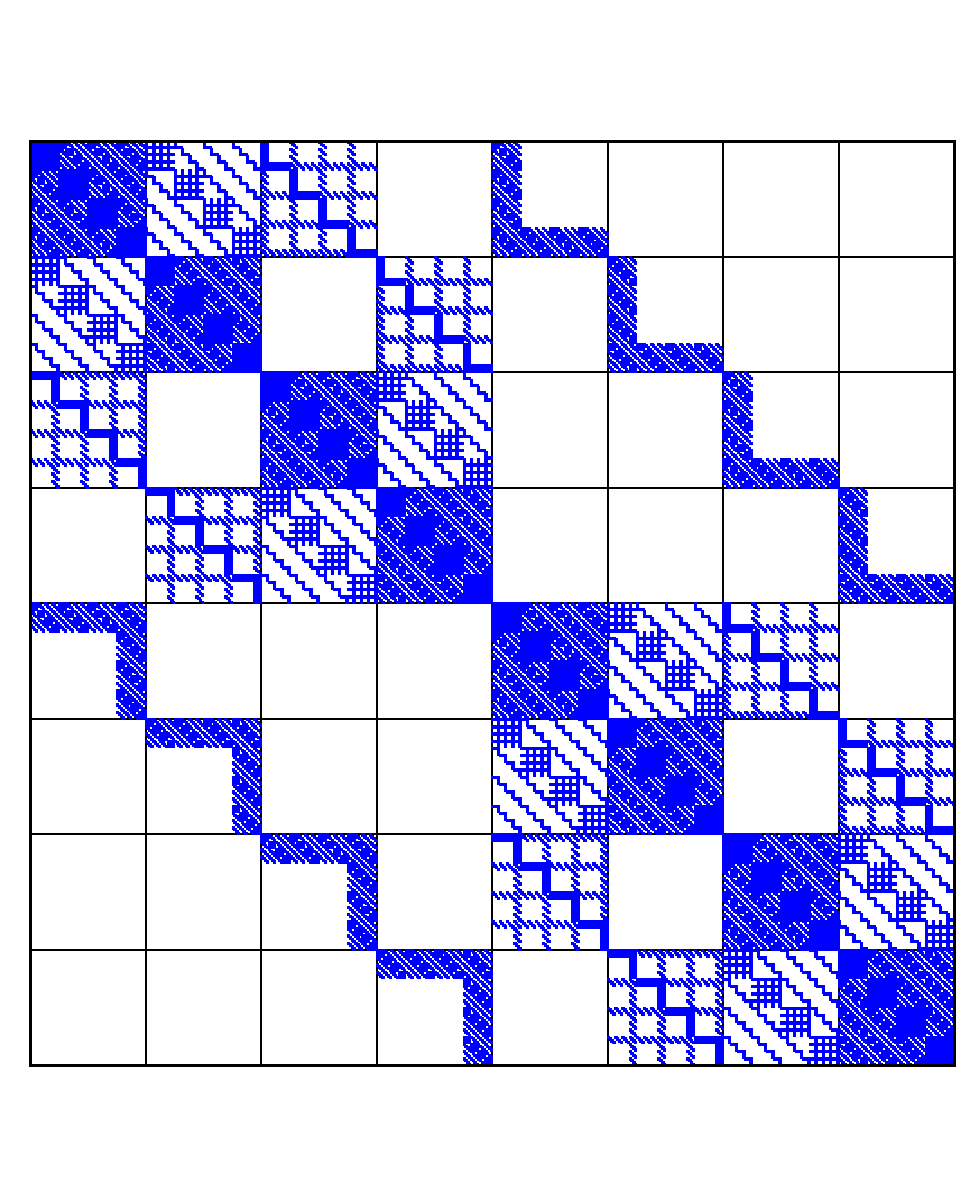}
}
\caption{Sparsity patterns of system matrices for 3D advection-diffusion scalar equations on the mesh of Figure \ref{fig:Static_3Dmesh}.} \label{fig:3DSparsityDiffusive}
\end{center}
\end{figure}

\begin{figure}[htbp]
\begin{center}
\subfigure[G-DGSEM]{
	\label{fig:3D_LG_DiffDetail} 
	\includegraphics[width=0.4\textwidth]{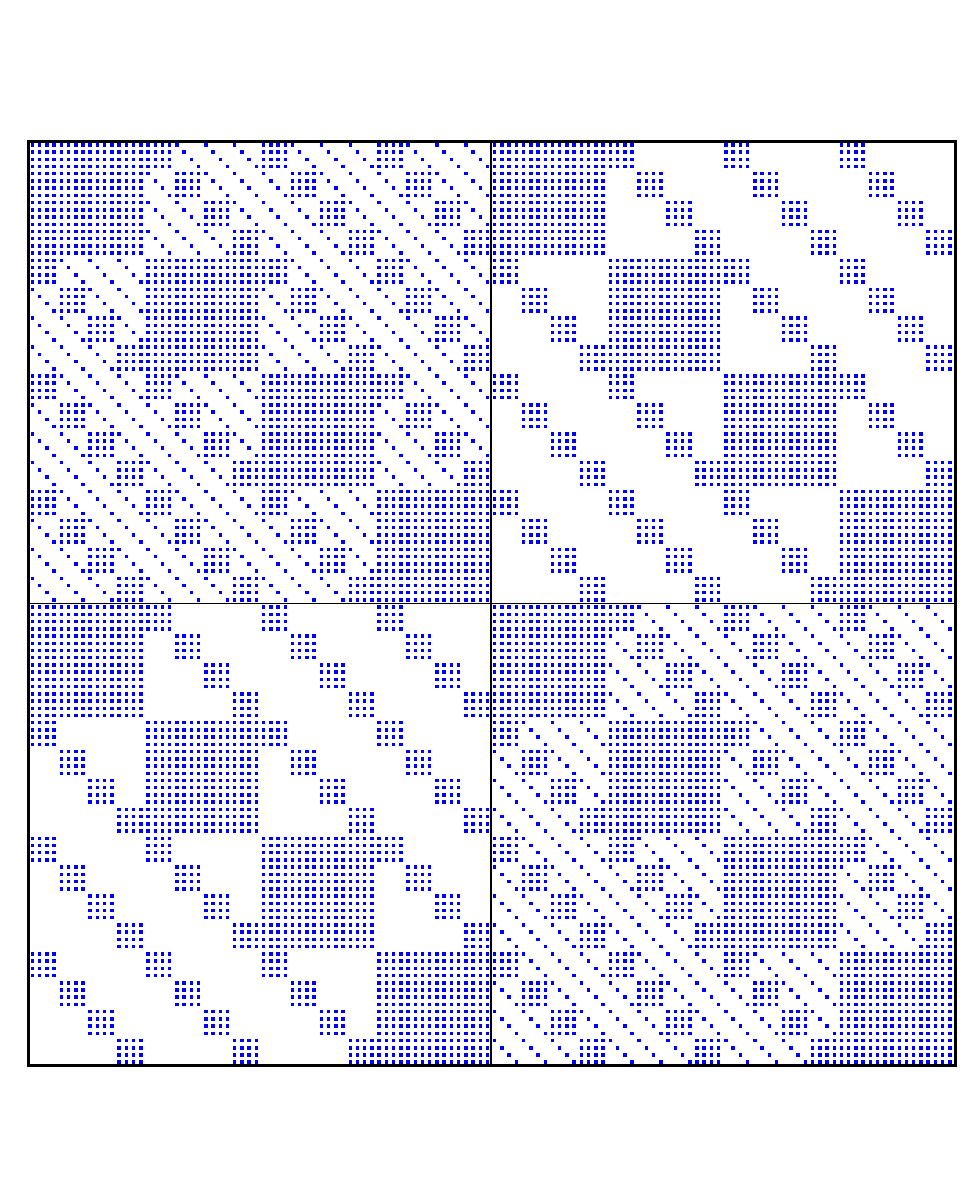}
} \qquad
\subfigure[GL-DGSEM]{
	\label{fig:3D_LGL_DiffDetail} 
	\includegraphics[width=0.4\textwidth]{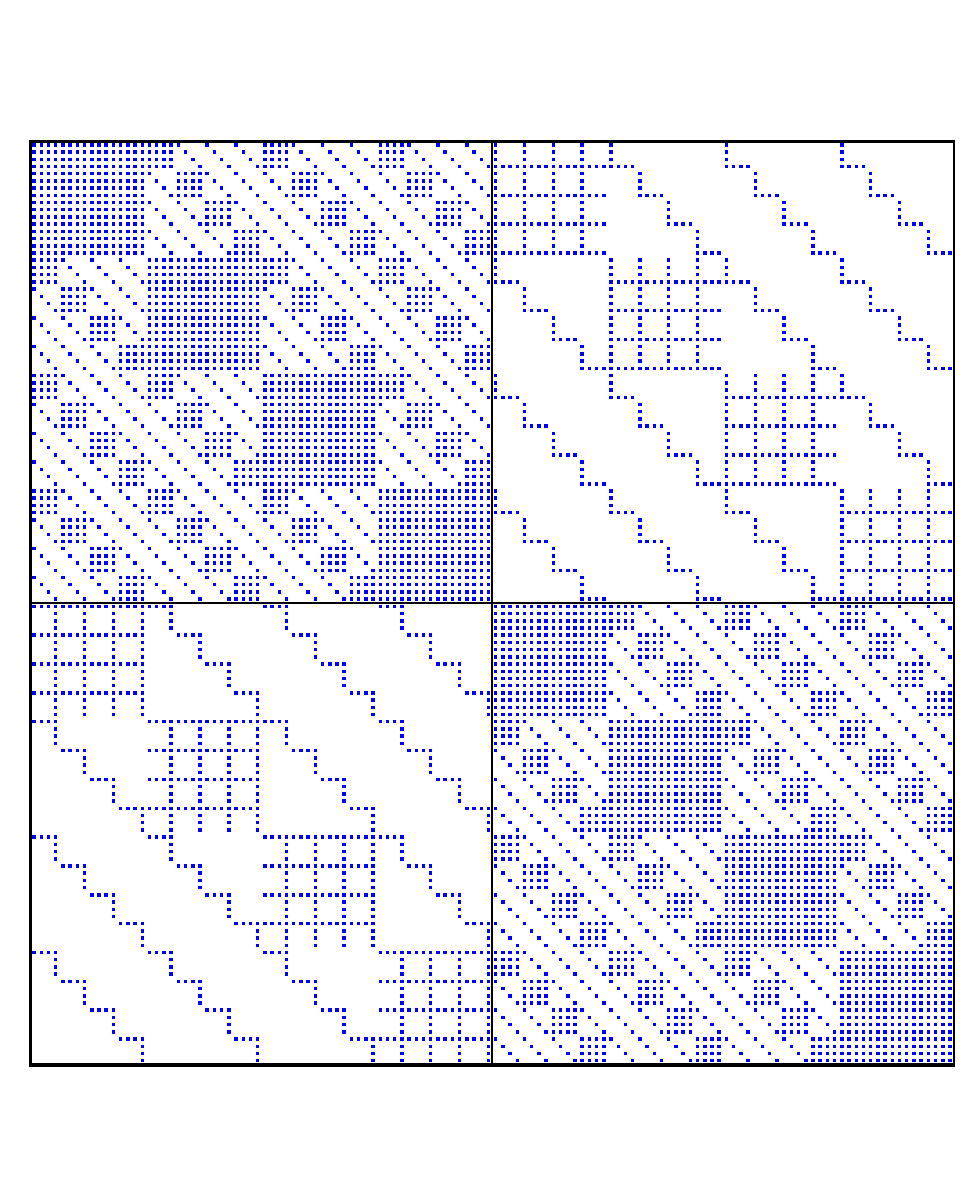}
}
\caption{Detail of the Jacobian blocks corresponding to elements 1 and 2.} \label{fig:3DSparsityDiffDetail}
\end{center}
\end{figure}

The 3D discretization is more complex than the 1D, but it retains similar properties. 
Figure \ref{fig:3DSparsityDiffusive} shows the sparsity pattern that is produced by the time-implicit DGSEM discretization of a scalar nonlinear advection-diffusion equation in the 3D mesh of Figure \ref{fig:Static_3Dmesh}.
As can be seen, both the diagonal and off-diagonal blocks are sparse because of the tensor-product basis expansions, but they are much denser than in the purely advective case.
This is a consequence of the additional spatial derivatives.
Moreover, it is again impossible to reorder the G-DGSEM matrix of Figure \ref{fig:3D_LG_Diff} to obtain a block-diagonal $\mat{A}_{ii}$ matrix.
In contrast, the matrix resulting from the Gauss-Lobatto discretization (Figure \ref{fig:3D_LGL_Diff}) is suitable for static condensation.
This is clearly seen from the appearance of some off-diagonal blocks, like the ones connecting elements $1$-$3$ or $1$-$5$.
However, the off-diagonal blocks that connect elements $1$-$2$ or $3$-$4$ seem to have a more complicated sparsity pattern that does not allow to obtain block-diagonal matrices when reordering.
This is just an artifice of the plotting. 
In fact, a detailed view of the part of the Jacobian that corresponds to elements $1$ and $2$ (Figure \ref{fig:3DSparsityDiffDetail}) reveals that only some of the degrees of freedom are coupled in the Gauss-Lobatto DGSEM.\\

In summary, the off-diagonal blocks of a GL-DGSEM discretization only take nonzero values if 
\begin{itemize}
\item $j$ \textbf{and} $r$ correspond to boundary degrees of freedom (advective case), or
\item $j$ \textbf{or} $r$ correspond to boundary degrees of freedom (diffusive case).
\end{itemize}

In either case, the system can be reorganized as a Schur complement problem with $\mat{A}_{ii}$ being a block-diagonal matrix.

We remark that, although the analysis of this section was made for the traditional GL-DGSEM, it can be easily extended to the split-form GL-DGSEM as the resulting sparsity pattern of the off-diagonal blocks is the same.

\subsection{Analysis and Implementation} \label{sec:Implementation}

As was shown in previous section, the linear system resulting from the GL-DGSEM discretization of an advection, diffusion or advection-diffusion conservation law can be reorganized and condensed to obtain a new system of the form,
\begin{equation}
\begin{bmatrix}
\mat{A}_{bb} - \mat{A}_{ib} \mat{A}_{ii}^{-1} \mat{A}_{bi} & \mathbf{0} \\
\mat{A}_{bi} & \mat{A}_{ii}
\end{bmatrix}
\begin{bmatrix}
\mathbf{Q}_b \\ \mathbf{Q}_i
\end{bmatrix}
=
\begin{bmatrix}
\mathbf{B}_b - \mat{A}_{ib} \mat{A}_{ii}^{-1} \mathbf{B}_i \\ 
\mathbf{B}_i
\end{bmatrix},
\end{equation}
where $\mat{A}_{ii}$ is a block-diagonal matrix. 
In addition, as was shown in Section \ref{sec:BackStaticCond}, this new linear system can be solved in two steps: the linear solve of the condensed system and the reconstruction of the solution on the inner degrees of freedom.

\begin{figure}[htbp]
\centering
\vcenteredhbox{\includegraphics[width=0.187\textwidth]{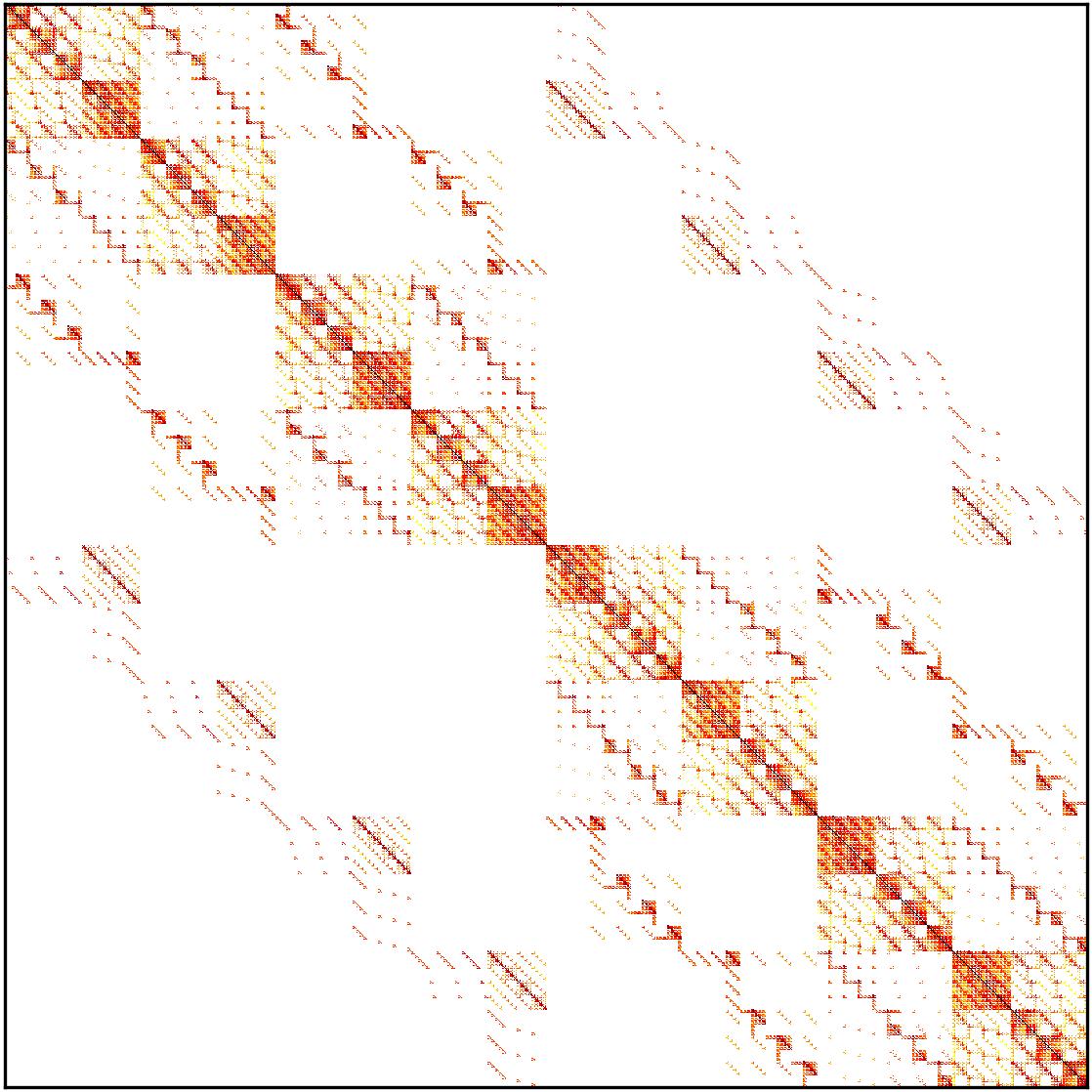} }
$-$
\vcenteredhbox{\includegraphics[width=0.138\textwidth]{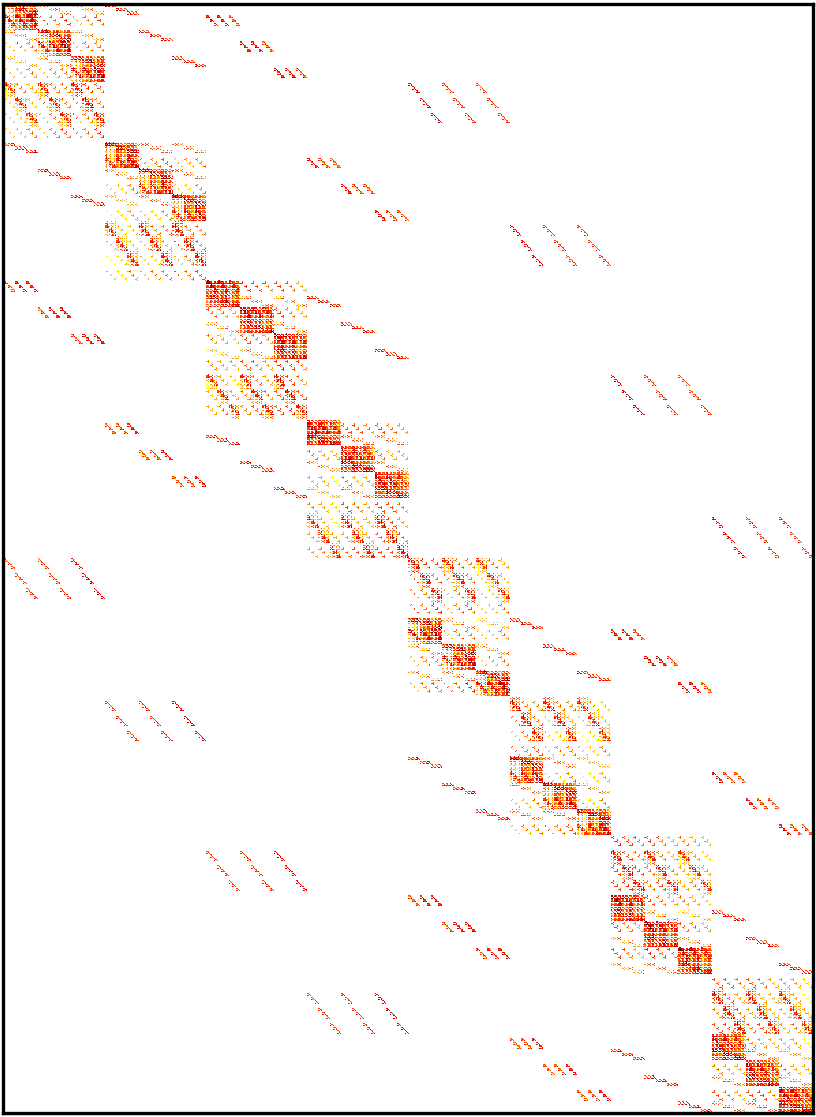} }
\vcenteredhbox{\includegraphics[width=0.138\textwidth]{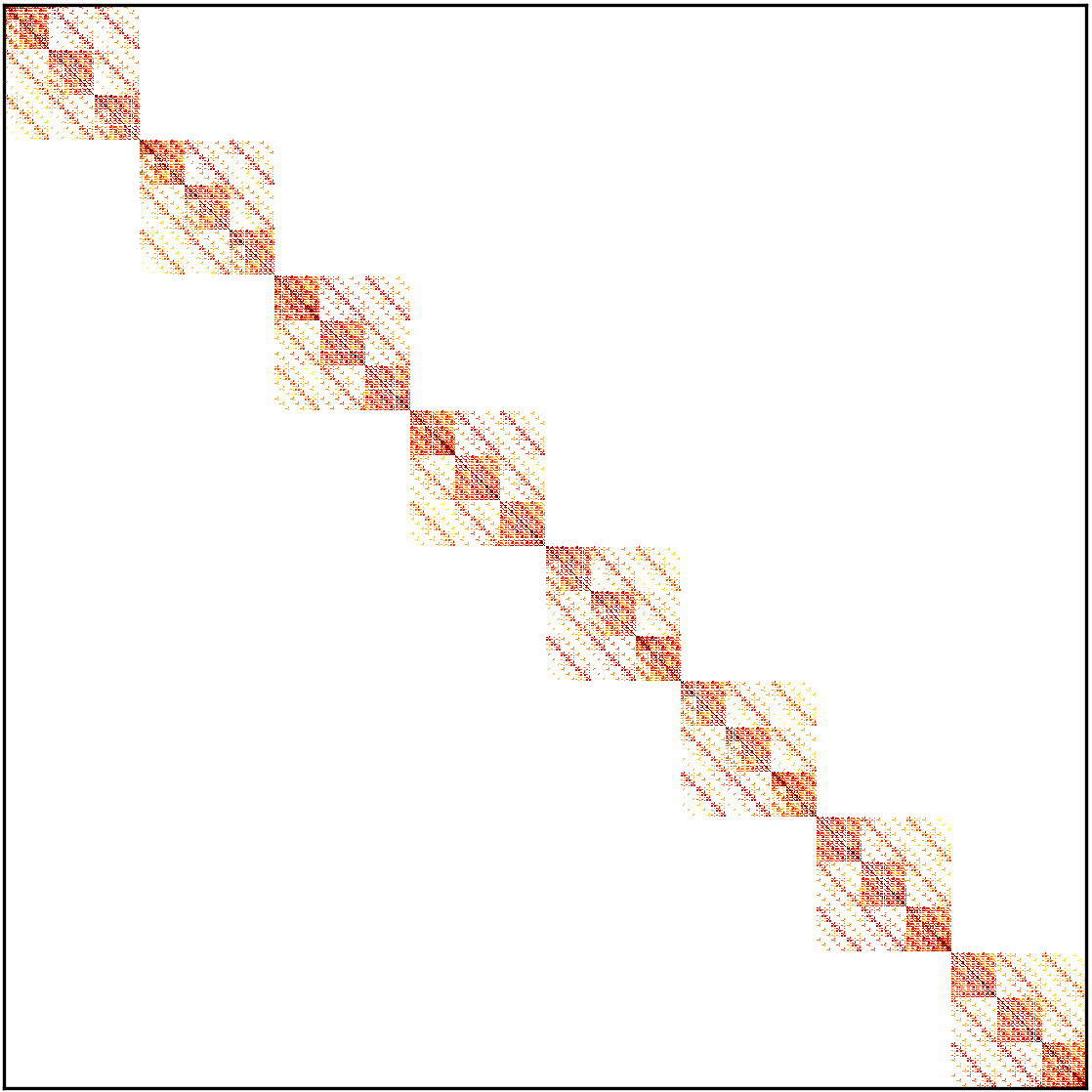} }
\addvbuffer[0.0cm 1.0cm]{$^{-1}$}
\vcenteredhbox{\includegraphics[width=0.187\textwidth]{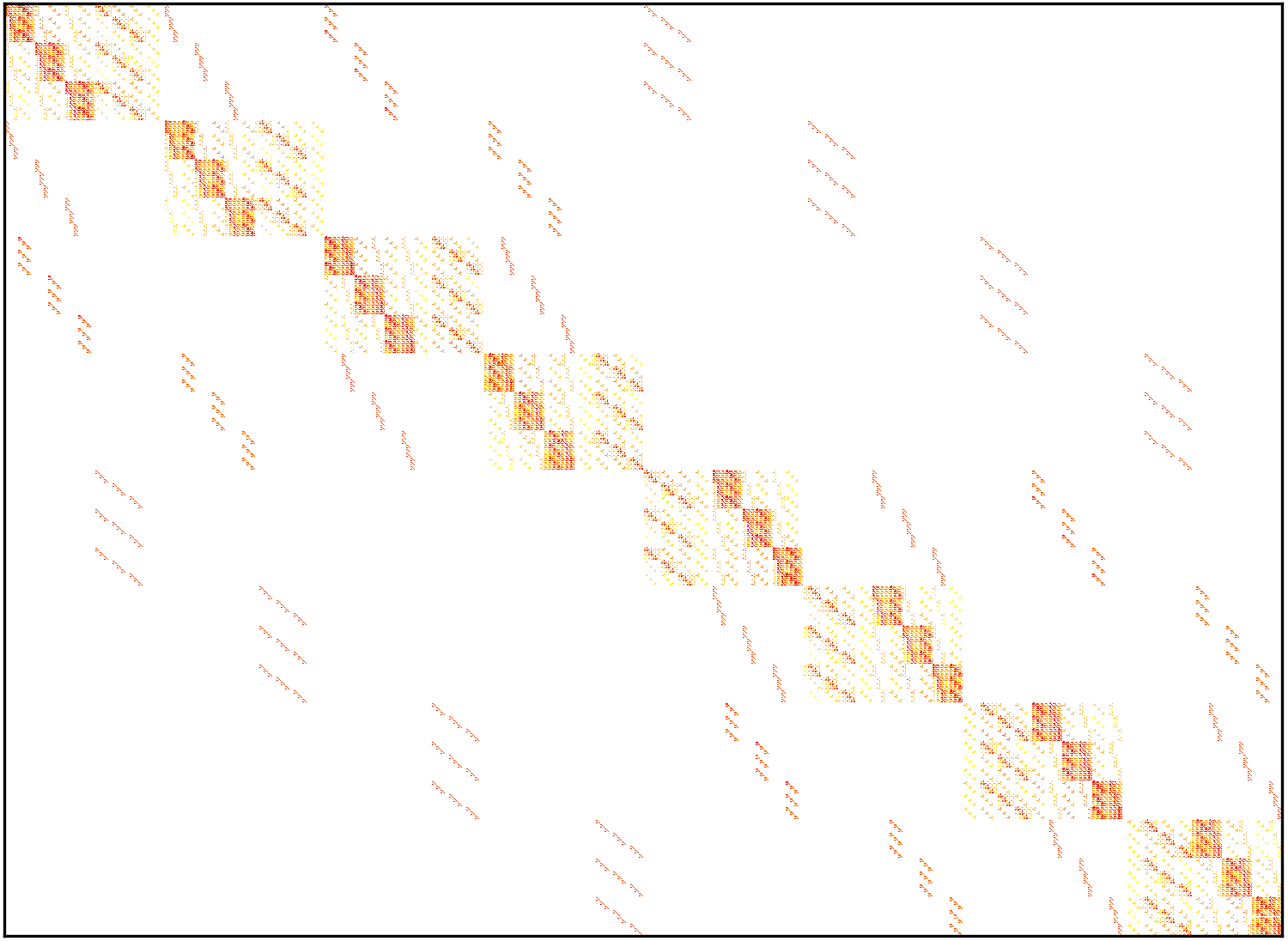} }
$=$
\vcenteredhbox{\includegraphics[width=0.187\textwidth]{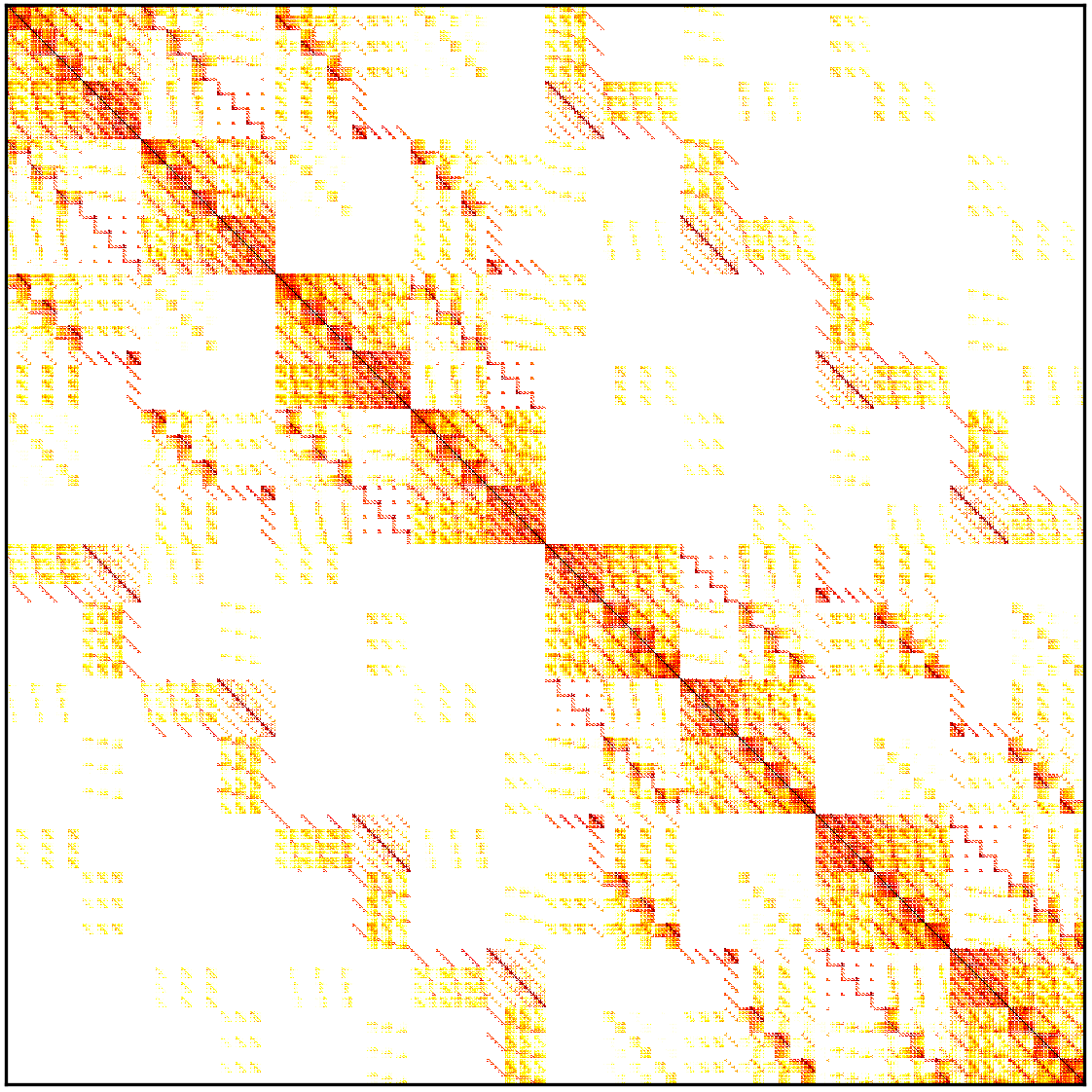} }

\includegraphics[width=0.5\textwidth]{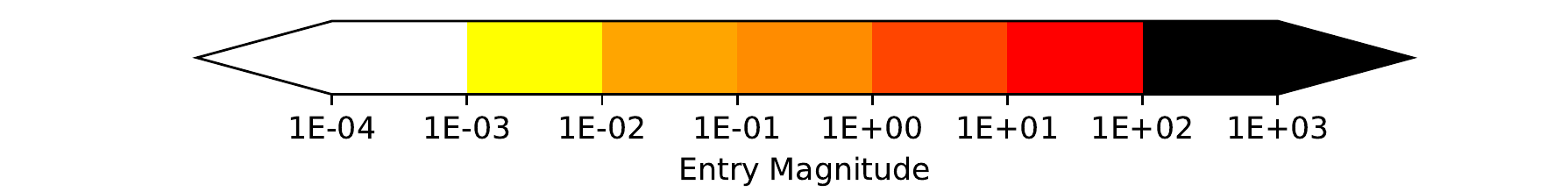} 
\caption{Matrix condensation operations for the 3D Navier-Stokes ($\ncons=5$) case: $\mat{A}_{bb} - \mat{A}_{ib} \mat{A}_{ii}^{-1} \mat{A}_{bi} = \mat{A}_C$. Every pixel corresponds to a matrix entry.} \label{fig:MatCondensation}
\end{figure}

The construction of the condensed system is graphically represented in Figure \ref{fig:MatCondensation} for the simple 3D mesh of Figure \ref{fig:Static_3Dmesh} and for the compressible Navier-Stokes equations ($\ncons = 5$).

A few remarks can be made. 
\begin{itemize}
\item The blocks of matrix $\mat{A}_{ii}$ keep the tensor-product sparsity and the whole matrix can be inverted locally (element by element).
\item The condensed matrix keeps the diagonal dominance with a seemingly denser structure.
\item The condensed system matrix exhibits element connectivities that were not spotted in the global matrices of last section: In spite of the use of a compact viscous numerical flux, there are off-diagonal entries that suggest a \textit{neighbors of neighbors} coupling. 
This behavior is not observed in the purely advective case (not shown here).
\end{itemize}

The ratio of the condensed system size ($n_1$) to the global system size ($n_1+n_2$) is a function of the polynomial order. 
Supposing a $p$-isotropic discretization and no physical boundaries, the ratio is
\begin{equation} \label{eq:ratio}
\frac{n_1}{n_1 + n_2} \bigg \rvert_{\max} = \frac{(N+1)^d - (N-1)^d}{(N+1)^d},
\end{equation}
where $d$ is the number of dimensions of the problem.
Table \ref{tab:ratio} shows the specific equations for $1 \le d \le 3$.
Note that an advantage in the system size is only observed for $N>1$.

\begin{table}[h]
\caption{Ratio of condensed system size to global system size}
\label{tab:ratio}
\begin{center}
\begin{tabular}{c|ccc}
$d=$	& $1$	& $2$	& $3$	\\
\hline
$\displaystyle \frac{n_1}{n_1 + n_2}$
		& $\displaystyle \frac{2}{N+1}$
			& $\displaystyle \frac{4N}{(N+1)^2}$
				& $\displaystyle \frac{6N^2+2}{(N+1)^3}$
\end{tabular}
\end{center}
\end{table}

When the simulation domain contains physical boundaries, the ratio of the condensed system size to the global system size can be lower than the value in \eqref{eq:ratio}.
Namely, from \eqref{eq:AdvOffDiag} and \eqref{eq:DiffOffDiag} it can be inferred that the degrees of freedom on physical boundaries do not contribute to the off-diagonal blocks and, therefore, it is not necessary to include them in the $\mathbf{Q}_b$ vector.
The ratio that is obtained by including the degrees of freedom on physical boundaries in the vector $\mathbf{Q}_i$ is called in this work the \textit{achievable ratio}.
The \textit{achievable ratio} is a problem-dependent quantity that is a function of how many element boundaries correspond to physical boundaries.

\begin{figure}[htbp]
\begin{center}
\includegraphics[width=0.5\textwidth]{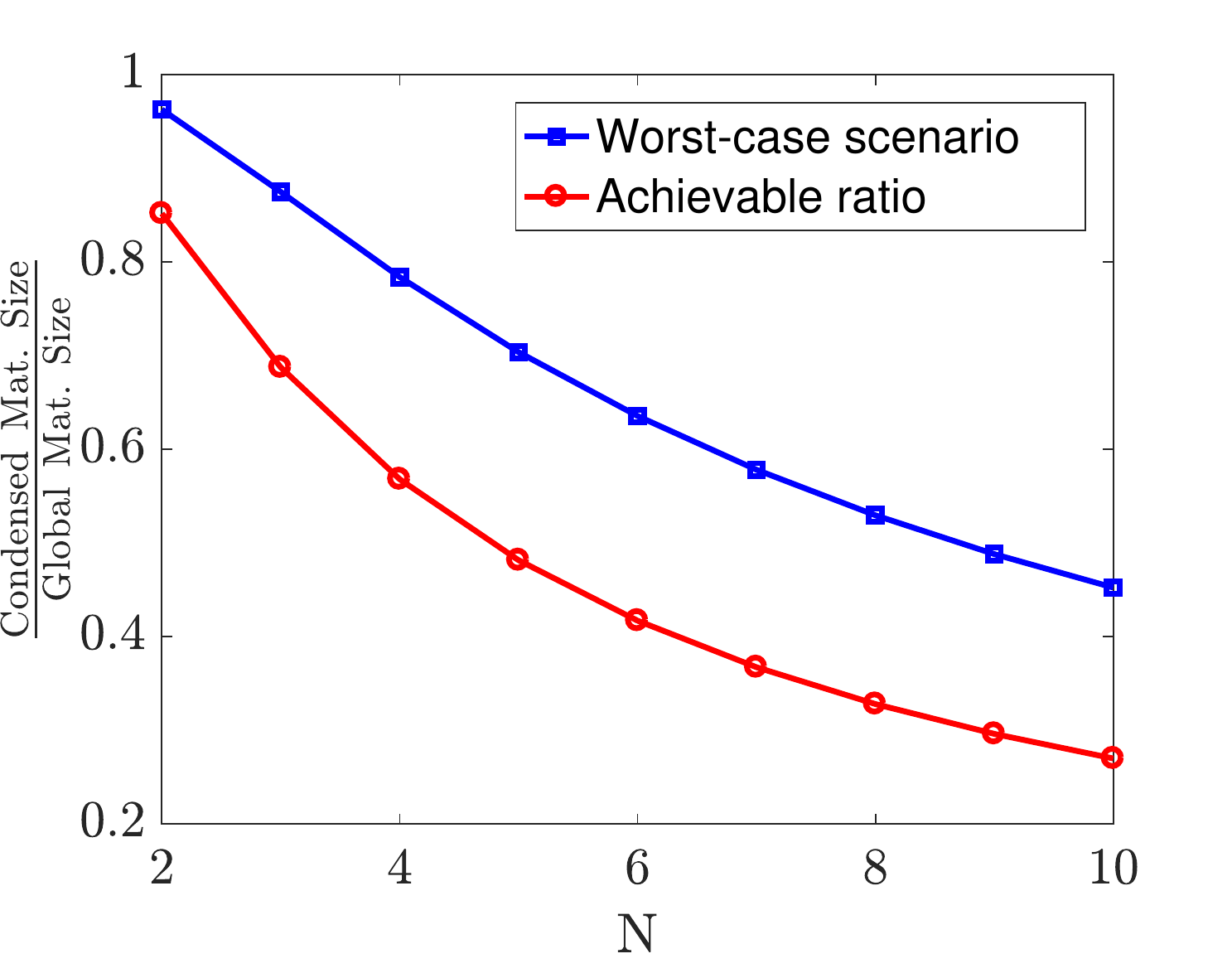} 
\caption{Ratio of condensed system size to global system size (3D diffuser example).} \label{fig:StaticSysSize}
\end{center}
\end{figure}

Figure \ref{fig:StaticSysSize} shows the ratio of the condensed system size to the global system size as a function of the polynomial order for $p$-isotropic discretizations in every element of the $d=3$ mesh of Figure \ref{fig:Static_3Dmesh}.
As expected, the static-condensation method provides increasing advantages as the polynomial order is incremented. 
The \textit{worst-case scenario} corresponds to the theoretical ratio that is computed with \eqref{eq:ratio}, supposing that the degrees of freedom on element boundaries are not condensed into $\mathbf{Q}_i$.
As can be seen, the \textit{achievable ratio} is much lower than the \textit{worst-case scenario} in this particular example.

In the following section, a numerical example is presented that was obtained using an implementation of the static-condensation method for the GL-DGSEM.
The Jacobian is computed analytically following the derivations of Section \ref{sec:ImplicitDGSEM} and stored in four matrices that correspond to $\mat{A}_{ii}$, $\mat{A}_{bi}$, $\mat{A}_{ib}$ and $\mat{A}_{bb}$.
To do that, the mesh connectivities are preprocessed to obtain appropriate permutation indexes for each degree of freedom.
The matrices $\mat{A}_{bi}$, $\mat{A}_{ib}$ and $\mat{A}_{bb}$ are stored in sparse CSR formats and the blocks of $\mat{A}_{ii}$ are stored as dense matrices. 
The matrix-matrix multiplications are performed with the routines provided by the BLAS libraries \cite{blackford2002updated} and the individual blocks of $\mat{A}_{ii}$ are inverted using the LU decomposition routines of the LAPACK \cite{lapack99} library with no regard of the tensor-product properties.

\section{Numerical Example} \label{sec:Static_Case}

\begin{figure}[htbp]
\begin{center}
\includegraphics[width=0.6\textwidth]{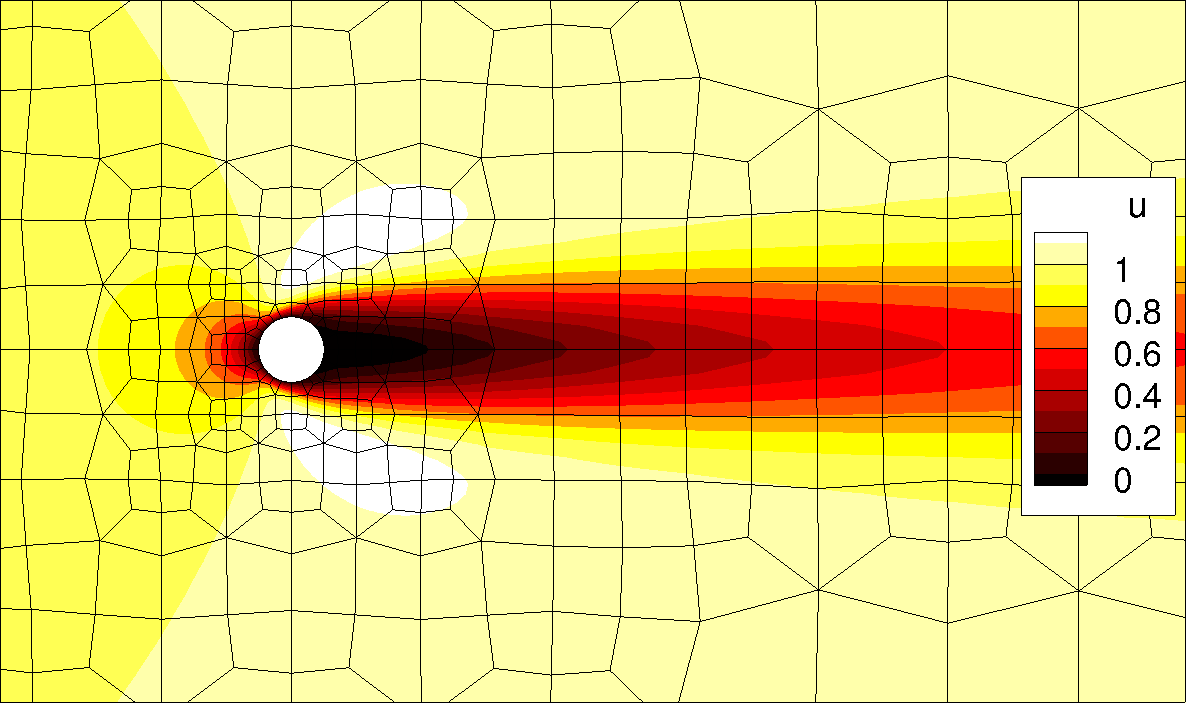}  
\caption{Cylinder test case.} \label{fig:CylStatic}
\end{center}
\end{figure}

In this section, we test the computational performance of the statically condensed time-implicit GL-DGSEM and compare it with the global (not statically condensed) time-implicit GL-DGSEM and a time-explicit GL-DGSEM.
The flow past a cylinder at $\Re = 30$ and $\Ma = 0.2$ is simulated using polynomial orders that range between $N=3$ and $N=7$. 
The Lax-Friedrichs flux is used as the advective numerical flux, $\numflux{f}^{\ a}$, and the symmetric interior penalty method is used as the diffusive numerical flux, $\numflux{f}^{\nu}$ and $\numflux{q}$.
All simulations were run on a $20$-core Intel(R) CPU $0000$ @ $2.20$GHz, and the Jacobians of the time-implicit simulations were computed analytically.
Figure \ref{fig:CylStatic} shows the horizontal velocity contours and the mesh used for this test case.

The time-explicit simulations use the Willamson's low-storage 3rd order Runge-Kutta method \cite{williamson1980low} as the time-marching scheme, where the time step is dynamically adapted according to the CFL condition \cite{Gassner2011}.
The time-implicit simulations use a BDF1 (backward Euler) scheme where the time-step size is fixed to $\Delta t = 1$ for simplicity, which corresponds to an approximate CFL of $\mathcal{O}(10^3)$.
In each case, the linear system (condensed and not) that results from the Newton linearized BDF1+DGSEM discretization is solved using the implementation of the parallel direct sparse solver (PARDISO) that is present in Intel's Math Kernel Library (MKL).

All simulations are restarted from an $N=3$ approximation after $10^4$ explicit RK3 time steps are taken, which is in turn started from a uniform flow condition.
The main reason is that that number of explicit time steps is computationally cheap to compute and that, after those $10^4$ time steps, the flow conditions are evolved enough to provide a Jacobian matrix that can be computed once and reused for multiple solves.
If the implicit simulations were started from a uniform flow condition, the Jacobian matrix would have to be computed several times at the beginning of the simulation, masking the performance of the implicit time-integration method.
All simulations are time-marched until reaching steady-state, which is assumed when the residual is $\norm{\mat{M}^{-1}\mathbf{H}}_{\infty} \le 10^{-9}$.

\begin{figure}[htbp]
\begin{center}
\includegraphics[width=0.6\textwidth]{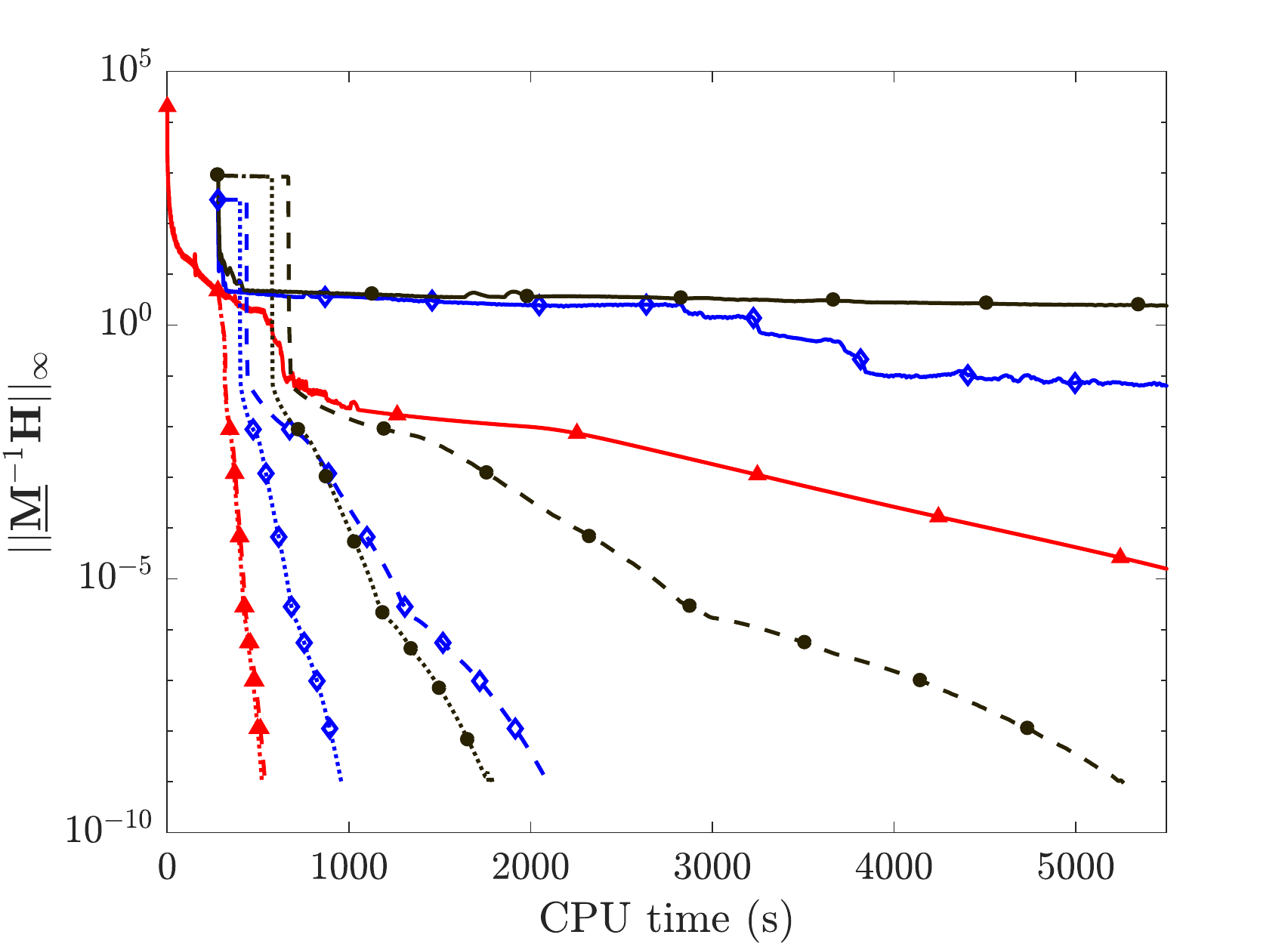}	
\begin{tabular}{c|ccc}
	  & RK3 & BDF1 & BDF1 + S.C. \\ \hline
$N=3$ & \drawline{red}{triangle*}{solid} & \drawline{red}{triangle*}{dashed} & \drawline{red}{triangle*}{dotted} \\
$N=5$ & \drawline{blue}{diamond}{solid}  & \drawline{blue}{diamond}{dashed}  & \drawline{blue}{diamond}{dotted}   \\
$N=7$ & \drawline{black}{*}{solid}       & \drawline{black}{*}{dashed}       & \drawline{black}{*}{dotted}       
\end{tabular}
\caption{Residual norm vs. CPU-time for the cylinder flow at $\Re=30$. Solid lines represent the purely explicit simulations (RK3), dashed lines are the implicit simulations solved globally (BDF1), and dotted lines are the implicit simulations solved with the static-condensation method (BDF1 + Static Condensation).} \label{fig:SCresidual}
\end{center}
\end{figure}

Figure \ref{fig:SCresidual} shows the evolution of the residual as a function of the elapsed CPU-time for the simulations of order $N=3$, $N=5$ and $N=7$.
Solid lines represent the purely explicit simulations (RK3), dashed lines are the implicit simulations solved globally (BDF1), and dotted lines are the implicit simulations solved with the static-condensation method (BDF1 + Static Condensation).
The convergence rate of the purely explicit simulations is very low, specially at high polynomial orders. 
In fact, the time-implicit methods are faster for all cases.
It is also noteworthy that there is a sudden increase in the residual after the high-order simulations are restarted from the $N = 3$ approximation. 
This increase corresponds to the unresolved high-order modes of the $N = 3$ approximation.

As can be seen in Figure \ref{fig:SCresidual}, there is a plateau after the sudden residual increase in the implicit simulations, which mainly corresponds to the Jacobian factorization times.
The Jacobian computation time also adds to the plateau, but it is negligible with respect to the computational cost of the LU decomposition.

Table \ref{tab:SpeedUp} shows the performance of the statically condensed and globally solved simulations with respect to the explicit simulations. 
It can be seen that speed-ups can be achieved with both methods, but that the statically condensed simulations excel when the polynomial order is increased.
The speed-up is as high as $207.8$ for $N=7$ with respect to the explicit method, which is about three times faster than the implicit method that solves the global system.

\begin{table}[htbp]
\begin{center}
\caption{Computation times and achieved speed-ups for implicit methods}
\label{tab:SpeedUp}
\begin{tabular}{c|cc|ccc}
  & \multicolumn{2}{c|}{BDF1} & \multicolumn{3}{c}{BDF1 + Static Condensation} \\ \hline
$N$ & CPU-Time(s) & Speed-up$^1$ & CPU-Time(s) & Speed-up$^1$ & Speed-up$^2$ \\ \hline
3 &  537.9 & 22.9 &  521.1 &  23.7 & 1.032 \\
5 & 2093.9 & 61.9 &  958.4 & 135.2 & 2.185 \\
7 & 5264.7 & 70.7 & 1790.1 & 207.8 & 2.941 \\ \hline
\multicolumn{6}{l}{\small $^1$ With respect to the explicit (RK3) simulation} \\
\multicolumn{6}{l}{\small $^2$ With respect to the implicit (BDF1) full Jacobian simulation}
\end{tabular}
\end{center}
\end{table}

The statically condensed simulations are computationally more efficient than their globally solved counterparts.
Namely, the extra computational resources that are needed for the condensation operations are rewarded with shorter computation times.
As can be seen in Table \ref{tab:Advantages}, it is more expensive to factorize the global Jacobian matrix than to statically condense it (i.e. to obtain the Schur complement) and then factorize it.
The main reason is that the traditional $LU$ factorization requires around $\mathcal{O}(n^3/3)$ operations \cite{trefethen1997numerical}. 
Although these operations are performed only once at the beginning of the simulation, a significant computational advantage is appreciated.

\begin{table} [htbp]
\begin{center}
\caption{Computation times in seconds for the operations that are performed once at the beginning of the simulation.}
\label{tab:Advantages}
\begin{tabular}{c|c|cc}
	& BDF1				& \multicolumn{2}{c}
	                            {BDF1 + Static Condensation}	\\ \hline
$N$	& LU factorization	& LU factorization	& Matrix cond. operations	\\ \hline
3	& 38.83				& 34.01				& 1.78  	\\
5	& 148.22			& 95.47 			& 13.95		\\
7	& 370.26			& 209.04			& 71.59		\\
\end{tabular}
\end{center}
\end{table}

The operations that are performed multiple times throughout the simulation are also cheaper for the statically condensed system.
Table \ref{tab:Advantages2} shows that the $LU$ solve (a forward and backward substitution used to obtain the solution of the linear system) is cheaper for the statically condensed system than for the global system.
The reason is that around $\mathcal{O}(n^2/2)$ operations are required for this operation \cite{trefethen1997numerical}.
In fact, the combined computational times of solving the condensed system and computing the condensed right hand sides of \eqref{eq:SC_step1} and \eqref{eq:SC_step2} are shorter than the single $LU$ solve for the global system.
As a result, the convergence rate, $\eta$, has a greater magnitude in the static condensation simulations.
The convergence rate is computed from the data in Figure \ref{fig:SCresidual} as the slope of the curves in the final part of the simulation,
\begin{equation}
\eta = \frac{\partial \log{\norm{\mat{M}^{-1}\mathbf{H}}_{\infty}}}{\partial \tau},
\end{equation}
where $\tau$ is the computational time.

\begin{table}
\begin{center}
\caption{Average CPU-times per time-step for the operations that are computed multiple times during the simulation, and resulting convergence rate.}
\label{tab:Advantages2}
\begin{tabular}{c|cc|ccc|c}
	& \multicolumn{2}{c}{BDF1}	
						& \multicolumn{2}{|c}
	                            {BDF1 + Static Condensation} &	\\ \hline
$N$	& LU solve time(s) 
			  & $\eta$ 
			  			& LU solve time(s)	
			  					 & Cond. Op. time(s) 
			  					 		  & $\eta_{S.C.}$
			  					 		  			& $\eta_{S.C}/\eta$	\\ \hline
3	& 0.7160  & -0.0361 & 0.5925 & 0.0302 & -0.0386 & 1.07\\ 
5	& 5.9638  & -0.0048 & 1.5634 & 0.2123 &	-0.0142	& 2.94\\ 
7	& 16.6578 & -0.0017 & 2.7071 & 0.9829 & -0.0066 & 3.92\\ 
\end{tabular}
\end{center}
\end{table}

\begin{figure}[htbp]
\begin{center}
\includegraphics[width=0.5\textwidth]{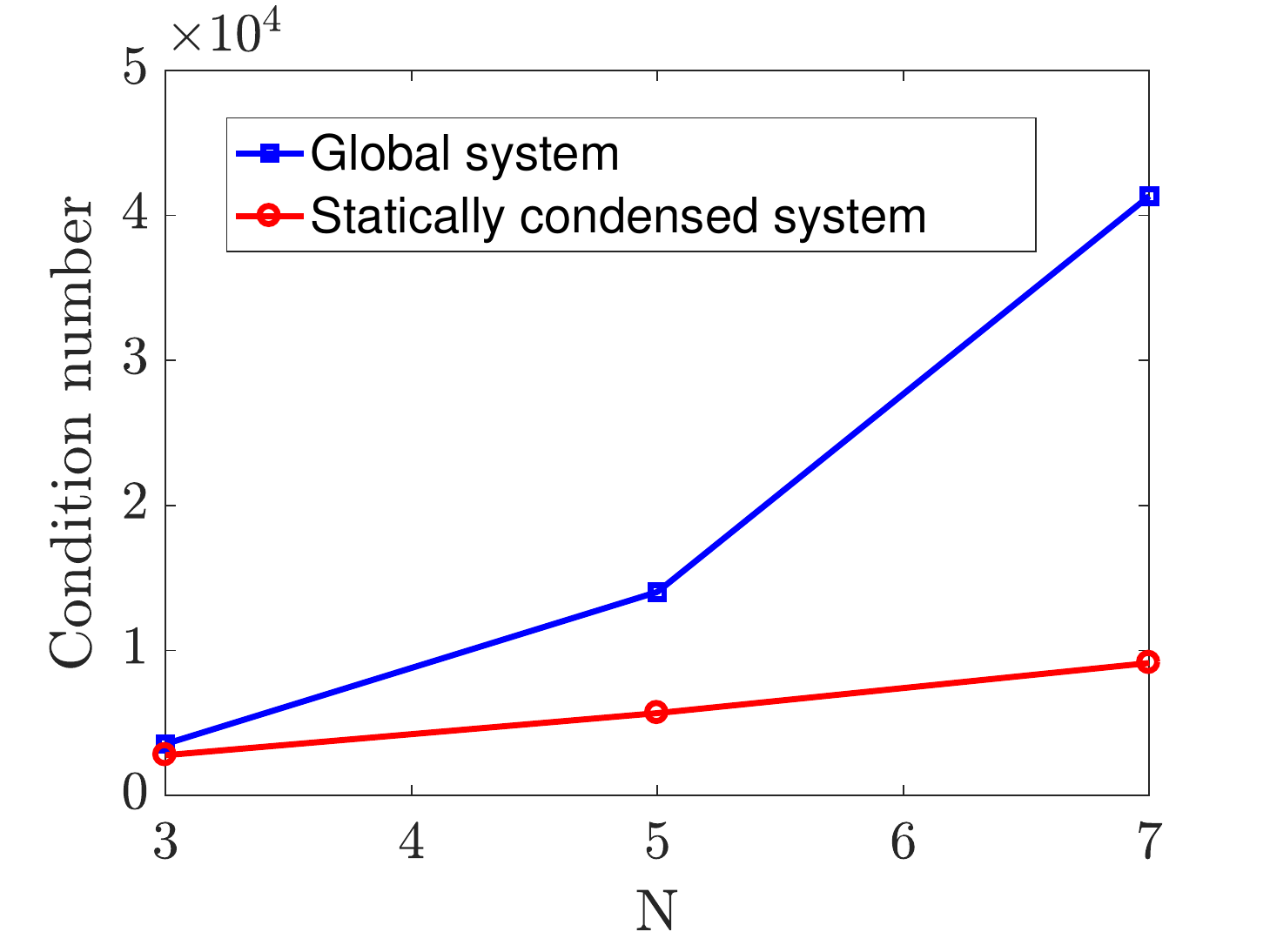} 
\caption{L2 condition number of the system matrix for the global system and the statically condensed one.} \label{fig:ConditionNumber}
\end{center}
\end{figure}

A further advantage of using static condensation is that the statically condensed system, besides being smaller in size, is better conditioned than the original global system.
This behavior was observed by Sherwin et al \cite{Sherwin2006} for their statically condensed DG method, and is shown in Figure \ref{fig:ConditionNumber} for the system matrices that come from the time-implicit GL-DGSEM discretization of the flow past a cylinder at $\Re=30$ and $\Ma=0.2$.

The $L_2$ condition number of the Jacobian matrices was approximated as the ratio of the largest and smallest eigenvalues, which were estimated using the shift-and-invert algorithm for sparse matrices that is implemented in the ARPACK library \cite{lehoucq1998arpack}.
In the example presented here, the maximum eigenvalue of the global and the statically condensed systems are very close.
The condition number of the latter is smaller, mainly because the minimum eigenvalue is moved to the left (away from the complex plane origin).

The lower condition number of the statically condensed linear systems suggests that the static-condensation method may also improve the convergence rate when using Krylov subspace linear solvers.
In fact, the convergence rate of a Krylov subspace method is directly related to the spectrum of the linear operator \cite{saad2003iterative}.

\begin{figure}[htbp]
\begin{center}
\subfigure[$N=3$]
	{
		\label{fig:SC_N3GMRES}
		\includegraphics[width=0.45\textwidth]{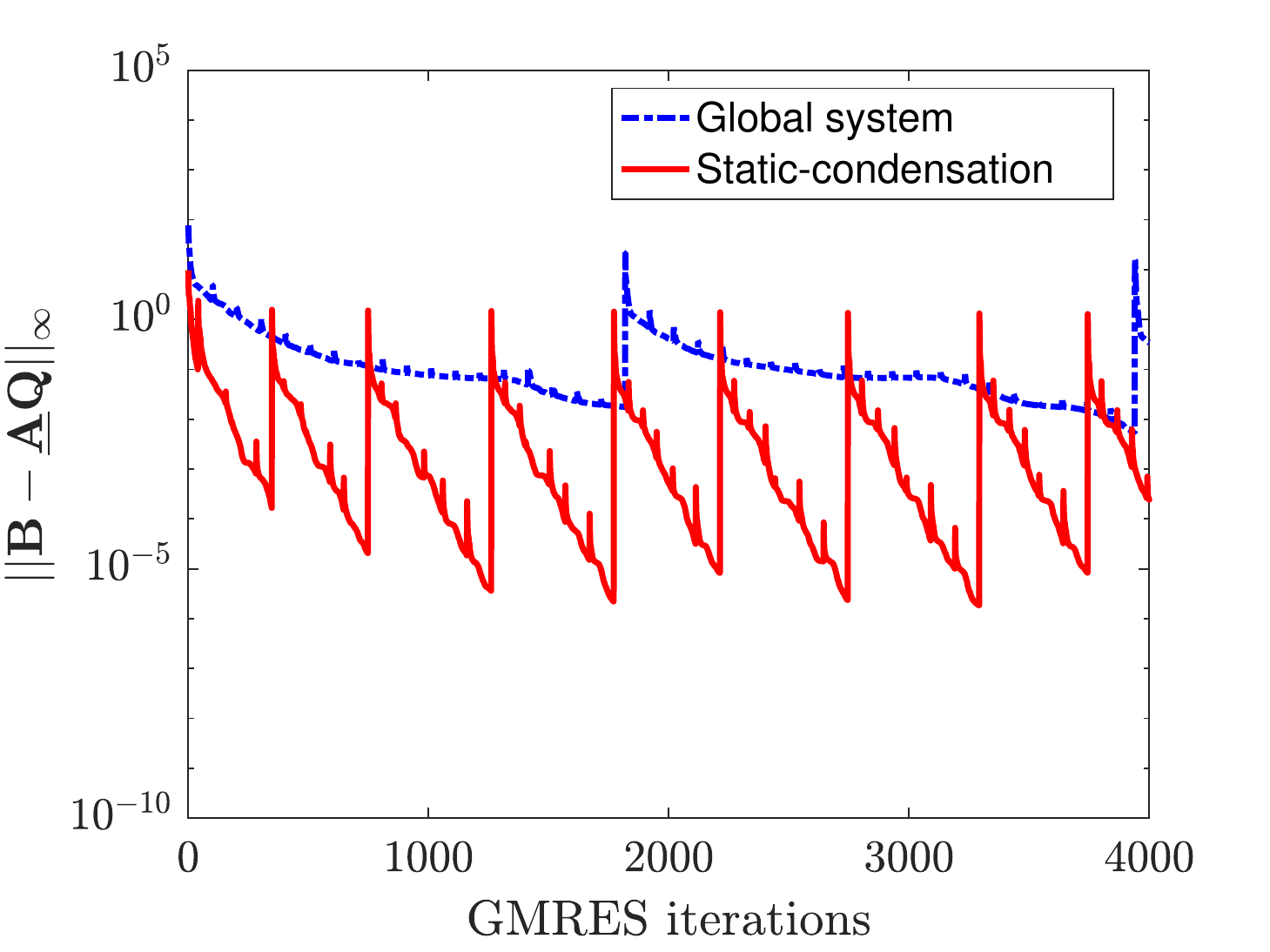}   	
	}
\subfigure[$N=5$]
	{
		\label{fig:SC_N5GMRES}
		\includegraphics[width=0.45\textwidth]{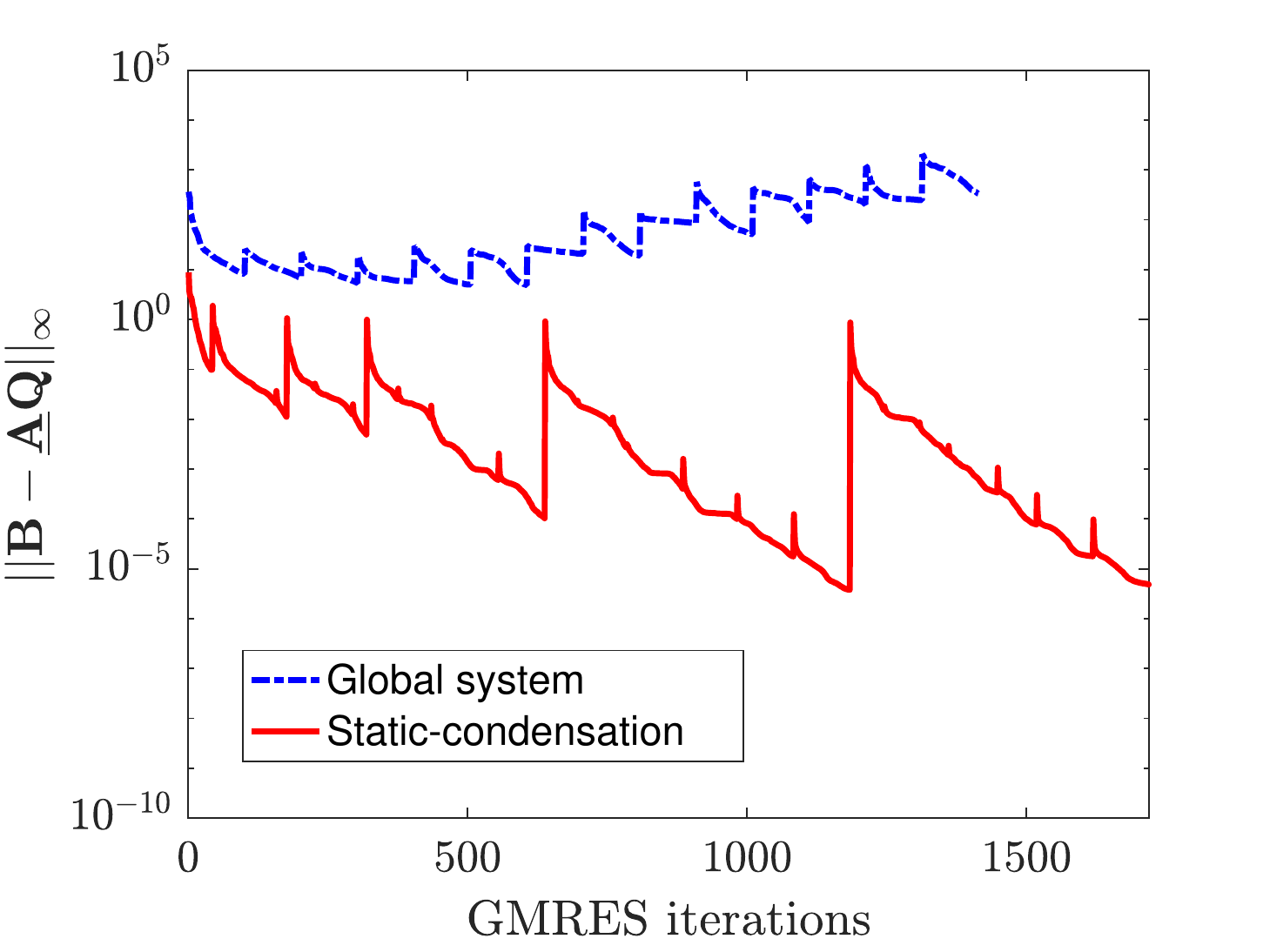}  
	}
\caption{Performance of the static-condensation method with a GMRES solver.} \label{fig:SC_GMRES}
\end{center}
\end{figure}

We next solve the linear system that arises in the cylinder test with the GMRES solver that is implemented in the PETSc library \cite{abhyankar2018petsc}, and a simple point Jacobi preconditioner.
Figure \ref{fig:SC_GMRES} shows the linear system residual as a function of the GMRES iterations for polynomial orders $N=3$ and $N=5$.
When a time step starts to be solved, a large increase in the linear system residual is exhibited, and every time a new Newton iteration starts there is a small increase in the linear system residual.

As can be seen in Figure  \ref{fig:SC_N3GMRES}, the $N=3$ statically condensed simulation takes several time steps while the globally solved one fails to converge the Newton method and reaches the maximum number of iterations allowed before starting the next time step.
This behavior is even more pronounced in the $N=5$ simulation, where the statically condensed simulation advances while the globally solved one diverges.

All in all, although a simple preconditioner is used (the point Jacobi preconditioner is known to be sub-optimal for high-order methods \cite{Ronquist1987,Persson2008}), the test shows that statically condensing the GL-DGSEM has a very positive impact in the convergence rate when using GMRES for the selected test case.

\section{Conclusions} \label{sec:Conclusions}

In this paper, we have shown that the DGSEM with Gauss-Lobatto nodes can be directly formulated as a Schur complement problem and solved in an efficient manner using static condensation, where the matrix that needs to be inverted to obtain the Schur complement is block-diagonal.
As a result, a statically condensed GL-DGSEM is presented that does not impose constraints on the choice of the basis functions or the form of the numerical fluxes (as do other statically condensed DG methods, such as Sherwin's \cite{Sherwin2006} or HDG \cite{Carrero2005}).

It is shown, by means of a numerical example with the compressible Navier-Stokes equations, that static condensation may produce speed-ups of up to $200$ when compared to the time-explicit GL-DGSEM, and speed-ups of up to three when compared with the time-implicit GL-DGSEM that solves the global system directly.

In addition, the statically condensed matrices of the examples presented are better conditioned than the global matrices from which they are constructed.
As a result, we can conclude that statically condensing the system provides robustness to the implicit time-discretization of the GL-DGSEM.

These findings constitute a further advantage of using GL-DGSEM over G-DGSEM. 
In summary, GL-DGSEM is computationally cheaper, simpler to implement, enables the formulation of provably stable \textit{uncrashable} schemes (as long as positivity is satisfied in nonlinear problems), allows larger time steps in time-explicit discretizations, enables unified formulations of certain viscous numerical fluxes, and we have now shown that it can be used to formulate a statically condensed DG method.

\section*{Acknowledgments}
This project has received funding from the European Union’s Horizon 2020 Research and Innovation Program under the Marie Skłodowska-Curie grant agreement No 675008 for the SSeMID project (Andrés M. Rueda-Ramírez and Eusebio Valero). 
This work was supported by a grant from the Simons Foundation ($\#426393$, David Kopriva).
This work has been partially supported by Ministerio de Economía y Competitividad under the research grant (EIN2019-103059, Gonzalo Rubio).
Andrés Rueda-Ramírez would like to thank Prof. Gustaaf Jacobs for his hospitality in San Diego, where part of this work was developed.

The authors acknowledge the computer resources and technical assistance provided by the \textit{Centro de Supercomputación y Visualización de Madrid} (CeSViMa).

\section*{References}
\bibliography{Biblio.bib}

\begin{appendices}
\section{Notation} \label{App:Notation}

We adopt the notation of \cite{Bohm2018,Gassner2018} to work with vectors of different nature. 
Spatial vectors are noted with an arrow on top (e.g. $\vec{x}=(x,y,z) \in \mathbb{R}^3$), state vectors are noted in bold (e.g. $\mathbf{q}=(\rho, \rho \vec{v}, \rho E)^T$), and block vectors, which contain a state vector in every spatial direction, are noted as
\begin{equation}
\blocktensor{f} =
\begin{bmatrix}
\mathbf{f}_1 \\ 
\mathbf{f}_2 \\
\mathbf{f}_3
\end{bmatrix} =
\mathbf{f}_1 \hat{\imath} + \mathbf{f}_2 \hat{\jmath} + \mathbf{f}_3 \hat{k}.
\end{equation}
Moreover, we note generic vectors with a bold uppercase letter. 
For instance, a vector that contains the state variables in all degrees of freedom is noted as $\mathbf{Q}=[\mathbf{q}_1, \mathbf{q}_2, \cdots , \mathbf{q}_{\NDOF}]^T$, where $\NDOF$ is the number of degrees of freedom.

The gradient of a state vector is a block vector:
\begin{equation}
\vec{\nabla} \mathbf{q} =
\begin{bmatrix}
\partial_x \mathbf{q} \\
\partial_y \mathbf{q} \\
\partial_z \mathbf{q} 
\end{bmatrix}
=
\partial_x \mathbf{q} \hat{\imath} + \partial_y \mathbf{q} \hat{\jmath} + \partial_z \mathbf{q} \hat{k},
\end{equation}
and the gradient of a spatial vector can be represented as a second order tensor, which can be written in matrix form using the outer product as
\begin{equation}
\mat{L} = 
\Nabla \vec{v} = 
\left( \Nabla \otimes \vec{v} \right)^T = 
\left( \Nabla \vec{v}^T \right)^T = 
\begin{bmatrix}
\bigpartialderiv{v_1}{x} & \bigpartialderiv{v_1}{y} & \bigpartialderiv{v_1}{z} \\
\bigpartialderiv{v_2}{x} & \bigpartialderiv{v_2}{y} & \bigpartialderiv{v_2}{z} \\
\bigpartialderiv{v_3}{x} & \bigpartialderiv{v_3}{y} & \bigpartialderiv{v_3}{z}
\end{bmatrix}.
\end{equation}
The underline is used throughout this work for second order tensors and matrices. 

Third order tensors are noted with a double underline, e.g. the flux derivative with respect to $\mathbf{q}$ can be expressed as
\begin{equation}
\bigpartialderiv{\blocktensor{f}}{\mathbf{q}} =
\thirdtensor{J} =
\begin{bmatrix}
\mat{J}_1 \\
\mat{J}_2 \\
\mat{J}_3
\end{bmatrix} 
=
\begin{bmatrix}
\bigpartialderiv{\mathbf{f}_1}{\mathbf{q}} \\
\bigpartialderiv{\mathbf{f}_2}{\mathbf{q}} \\
\bigpartialderiv{\mathbf{f}_3}{\mathbf{q}} \\
\end{bmatrix}.
\end{equation}

Similarly, fourth order tensors are noted with a triple underline. For example, the derivative of the flux with respect to $\Nabla \mathbf{q}$ can be written as
\begin{equation}
\bigpartialderiv{\blocktensor{f}}{ (\Nabla \mathbf{q}) } =
\fourthtensor{G} =
\begin{bmatrix}
\thirdtensor{G}{}_1 \\
\thirdtensor{G}{}_2 \\
\thirdtensor{G}{}_3
\end{bmatrix} 
=
\begin{bmatrix}
\bigpartialderiv{\mathbf{f}_1}{(\Nabla \mathbf{q})} \\
\bigpartialderiv{\mathbf{f}_2}{(\Nabla \mathbf{q})} \\
\bigpartialderiv{\mathbf{f}_3}{(\Nabla \mathbf{q})} 
\end{bmatrix}
=
\begin{bmatrix}
\bigpartialderiv{\mathbf{f}_1}{(\partial_x \mathbf{q})} 
	& \bigpartialderiv{\mathbf{f}_1}{(\partial_y \mathbf{q})} 
		& \bigpartialderiv{\mathbf{f}_1}{(\partial_z \mathbf{q})} \\
\bigpartialderiv{\mathbf{f}_2}{(\partial_x \mathbf{q})} 
	& \bigpartialderiv{\mathbf{f}_2}{(\partial_y \mathbf{q})} 
		& \bigpartialderiv{\mathbf{f}_2}{(\partial_z \mathbf{q})} \\
\bigpartialderiv{\mathbf{f}_3}{(\partial_x \mathbf{q})}  
	& \bigpartialderiv{\mathbf{f}_3}{(\partial_y \mathbf{q})}
		& \bigpartialderiv{\mathbf{f}_3}{(\partial_z \mathbf{q})} \\
\end{bmatrix}
=
\begin{bmatrix}
\mat{G}_{11} & \mat{G}_{12} & \mat{G}_{13} \\
\mat{G}_{21} & \mat{G}_{22} & \mat{G}_{23} \\
\mat{G}_{31} & \mat{G}_{32} & \mat{G}_{33}
\end{bmatrix}
.
\end{equation}

The dot (inner) product of two block vectors is a scalar,
\begin{equation}
\blocktensor{f} \cdot \blocktensor{g} 
= \sum_{i=1}^d \mathbf{f}_i \cdot \mathbf{g}_i,
= \sum_{i=1}^d \mathbf{f}_i^T \mathbf{g}_i.
\end{equation}
Moreover, the dot product of a spatial vector with a block vector is a state vector,
\begin{equation}
\vec{v} \cdot \blocktensor{f} = \sum_{i=1}^d v_i \mathbf{f}_i, \ \ 
\Nabla  \cdot \blocktensor{f} = \sum_{i=1}^d \partial_i \mathbf{f}_i.
\end{equation}

Finally, the product of a third order tensor by a state vector is a block vector,
\begin{equation}
\thirdtensor{J} \mathbf{q} = 
\begin{bmatrix}
\mat{J}_1 \\
\mat{J}_2 \\
\mat{J}_3
\end{bmatrix} 
\mathbf{q}
=
\begin{bmatrix}
\mat{J}_1 \mathbf{q} \\
\mat{J}_2 \mathbf{q} \\
\mat{J}_3 \mathbf{q}
\end{bmatrix},
\end{equation}
and the product of a fourth order tensor of $d$ columns by a block vector is also a block vector, for example,
\begin{equation}
\fourthtensor{G} \blocktensor{g} =
\fourthtensor{G} \Nabla \mathbf{q} =
\begin{bmatrix}
\mat{G}_{11} & \mat{G}_{12} & \mat{G}_{13} \\
\mat{G}_{21} & \mat{G}_{22} & \mat{G}_{23} \\
\mat{G}_{31} & \mat{G}_{32} & \mat{G}_{33}
\end{bmatrix}
\begin{bmatrix}
\mathbf{g}_1 \\
\mathbf{g}_2 \\
\mathbf{g}_3 
\end{bmatrix} =
\begin{bmatrix}
\mat{G}_{11} \mathbf{g}_1 + \mat{G}_{12} \mathbf{g}_2 + \mat{G}_{13} \mathbf{g}_3 \\
\mat{G}_{21} \mathbf{g}_1 + \mat{G}_{22} \mathbf{g}_2 + \mat{G}_{23} \mathbf{g}_3\\
\mat{G}_{31} \mathbf{g}_1 + \mat{G}_{32} \mathbf{g}_2 + \mat{G}_{33} \mathbf{g}_3
\end{bmatrix}.
\end{equation}

\section{The Compressible Navier-Stokes Equations} \label{App:NS}

The compressible Navier-Stokes equations in conservative form can be written in nondimensional form as
\begin{equation}
\partial_t \mathbf{q} + \Nabla \cdot \left( \blocktensor{f}^a - \blocktensor{f}^{\nu} \right) = \mathbf{s}, 
\end{equation}
where the conserved quantities are the mass, momentum and energy (per unit of volume), $\mathbf{q} = \left[ \rho, \rho \vec{v}, \rho E  \right]^T$, $\mathbf{s}$ is an external source term, and $\blocktensor{f}^a$ and $\blocktensor{f}^{\nu}$ are the advective and diffusive flux block vectors, respectively. 
The Euler equations of gas dynamics are defined in the same way, but omitting the viscous flux. The flux tensors can be written in compact form as
\begin{equation}
\blocktensor{f}^a (\mathbf{q})=
\begin{bmatrix}
\rho \vec{v} \\
\rho \vec{v} \otimes \vec{v} + \mat{I} p \\
\vec{v} (\rho E + p)
\end{bmatrix}
, \ \
\blocktensor{f}^{\nu} (\mathbf{q},\Nabla \mathbf{q}) = \frac {1} {\textrm{Re}_{\infty}}
\begin{bmatrix}
\vec{0} \\
\mat{\tau} \\
\mat{\tau} \vec{v} + \kappa \Nabla T
\end{bmatrix}.
\end{equation} 

Here, $\rho$ is the fluid's density, $\vec{v}$ is the velocity vector, $p$ is the (static) pressure, $E$ is specific the total energy (internal energy plus kinetic energy), $\mat{\tau}$ is the stress tensor, $T$ is the temperature, $\textrm{Re}_{\infty}$ is the reference Reynolds number, and $\kappa$ is a non-dimensional thermal conductivity.

The nondimensionalization is performed using reference values for the density ($\rho_{\infty}$), velocity ($V_{\infty}$), length ($L_{\infty}$), dynamic viscosity ($\mu_{\infty}$), and temperature ($T_{\infty}$), which give the definition of the Reynolds number,
\begin{equation}
\textrm{Re}_{\infty}=V_{\infty} L_{\infty} \rho_{\infty}/ \mu_{\infty},
\end{equation}
and of the nondimensional thermal conductivity,
\begin{equation}
\kappa = \frac{\mu}{(\gamma - 1) \Pr \ \Ma},
\end{equation}
where $\gamma=c_p/c_v$ is the heat capacity ratio, $c_p$ and $c_v$ are the heat capacities at constant pressure and constant volume, respectively, $\Ma = V_{\infty}/c_{\infty}$ is the reference Mach number, $c=\sqrt{\gamma p / \rho}$ is the speed of sound, $\Pr=c_p \mu / \kappa_{th}$ is the Prandtl number (assumed to be constant), and $\kappa_{th}$ is the thermal conductivity of the fluid. 

The pressure, $p$, is computed using the calorically perfect gas approximation,
\begin{equation}
p = (\gamma - 1) \rho e,
\end{equation}
where $e=E-\norm{\vec{v}}^2/2$ is the specific internal energy, and the stress tensor is computed using the Stokes hypothesis,
\begin{equation}
\mat{\tau} 
= \mu \left( \left( \Nabla \vec{v} \right)^T + \Nabla \vec{v} \right) 
- \lambda \Nabla \cdot \vec{v} \mat{I},
\end{equation}
with $\lambda = -\frac{2}{3} \mu$ the bulk viscosity coefficient and $\mat{I}$ a $3 \times 3$ identity matrix. 
In this paper, we choose the typical parameters for air: $\Pr = 0.72$, $\gamma = 1.4$, while $\mu$ is calculated using Sutherland law,
\begin{equation}
\mu = \frac{1 + T_{suth}/T_{\infty}} {T + T_{suth}/T_{\infty}} T^{\frac{3}{2}}
\end{equation}
where $T_{suth}=110.4 \rm{K}$ is the Sutherland's temperature.

For three dimensional flows, $\vec{v}=[u, v, w]^T$, the compressible Navier-Stokes equations imply the conservation of $\mathbf{q} = \left[ \rho, \rho u, \rho v , \rho w, \rho E  \right]^T$ under the influence of the advective fluxes,
\begin{equation}
\mathbf{f\ }_1^a =
          \begin{bmatrix}
           \rho u      \\           
           p + \rho u^2\\
           \rho u v    \\
           \rho u w    \\
           u (\rho E + p)
          \end{bmatrix},
          		\mathbf{f\ }_2^a = 
        				\begin{bmatrix}
				           \rho v      \\           
				           \rho u v    \\
           				   p + \rho v^2\\
				           \rho v w    \\
				           v (\rho E + p)
        				\end{bmatrix},
          						\mathbf{f\ }_3^a = 
          								\begin{bmatrix}
								           \rho w      \\           
								           \rho u w    \\
				           				   \rho v w    \\
								           p + \rho w^2\\
								           w (\rho E + p)
			          					\end{bmatrix},
\end{equation}
and the diffusive fluxes,
\begin{equation}
\mathbf{f\ }^{\nu}_1 = \frac {1} {\textrm{Re}_{\infty}}
          \begin{bmatrix}
           0           \\           
           \tau_{xx}   \\
           \tau_{xy}   \\
           \tau_{xz}   \\
           v_i \tau_{1i} + \kappa \partial_x T 
          \end{bmatrix}, 
          		\mathbf{f\ }^{\nu}_2 = \frac {1} {\textrm{Re}_{\infty}}
        				\begin{bmatrix}
				           0           \\           
				           \tau_{yx}   \\
				           \tau_{yy}   \\
				           \tau_{yz}   \\
				           v_i \tau_{2i} + \kappa \partial_y T \\
        				\end{bmatrix}, 
          						\mathbf{f\ }^{\nu}_3 = \frac {1} {\textrm{Re}_{\infty}}
          								\begin{bmatrix}
								           0           \\           
								           \tau_{zx}   \\
								           \tau_{zy}   \\
								           \tau_{zz}   \\
								           v_i \tau_{3i} + \kappa \partial_z T
			          					\end{bmatrix}.
\end{equation}

\subsection{Jacobians} \label{sec:NSJacobians}
This section contains the flux Jacobians for the 3D compressible Navier-Stokes equations.

\subsubsection{Advective Flux} \label{sec:NSAdvJacobians}
The derivative of the advective flux with respect to the conserved state is the third order tensor,
\begin{equation}
\thirdtensor{J}^{a}(\mathbf{q}) =
\bigpartialderiv{\blocktensor{f}^{a}}{ \mathbf{q}} 
\end{equation}
whose components are:
\begin{equation}
\mat{J}^{a}_{1} =
\begin{bmatrix}
 0		& 1		& 0 	& 0 	& 0 \\
-u^2+\frac{1}{2} + (\gamma - 1) \norm{\vec{v}}^2
		&  (3 - \gamma) u  
				& -(\gamma - 1) v 
						& -(\gamma - 1) w
								& (\gamma - 1) \\
 -uv 	& v 	& u 	& 0 	& 0 \\ 
 -uw 	& w 	& 0 	& u 	& 0 \\ 
u \left( \frac{1}{2} (\gamma - 1) \norm{\vec{v}}^2 - H \right)
 		& H - (\gamma - 1) u^2
 			 	& -(\gamma - 1) uv  
 			 			& -(\gamma - 1) uw 
 			 					& \gamma u \\
\end{bmatrix},
\end{equation}

\begin{equation}
\mat{J}^{a}_{2} =
\begin{bmatrix}
 0		& 0		& 1 	& 0 	& 0 \\
 -uv 	& v 	& u 	& 0 	& 0 \\ 
-v^2+\frac{1}{2} + (\gamma - 1) \norm{\vec{v}}^2
		&  -(\gamma - 1) u  
				& (3 - \gamma) v 
						& -(\gamma - 1) w
								& (\gamma - 1) \\
 -vw 	& 0 	& w 	& v 	& 0 \\ 
v \left( \frac{1}{2} (\gamma - 1) \norm{\vec{v}}^2 - H \right)
 		& -(\gamma - 1) uv  
 			 	& H - (\gamma - 1) v^2
 			 			& -(\gamma - 1) vw 
 			 					& \gamma v \\
\end{bmatrix},
\end{equation}

\begin{equation}
\mat{J}^{a}_{3} =
\begin{bmatrix}
 0		& 0		& 0 	& 1 	& 0 \\
 -uw 	& w 	& 0 	& u 	& 0 \\ 
 -vw 	& 0 	& w 	& v 	& 0 \\ 
-w^2+\frac{1}{2} + (\gamma - 1) \norm{\vec{v}}^2
		& -(\gamma - 1) u  
				& -(\gamma - 1) v 
						& (3 - \gamma) w
								& (\gamma - 1) \\
w \left( \frac{1}{2} (\gamma - 1) \norm{\vec{v}}^2 - H \right)
 		& -(\gamma - 1) uw 
 			 	& -(\gamma - 1) vw
 			 			& H - (\gamma - 1) w^2
 			 					& \gamma w \\
\end{bmatrix},
\end{equation}
where $H$ is the specific stagnation enthalpy:

\begin{equation}
H=E + \frac{p}{\rho}
\end{equation}

\subsubsection{Viscous Flux} \label{sec:NSViscJacobians}

Since the viscous flux, $\blocktensor{f}^{\nu} (\mathbf{q}, \Nabla \mathbf{q})$, depends on both the conserved variables, $\mathbf{q}$, and their gradients, $\Nabla \mathbf{q}$, it will be useful to define two kinds of gradients, $\thirdtensor{J}$ and $\fourthtensor{G}$.

Let us first consider the Jacobian with respect to $\Nabla \mathbf{q}$, when $\mathbf{q}$ is constant. The viscous flux of the compressible Navier-Stokes equations has a linear dependence on $\Nabla \mathbf{q}$, such that,
\begin{equation}
\blocktensor{f}^{\nu} (\mathbf{q}, \Nabla  \mathbf{q}) = \mat{G}_{ij}(\mathbf{q}) \frac{\partial \mathbf{q}}{\partial x_j} \hat{\imath}_i.
\end{equation}

Here, the matrices $\mat{G}_{ij}$ are components of the fourth order tensor,
\begin{equation}
\fourthtensor{G (\mathbf{q}) }  = \frac{\partial \blocktensor{f}^{\nu}}{\partial (\Nabla \mathbf{q})} \bigg\rvert_{\mathbf{q}=const}.
\end{equation}

The Jacobians for the flux in the $x$-direction are,

\begin{equation}
\mat{G}_{11}=\frac{\mu}{\rho Re}
\begin{bmatrix}
0 				& 0 			& 0 & 0 & 0 \\
-\frac{4}{3}u 	&  \frac{4}{3}  & 0 & 0 & 0 \\
      -v 		& 0 			& 1 & 0 & 0 \\
      -w 		& 0 			& 0 & 1 & 0 \\
 -(\frac{1}{3} u^2 + \norm{\vec{v}}^2 + \frac {\gamma}{\Pr}( E - \norm{\vec{v}}^2)) & u  (  \frac{4}{3} - \frac {\gamma}{\Pr}) & v (1 - \frac {\gamma}{\Pr}) & w (1 - \frac {\gamma}{\Pr}) & \frac {\gamma}{\Pr} \\
\end{bmatrix},
\end{equation}

\begin{equation}
\mat{G}_{12}=\frac{\mu}{\rho Re}
\begin{bmatrix}
 0 		  & 0 & 0 		& 0 & 0 \\
  \frac{2}{3} v  & 0 & -\frac{2}{3}   & 0 & 0 \\
      -u  & 1 & 0	 	& 0 & 0 \\
 0		  & 0 & 0		& 0 & 0 \\
 -\frac{1}{3} u v& v & -\frac{2}{3} u	& 0 & 0 \\
\end{bmatrix}
,
\mat{G}_{13}=\frac{\mu}{\rho Re}
\begin{bmatrix}
0&0&0&0&0 \\
  \frac{2}{3} w &0&0& -\frac{2}{3}   &0 \\
0&0&0&0&0 \\
      -u &1&0&0&0  \\
 -\frac{1}{3} u w &       w   &0& -\frac{2}{3} u   &0 \\
\end{bmatrix},
\end{equation}
the Jacobians for the flux in the $y$-direction are:
\begin{equation}
\mat{G}_{22}=\frac{\mu}{\rho Re}
\begin{bmatrix}
0 				& 0 & 0 			& 0 & 0 \\
      -u	 	& 1 & 0 			& 0 & 0 \\
  -\frac{4}{3}v	& 0 & \frac{4}{3} 	& 0 & 0 \\
      -w 		& 0 & 0 			& 1 & 0 \\
 -(\frac{1}{3} v^2 + \norm{\vec{v}}^2 + \frac {\gamma}{\Pr}( E - \norm{\vec{v}}^2)) 
	& u  (  1 - \frac {\gamma}{\Pr}) 
		& v ( \frac{4}{3} - \frac {\gamma}{\Pr}) 
			& w ( 1 - \frac {\gamma}{\Pr}) 
				& \frac {\gamma}{\Pr} \\
\end{bmatrix},
\end{equation}

\begin{equation}
\mat{G}_{21}=\frac{\mu}{\rho Re}
\begin{bmatrix}
 0 		  		& 0 			& 0 & 0 & 0 \\
 -v		  		& 0 			& 1 & 0 & 0 \\
 \frac{2}{3} u  & -\frac{2}{3}  & 0	& 0 & 0 \\
 0		  		& 0				& 0	& 0 & 0 \\
-\frac{1}{3} uv & -\frac{2}{3}v & u	& 0 & 0 \\
\end{bmatrix}
,
\mat{G}_{23}=\frac{\mu}{\rho Re}
\begin{bmatrix}
 0 		  		& 0 & 0 & 0 			& 0 \\
 0		  		& 0 & 0 & 0 			& 0 \\
 \frac{2}{3} w  & 0 & 0 & -\frac{2}{3}  & 0 \\
 -v		  		& 0 & 1 & 0				& 0 \\
-\frac{1}{3} vw & 0 & w & -\frac{2}{3}v & 0 \\
\end{bmatrix},
\end{equation}
and the Jacobians for the flux in the $z$-direction are,
\begin{equation}
\mat{G}_{31}=\frac{\mu}{\rho Re}
\begin{bmatrix}
 0 		  		& 0 			& 0 & 0 & 0 \\
 -w		  		& 0 			& 0 & 1 & 0 \\
 0  			& 0			 	& 0	& 0 & 0 \\
 \frac{2}{3} u	& -\frac{2}{3} 	& 0	& 0 & 0 \\
-\frac{1}{3} uw & -\frac{2}{3}w & 0	& u & 0 \\
\end{bmatrix}
,
\mat{G}_{32}=\frac{\mu}{\rho Re}
\begin{bmatrix}
 0 		  		& 0 &  0 			 & 0 & 0 \\
 0		  		& 0 &  0 			 & 0 & 0 \\
 -w		  	 	& 0 &  0			 & 1 & 0 \\
 \frac{2}{3} v 	& 0 &  -\frac{2}{3}  & 0 & 0 \\
-\frac{1}{3} vw & 0 &  -\frac{2}{3}w & v & 0 \\
\end{bmatrix},
\end{equation}

\begin{equation}
\mat{G}_{33}=\frac{\mu}{\rho Re}
\begin{bmatrix}
0 				& 0 & 0 			& 0 & 0 \\
      -u	 	& 1 & 0 			& 0 & 0 \\
      -v 		& 0 & 1 			& 0 & 0 \\
 -\frac{4}{3}w 	& 0 & 0 			& \frac{4}{3} & 0 \\
 -(\frac{1}{3} w^2 + \norm{\vec{v}}^2 + \frac {\gamma}{\Pr}( E - \norm{\vec{v}}^2)) 
	& u  (  1 - \frac {\gamma}{\Pr}) 
		& v ( 1 - \frac {\gamma}{\Pr}) 
			& w ( \frac{4}{3} - \frac {\gamma}{\Pr}) 
				& \frac {\gamma}{\Pr} \\
\end{bmatrix}.
\end{equation}

On the other hand, the Jacobian with respect to $\mathbf{q}$ is the third order tensor,
\begin{equation}
\thirdtensor{J}^{\nu}(\mathbf{q},\Nabla \mathbf{q}) =
\bigpartialderiv{\blocktensor{f}^{\nu}}{ \mathbf{q}} \bigg\rvert_{\Nabla \mathbf{q}=const},
\end{equation}
which is defined in terms of the derivative of Sutherland's law, 
\begin{equation}
\frac{\partial \mu}{\partial \mathbf{q}} = 
\frac{1 + T_{suth}/T_{\infty}} {T + T_{suth}/T_{\infty}} \sqrt{T} \left( \frac{3}{2} - \frac{T}{T + T_{suth}/T_{\infty} }\right) \frac{\partial T}{\partial \mathbf{q}},
\end{equation}
where
\begin{equation}
\frac{\partial T}{\partial \mathbf{q}} = 
\begin{bmatrix}
\norm{\vec{v}}^2 - E & -u & -v & -w & 1
\end{bmatrix}.
\end{equation}

The components of $\thirdtensor{J}^{\nu}$ are written in equations \eqref{eq:Jnu1}, \eqref{eq:Jnu2} and \eqref{eq:Jnu3}.

\begin{landscape}%
\begin{equation} \label{eq:Jnu1} 
\mat{J}^{\nu}_1 = \frac{\mu}{\rho^2 \Re}
\left[
\begin{array}{c;{1pt/1pt}c;{1pt/1pt}c;{1pt/1pt}c;{1pt/1pt}c}
0 & 0 & 0 & 0 & 0\\ \hdashline[1pt/1pt]
2 \lambda \left( \rho \Nabla \cdot \vec{v} -\vec{v} \cdot \Nabla \rho \right) + 2 \left( u \rho_x - \rho u_x \right) 
	& - 4 \lambda \rho_x 
		& 2 \lambda \rho_y 
			& 2 \lambda \rho_z 
				& 0 \\  \hdashline[1pt/1pt]
u \rho_y  + v \rho_x - \rho \left( u_y + v_x \right) 
	& - \rho_y 
		& -\rho_x 
			& 0 
				& 0 \\ \hdashline[1pt/1pt]
u \rho_z  + w \rho_x - \rho \left( u_z + w_x \right) 
	& - \rho_z 
		& 0 
			& -\rho_x 
				& 0 \\ \hdashline[1pt/1pt]
\begin{smallmatrix} 
\left( 1-\frac{\gamma}{\Pr} \right) \left( \norm{\vec{v}}^2 \rho_x - 2 \rho \vec{v} \cdot \vec{v}_x \right) + \lambda u \left( 4\rho \Nabla \cdot \vec{v} + \vec{v} \cdot \Nabla \rho \right) \\ - 2 \rho \vec{v} \cdot \Nabla u + \frac{\gamma}{\Pr} \left( E \rho_x - \rho E_x \right) 
\end{smallmatrix}

	& 
	\begin{smallmatrix} -u \left( \lambda - \frac{\gamma}{\Pr} \right) \rho_x + \rho \left( 2 - \frac{\gamma}{\Pr} \right) u_x \\ - \vec{v} \cdot 	\Nabla \rho - 2 \lambda \rho \Nabla \cdot \vec{v} 
	\end{smallmatrix}
	
		& 
		\begin{smallmatrix} \left( 1 - \frac{\gamma}{\Pr} \right) \left( \rho v_x - v \rho_x \right) \\ + \rho u_y + 2 \lambda u \rho_y  
		\end{smallmatrix}
			& 
			\begin{smallmatrix} \left( 1 - \frac{\gamma}{\Pr} \right) \left( \rho w_x - w \rho_x \right) \\ + \rho u_z + 2 \lambda u \rho_z  
			\end{smallmatrix}
				& -\frac{\gamma}{\Pr} \rho_x
			
\end{array}
\right]
+ \mathbf{f\ }_1^{\nu} \otimes \frac{\partial \mu}{\partial \mathbf{q}},
\end{equation}

\begin{equation} \label{eq:Jnu2} 
\mat{J}^{\nu}_2 = \frac{\mu}{\rho^2 \Re}
\left[
\begin{array}{c;{1pt/1pt}c;{1pt/1pt}c;{1pt/1pt}c;{1pt/1pt}c}
0 & 0 & 0 & 0 & 0\\ \hdashline[1pt/1pt]
v \rho_x  + u \rho_y - \rho \left( v_x + u_y \right) 
	& -\rho_y
		& - \rho_x
			& 0
				& 0 \\  \hdashline[1pt/1pt]
2 \lambda \left( \rho \Nabla \cdot \vec{v} -\vec{v} \cdot \Nabla \rho \right) + 2 \left( v \rho_y - \rho v_y \right)
	& 2 \lambda \rho_x
		& - 4 \lambda \rho_y
			& 2 \lambda \rho_z 
				& 0 \\ \hdashline[1pt/1pt]
v \rho_z  + w \rho_y - \rho \left( v_z + w_y \right)
	& 0
		& - \rho_z  
			& -\rho_y
				& 0 \\ \hdashline[1pt/1pt]
				
\begin{smallmatrix} \left( 1-\frac{\gamma}{\Pr} \right) \left( \norm{\vec{v}}^2 \rho_y - 2 \rho \vec{v} \cdot \vec{v}_y \right) + \lambda v \left( 4\rho \Nabla \cdot \vec{v} + \vec{v} \cdot \Nabla \rho \right) \\ - 2 \rho \vec{v} \cdot \Nabla v + \frac{\gamma}{\Pr} \left( E \rho_y - \rho E_y \right) \end{smallmatrix}

	& \begin{smallmatrix} \left( 1 - \frac{\gamma}{\Pr} \right) \left( \rho u_y - u \rho_y \right) \\ + \rho v_x + 2 \lambda v \rho_x  \end{smallmatrix}
	
		& \begin{smallmatrix} -v \left( \lambda - \frac{\gamma}{\Pr} \right) \rho_y + \rho \left( 2 - \frac{\gamma}{\Pr} \right) v_y \\ - \vec{v} \cdot 	\Nabla \rho - 2 \lambda \rho \Nabla \cdot \vec{v} \end{smallmatrix}
		
			& \begin{smallmatrix} \left( 1 - \frac{\gamma}{\Pr} \right) \left( \rho w_y - w \rho_y \right) \\ + \rho v_z + 2 \lambda v \rho_z  \end{smallmatrix}
				& -\frac{\gamma}{\Pr} \rho_y
			
\end{array}
\right]
+ \mathbf{f\ }_2^{\nu} \otimes \frac{\partial \mu}{\partial \mathbf{q}},
\end{equation}

\begin{equation} \label{eq:Jnu3} 
\mat{J}^{\nu}_3 = \frac{\mu}{\rho^2 \Re}
\left[
\begin{array}{c;{1pt/1pt}c;{1pt/1pt}c;{1pt/1pt}c;{1pt/1pt}c}
0 & 0 & 0 & 0 & 0\\ \hdashline[1pt/1pt]
w \rho_x  + u \rho_z - \rho \left( w_x + u_z \right) 
	& -\rho_z
		& 0
			& - \rho_x
				& 0 \\  \hdashline[1pt/1pt]

w \rho_y  + v \rho_z - \rho \left( w_y + v_z \right)
	& 0
		& -\rho_z
			& - \rho_y
				& 0 \\ \hdashline[1pt/1pt]

2 \lambda \left( \rho \Nabla \cdot \vec{v} -\vec{v} \cdot \Nabla \rho \right) + 2 \left( w \rho_z - \rho w_z \right)
	& 2 \lambda \rho_x
		& 2 \lambda \rho_y 
			& - 4 \lambda \rho_z
				& 0 \\ \hdashline[1pt/1pt]

\begin{smallmatrix} 
\left( 1-\frac{\gamma}{\Pr} \right) \left( \norm{\vec{v}}^2 \rho_z - 2 \rho \vec{v} \cdot \vec{v}_z \right) + \lambda w \left( 4\rho \Nabla \cdot \vec{v} + \vec{v} \cdot \Nabla \rho \right) \\ - 2 \rho \vec{v} \cdot \Nabla w + \frac{\gamma}{\Pr} \left( E \rho_z - \rho E_z \right) 
\end{smallmatrix}

	& \begin{smallmatrix} \left( 1 - \frac{\gamma}{\Pr} \right) \left( \rho u_z - u \rho_z \right) \\ + \rho w_x + 2 \lambda w \rho_x  \end{smallmatrix}
	
		& \begin{smallmatrix} \left( 1 - \frac{\gamma}{\Pr} \right) \left( \rho v_z - v \rho_z \right) \\ + \rho w_y + 2 \lambda w \rho_y  \end{smallmatrix}
		
			& \begin{smallmatrix} -w \left( \lambda - \frac{\gamma}{\Pr} \right) \rho_z + \rho \left( 2 - \frac{\gamma}{\Pr} \right) w_z \\ - \vec{v} \cdot 	\Nabla \rho - 2 \lambda \rho \Nabla \cdot \vec{v} \end{smallmatrix}
			
				& -\frac{\gamma}{\Pr} \rho_z
			
\end{array}
\right]
+ \mathbf{f\ }_3^{\nu} \otimes \frac{\partial \mu}{\partial \mathbf{q}}.
\end{equation}

\end{landscape}

\section{The Discontinuous Galerkin Spectral Element Method} \label{App:DGSEM}

In this section, we obtain the DGSEM discretization of the advection-diffusion system, \eqref{eq:SystemAdvDiff},
\begin{subequations}\label{eq:SystemAdvDiff2}
\begin{empheq}[left=\empheqlbrace]{align} 
\partial_t \mathbf{q} 
+ \Nabla \cdot \left( \blocktensor{f}^a (\mathbf{q}) - \blocktensor{f}^{\nu} (\mathbf{q}, \blocktensor{g}) \right)&= \mathbf{0} \ , \ \text{in } \Omega,  \label{eq:OuterEq2} \\
\Nabla \mathbf{q}
&=  \blocktensor{g}  , \  \text{in } \Omega. \label{eq:InnerEq2}
\end{empheq}
\end{subequations}

We start by multiplying \eqref{eq:OuterEq2} by an arbitrary and smooth test function, $\mathbf{v}$, and integrating by parts over the domain, $\Omega$,
\begin{equation} \label{eq:weak}
\int_{\Omega} \partial_t \mathbf{q} \cdot \mathbf{v} \textrm{d} \Omega 
- \int _{\Omega} \blocktensor{f} \cdot \Nabla \mathbf{v} \textrm{d} \Omega 
+ \int_{\partial \Omega} \left( \blocktensor{f} \cdot \vec{n} \right) \cdot \mathbf{v} \textrm{d} S 
= \mathbf{0},
\end{equation}
where $\vec{n}$ is the normal unit vector on the boundary $\partial \Omega$. 

Let the domain $\Omega$ be approximated by a tessellation $\mathscr{T} = \lbrace e \rbrace$, i.e. a combination of $K$ spectral elements $e$ of domain $\Omega^e$ and boundary $\partial \Omega^e$. 
Moreover, let $\mathbf{q}$, $\blocktensor{f}$ and $\mathbf{v}$ be approximated by piece-wise polynomial functions $\mathbf{q}^N$, $\blocktensor{f}^N$ and $\mathbf{v}^N$ (which are continuous in each element) defined in the space of $L^2$ functions
\begin{equation}
\mathscr{V}^N = \lbrace \mathbf{v}^N \in L^2(\Omega) : \mathbf{v}^N\vert_{\Omega^e} \in \mathscr{P}^N(\Omega^e) \ \ \forall \ \Omega^e \in \mathscr{T} \rbrace,
\end{equation}
where $\mathscr{P}^N(\Omega^e)$ is the space of polynomials of degree at most $N$ defined in $\Omega^e$, the domain of element $e$. 

Since the functions in $\mathscr{V}^N$ may be discontinuous at element interfaces, the quantity $\blocktensor{f}^N \cdot \vec{n}$ is not uniquely defined at the element traces. 
Therefore, it is replaced by a numerical flux function, 
\begin{equation}
\blocktensor{f}^N \cdot \vec{n} \leftarrow \numflux{f} = \numflux{f}^a - \numflux{f}^{\nu},
\end{equation}
which allows one to uniquely define the flux at the element interfaces and to weakly prescribe the boundary data as a function of the normal vector and the state on both sides of the boundary/interface.
Equation \eqref{eq:weak} can then be rewritten for each element as
\begin{equation} \label{eq:weak2}
\int _{\Omega^e} {\partial_t \mathbf{q}}^N \cdot {\mathbf{v}}^N \textrm{d} \Omega^e 
- \int_{\Omega^e} {\blocktensor{f}}^N \cdot \Nabla {\mathbf{v}}^N \textrm{d} \Omega^e 
+ \int_{\partial \Omega^e} \numflux{f}  \cdot {\mathbf{v}}^N \textrm{d} S^e 
= \mathbf{0}.
\end{equation}  

The quantities $\mathbf{q}^N$, $\mathbf{v}^N$ and $\blocktensor{f}^N$ belong to the polynomial space $\mathscr{V}^N$. 
Therefore, it is possible to represent them inside every element as a linear combination of basis functions, $\phi_j \in \mathscr{P}^N(\Omega^e)$, so that
\begin{equation}\label{eq:PolExp}
\mathbf{q}^N \rvert_{\Omega^e} = \sum_{j=1}^{\NDOF^e} \mathbf{q}_j^N \phi_j (\mathbf{x}),  \ \ \ 
\mathbf{v}^N \rvert_{\Omega^e} = \sum_{j=1}^{\NDOF^e} \mathbf{v}_j^N \phi_j (\mathbf{x}), \ \ \ 
\blocktensor{f}^N \rvert_{\Omega^e} = \sum_{j=1}^{\NDOF^e} \blocktensor{f}_j^N \phi_j (\mathbf{x}),
\end{equation}
where the (spatial) number of degrees of freedom ($\NDOF$) in hexahedral elements depends on the polynomial order of the approximation, $\NDOF^e = (N + 1)^d$, where $d$ is the number of spatial dimensions.

Since the test function $\mathbf{v}^N$ is an arbitrary polynomial, \eqref{eq:weak2} must hold for every basis function $\phi_j$.
Therefore, the DG discretization of \eqref{eq:OuterEq2} becomes
\begin{equation} \label{eq:weakBasis}
\int _{\Omega^e} {\partial_t \mathbf{q}}^N {\phi_j} \textrm{d} \Omega^e 
- \int_{\Omega^e} {\blocktensor{f}}^N \cdot \Nabla {\phi_j} \textrm{d} \Omega^e 
+ \int_{\partial \Omega^e} \numflux{f} {\phi_j} \textrm{d} S^e 
= \mathbf{0},
\end{equation}

Following a similar procedure, we can obtain the DG discretization of \eqref{eq:InnerEq2} as
\begin{align} \label{eq:ViscousWeakWeak}
\int _{\Omega^e} \blocktensor{g}^N \phi_j \d \Omega^e 
= - \int _{\Omega^e} \mathbf{q}^N \Nabla \phi_j \d \Omega^e 
+ \int_{\partial \Omega^e} \phi_j \hat{\mathbf{q}} \vec{n} \d S^e
\end{align}
where $\hat{\mathbf{q}}$ is a numerical \textit{flux} (actually the numerical trace of the solution) that corresponds to the interface value assumed by the state vector $\mathbf{q}$. 
Note that in the notation used here, the definition of the numerical trace of $\mathbf{q}$ does not include the action of the normal vector. 

The advective numerical flux, $\numflux{f}^{\ a} (\mathbf{q}^+,\mathbf{q}^-,\vec{n})$, is a function of the normal vector to the surface $\partial \Omega^e$, $\vec{n}$, of the discrete solution on element $e$,
\begin{equation}
\mathbf{q}^+ = \sum_{j=1}^{\NDOF^e} \mathbf{q}^N_j \phi_j,
\end{equation}
and of the outer solution, $\mathbf{q}^-$, which can be a Dirichlet boundary condition that depends on the inner state, $\mathbf{q}^-(\mathbf{q}^+)$, or the discrete solution on a neighbor element,
\begin{equation}
\mathbf{q}^- = \sum_{j=1}^{\NDOF^{\bunderline{e}}} \mathbf{q}^{\bunderline{N}}_j \phi^-_j,
\end{equation}
where the quantities $\mathbf{q}^{\bunderline{N}}_j$, $\phi^-_j$ and $\NDOF^{\bunderline{e}}$ correspond to the coefficients of the solution, the values of the basis functions, and the number of degrees of freedom on a neighbor element, respectively.
Several advective numerical fluxes are available in the finite volume literature \cite{toro2013riemann}.

When solving advection-diffusion PDEs, such as the Navier-Stokes equations, we also need a way to define the numerical trace of the solution, $\numflux{q}$, and the diffusive numerical flux, $\numflux{f}^{\nu}$.
The former usually has a linear dependency on the solution on both sides of the interface \cite{Arnold2002}, $\numflux{q} (\mathbf{q}^+,\mathbf{q}^-)$, and the latter can be classified as compact or non-compact depending on its dependencies.

The diffusive numerical flux is said to be compact if, besides depending on $\mathbf{q}^+$, $\mathbf{q}^-$ and $\vec{n}$, it also depends on the local gradients of the discrete solution on both sides of the surface $\partial \Omega^e$,
\begin{equation}
\numflux{f}^{\nu}  = 
\numflux{f}^{\nu} (\mathbf{q}^+, \Nabla \mathbf{q}^+, \mathbf{q}^-, \Nabla \mathbf{q}^-, \vec{n}).
\end{equation}
Several compact fluxes are available in the literature, such as the Interior Penalty (IP) flux \cite{Douglas1976} or the Bassi-Rebay 2 (BR2) flux \cite{bassi1997high}.
On the other hand, the diffusive numerical flux is said to be non-compact if it depends on the DG discretized gradients on both sides of the surface $\partial \Omega^e$,
\begin{equation}
\numflux{f}^{\nu}  = 
\numflux{f}^{\nu} (\mathbf{q}^+, \blocktensor{g}^+, \mathbf{q}^-, \blocktensor{g}^-, \vec{n}).
\end{equation}
The Bassi-Rebay 1 (BR1) flux \cite{Bassi1997} is an example of a non-compact diffusive numerical flux.

In this paper, we restrict the analysis to compact diffusive numerical fluxes, as they link each element only with its neighbors. 
Non-compact diffusive numerical fluxes link each element with the neighbors of its neighbors.
As a result, compact viscous numerical fluxes lead to sparser matrices that need less storage and whose matrix operations require fewer floating point operations.\\

In the DGSEM \cite{Black1999,Kopriva2009implementing}, the tessellation is performed with non-overlapping quadrilateral ($d=2$) or hexahedral ($d=3$) elements, whose physical coordinates are obtained from a reference element in $[-1,1]^d$ with a high-order mapping of order $M$,
\begin{eqnarray} \label{eq:mapping}
\vec{x}^e = \vec{x}^e \left( \vec{\xi} \right) \in \mathscr{P}^M, & & \vec{\xi} = \left( \xi, \eta, \zeta \right) \in \left[ -1, 1 \right]^3.
\end{eqnarray}

The high-order mapping allows one to describe curved boundaries accurately, and to evaluate the integrals numerically by means of a Gaussian quadrature rule. 
In the DGSEM, we use a \textit{non-overintegrated} quadrature of order $N$,
\begin{equation} \label{eq:quadRule}
\int_{\Omega^e} f \d \Omega^e
=
\int_{-1}^1 \int_{-1}^1 \int_{-1}^1 J f \d \xi \d \eta \d \zeta
\approx 
\int_{\Omega^e}^N f \d \Omega^e
= \sum_{i,j,k=0}^N J_{ijk} w_i w_j w_k f (\vec{\xi}_{i,j,k}),
\end{equation}
where $w_i$, $w_j$ and $w_k$ are the quadrature weights, $\vec{\xi}_{ijk}$ are the quadrature nodes, and $J$ is the Jacobian of the transformation \cite{Kopriva2009implementing}.

Furthermore, in the DGSEM the polynomial basis functions, $\phi_j$, are tensor-product reconstructions of Lagrange interpolating polynomials on the quadrature points in each local coordinate direction:
\begin{equation}
\mathbf{q}^N (\vec{\xi}) 
= 
\sum_{n=1}^{\NDOF^e} \mathbf{q}_n^N \phi_n (\vec{\xi}) 
= 
\sum_{i,j,k=0}^{N} \mathbf{q}_{i,j,k}^{N} \ell^{\xi}_i (\xi) \ell^{\eta}_j (\eta) \ell^{\zeta}_k (\zeta),
\end{equation}
and the Lagrange polynomials are
\begin{equation}
\ell_i^{\xi} (\xi) = \prod_{\substack{m=0 \\ m \ne i}}^{N_1} \frac{\xi - \xi_m}{\xi_i - \xi_m}.
\end{equation}

\begin{figure}[htbp]
\centering
\subfigure[Legendre-Gauss.]{\label{fig:GaussPoints} \includegraphics[width=0.3\textwidth]{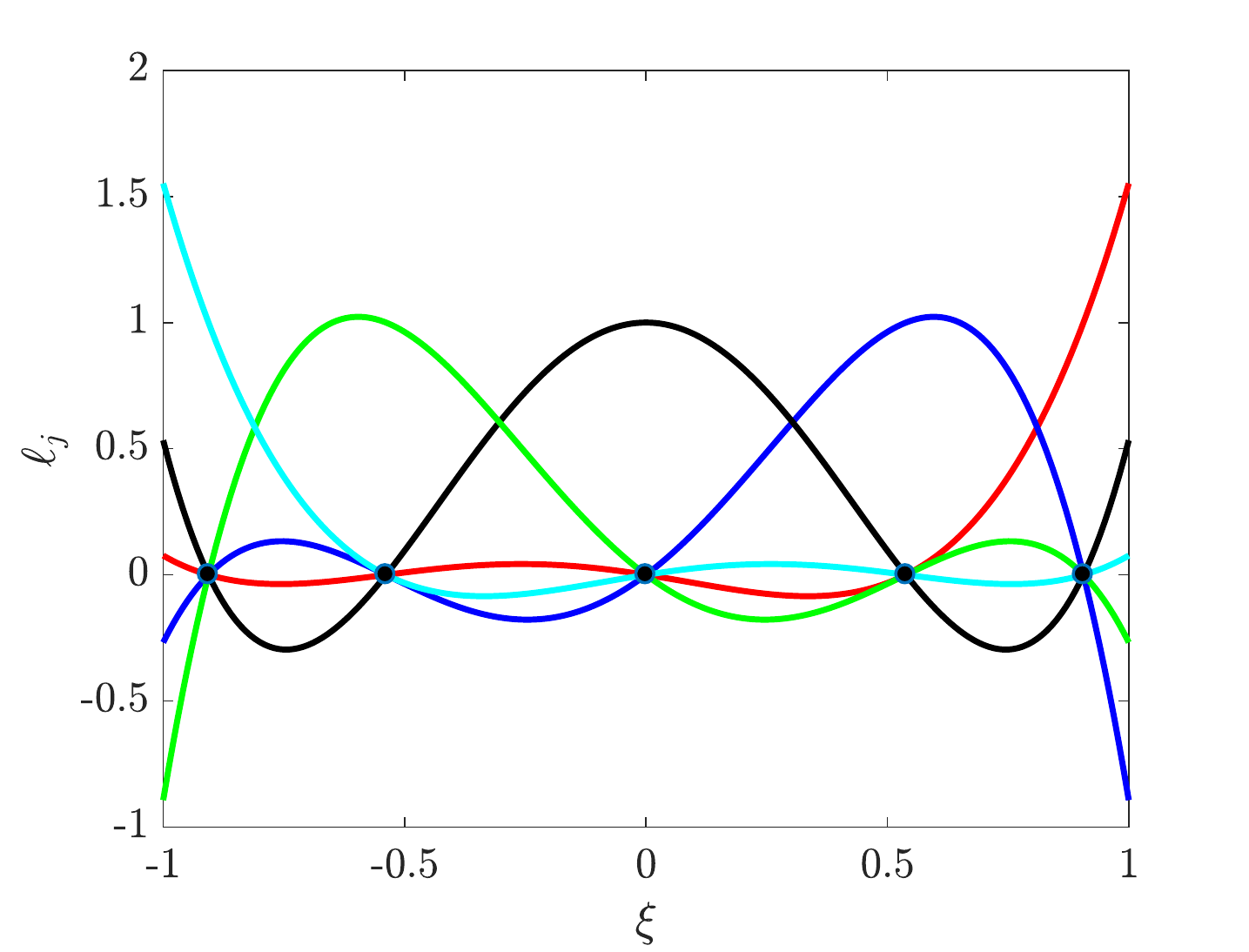}}\qquad
\subfigure[Legendre-Gauss-Lobatto.]{\label{fig:GaussLobattoPoints} \includegraphics[width=0.3\textwidth]{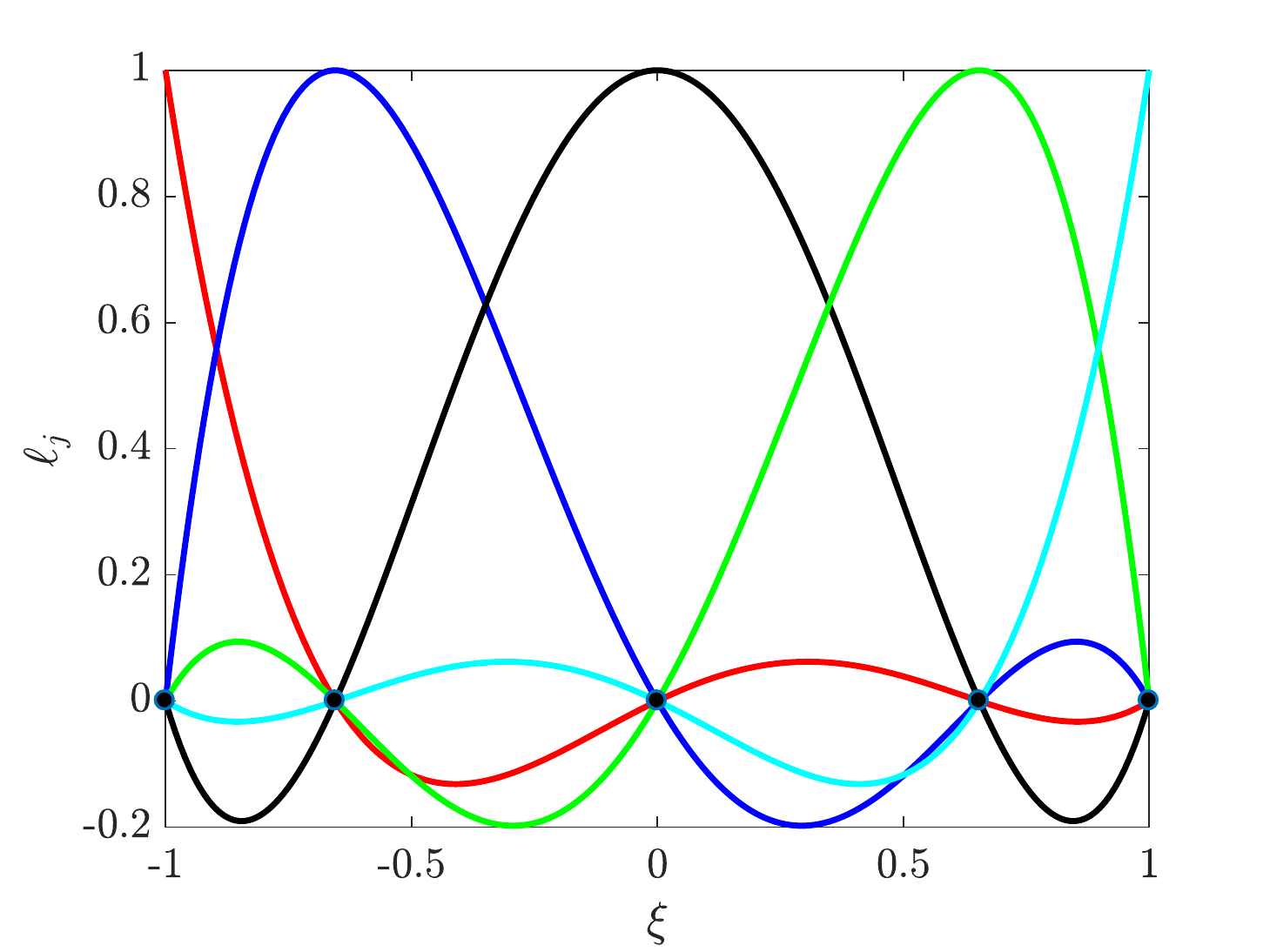}}

\caption{Lagrange interpolating polynomials on Legendre-Gauss and Legendre-Gauss-Lobatto points.}\label{fig:GaussAndGaussLobatto}
\end{figure}

The standard choices for the quadrature rule are the Legendre-Gauss and the Legendre-Gauss-Lobatto nodes \cite{Kopriva2009implementing} (usually called only Gauss or Gauss-Lobatto points, respectively). 
Figure \ref{fig:GaussAndGaussLobatto} shows the Lagrange interpolating polynomials on both quadrature rules for a $d=1$ discretization with $N=4$.
Note that the Lagrange interpolating polynomials are discretely orthogonal, 
\begin{equation}
\ell^{\xi}_i(\xi_j) = \delta_{ij}.
\end{equation}
Therefore, all basis functions take a nonzero value on the boundary when using Gauss nodes, whereas only one basis function takes a value on each boundary when using Gauss-Lobatto nodes.

Replacing the integrals by quadrature rules and the basis functions by tensor-product Lagrange polynomials  in \eqref{eq:weakBasis} and \eqref{eq:ViscousWeakWeak}, the DGSEM version of system \eqref{eq:SystemAdvDiff2} reads
\begin{subequations}\label{eq:DGSEMsystem2}
\begin{empheq}[left=\empheqlbrace]{align} 
J_j w_j \partial_t \mathbf{q}^N_j 
- \int_{\Omega^e}^N {\blocktensor{f}}^N \cdot \Nabla {\phi_j} \d \Omega^e 
+ \int_{\partial \Omega^e}^N \numflux{f} {\phi_j} \textrm{d} S^e 
&= \mathbf{0}, \label{eq:DGSEMsystem2:1}\\
- \int _{\Omega^e}^N \mathbf{q}^N \Nabla \phi_j \d \Omega^e 
+ \int_{\partial \Omega^e}^N \phi_j \hat{\mathbf{q}} \vec{n} \d S^e
&=
J_j w_j  \blocktensor{g}^N_j. \label{eq:DGSEMsystem2:2}
\end{empheq}
\end{subequations}

\section{State-of-Art Static Condensation for Continuous and Discontinuous Galerkin Methods} \label{App:Static}

In this appendix, we provide a brief review of how static condensation has been applied to time-implicit discretizations in continuous and discontinuous Galerkin methods.

\subsection{Continuous Galerkin Methods} \label{sec:CGStatic}
Static condensation was first used by Fraeijs in 1965 \cite{fraeijs1965displacement} to reduce the size of the linear system that results from time-implicit Finite Element discretizations. In general, in medium-to-high order ($N \ge 2$) continuous Galerkin (CG) discretizations, it is easy to reorganize the linear system \eqref{eq:SC_linsys} as 
\begin{equation} \label{eq:Static_QbQiSystem}
\begin{bmatrix}
\mat{A}_{bb} & \mat{A}_{ib} \\
\mat{A}_{bi} & \mat{A}_{ii}
\end{bmatrix}
\begin{bmatrix}
\mathbf{Q}_b \\ \mathbf{Q}_i
\end{bmatrix}
=
\begin{bmatrix}
\mathbf{B}_b \\ \mathbf{B}_i
\end{bmatrix},
\end{equation}
where $\mathbf{Q}_b$ is the solution on the degrees of freedom that sit on the element boundaries (interfaces), and $\mathbf{Q}_i$ is the solution on the inner degrees of freedom. Moreover, $\mat{A}_{bb}$ is the boundary-to-boundary matrix, $\mat{A}_{ib}$ is the interior-to-boundary matrix, $\mat{A}_{bi}$ is the boundary-to-interior matrix, and $\mat{A}_{ii}$ is the interior-to-interior matrix.

Note that system \eqref{eq:Static_QbQiSystem} is equivalent to system  \eqref{eq:BlockSystem}, for
\begin{equation} \label{eq:SC_CGequivalence}
\begin{bmatrix}
\mat{B} & \mat{C} \\
\mat{D} & \mat{E}
\end{bmatrix}
\leftrightarrow
\begin{bmatrix}
\mat{A}_{bb} & \mat{A}_{ib} \\
\mat{A}_{bi} & \mat{A}_{ii}
\end{bmatrix}, \ \
\begin{bmatrix}
\mathbf{X}_1 \\ \mathbf{X}_2
\end{bmatrix}
\leftrightarrow
\begin{bmatrix}
\mathbf{Q}_b \\ \mathbf{Q}_i
\end{bmatrix}, \ \
\begin{bmatrix}
\mathbf{F}_1 \\ \mathbf{F}_2
\end{bmatrix}
\leftrightarrow
\begin{bmatrix}
\mathbf{B}_b \\ \mathbf{B}_i
\end{bmatrix},
\end{equation}
and it can be solved using the same two-step procedure.
Furthermore, since the coupling between elements in CG occurs only through the (shared) degrees of freedom on element interfaces, the matrix $\mat{A}_{ii}$ has a block diagonal structure, which makes it easy to invert in a local manner.

Static condensation has also been used by Karniadakis and Sherwin \cite{karniadakis2013spectral} and Vos et al. \cite{Vos2010} for high-order CG methods, who have shown that the computational efficiency is increased when the order of the approximation ($N$) is increased because the relative size of the condensed system, $n_1 / (n_1 + n_2)$, decreases with $N$.

\subsection{Discontinuous Galerkin Methods}

We now discuss how static condensation has been used with high-order DG methods.
Unlike CG methods, DG methods may couple all the degrees of freedom of an element with the degrees of freedom of its neighbors (see \eqref{eq:AdvImpDG} and \eqref{eq:DiffImpDGouter}), or neighbors of neighbors (if a non-compact DG method is used), through the numerical flux functions. As a result, the static-condensation method is in general not directly applicable.

The first implementation of a static-condensation DG scheme was presented by Sherwin et al. \cite{Sherwin2006}, who were able to make a modal DG scheme suitable for static condensation by using $C^0$-type expansions for the basis functions on element boundaries and bubble functions for the inner modes. 
This choice of basis resembles the one used in $p$-FEM, a type of continuous Galerkin methods. 
Consequently, the linear system that is produced from the implicit time-integration of the modified DG scheme can also be arranged as the system \eqref{eq:Static_QbQiSystem}, where the matrix $\mat{A}_{ii}$ is also block diagonal. 
This proved to be advantageous since the statically condensed system was shown to be not only smaller in size but also cheap-to-compute and better conditioned than the global system. 

Sherwin et al. \cite{Sherwin2006} allege that an additional advantage of their statistically condensable DG is that the boundary conditions can be imposed through global lifting, as in continuous Galerkin methods, hence reducing the number of degrees of freedom of the problem. This is useful to treat elliptic problems but it may need stabilization for hyperbolic equations.

The only drawback of Sherwin's approach is that the new set of basis functions (bubble function $ + C^0$) are neither orthogonal expansions nor tensor-product bases. 
Therefore, several advantages of such basis functions cannot be kept, such as the existence of diagonal mass matrices, sparser Jacobians, the possibility to evaluate the anisotropic truncation error estimator of \cite{RuedaRamirez2019}, the ability to perform anisotropic $p$-adaptation \cite{RuedaRamirez2019a}, among others.

Another approach to perform static condensation in DG methods was developed simultaneously and independently from Sherwin's approach by Carrero and Cockburn et al. \cite{Carrero2005,Cockburn2009} and is known as the Hybridizable Discontinuous Galerkin (HDG) method. 
This method imposes no restrictions on the choice of basis functions and has gained increased popularity in recent years \cite{Fernandez2017,Soon2009}. 

The HDG method expands the original DG system with a new unknown variable $\pmb{\lambda}$ (typically the numerical trace of the solution, $\pmb{\lambda} = \numflux{q}$) that lives only on the mesh skeleton, with which it is possible to statically condense the system.
The expanded system is
\begin{equation} \label{eq:Static_HDG}
\begin{bmatrix}
\mat{B} & \mat{C} \\
\mat{D} & \mat{A}_{B}
\end{bmatrix}
\begin{bmatrix}
\pmb{\Lambda} \\ \mathbf{Q}
\end{bmatrix}
=
\begin{bmatrix}
\mathbf{F}_1 \\ \mathbf{B}
\end{bmatrix},
\end{equation}
where $\mat{A}_B$ is a matrix formed by the diagonal blocks of matrix $\mat{A}$, $\pmb{\Lambda}$ is the sampled version of $\pmb{\lambda}$, and $\mat{B}$, $\mat{C}$, $\mat{D}$ and $\mathbf{F}_1$ are additional terms that contain the scattered information of the off-diagonal blocks of $\mat{A}$. 
A similar approach is done by Petersen \cite{Petersen2009} for a space-time DG, where $\pmb{\Lambda}$ are Lagrange multipliers.

Note that we kept almost the same notation of \eqref{eq:SC_linsys} in \eqref{eq:Static_HDG}, but some variables where changed by the names they usually have in the HDG community. Therefore we have 
\begin{equation} \label{eq:SC_HDGequivalence}
\begin{bmatrix}
\mat{B} & \mat{C} \\
\mat{D} & \mat{E}
\end{bmatrix}
\leftrightarrow
\begin{bmatrix}
\mat{B} & \mat{C} \\
\mat{D} & \mat{A}_{B}
\end{bmatrix}, \ \
\begin{bmatrix}
\mathbf{X}_1 \\ \mathbf{X}_2
\end{bmatrix}
\leftrightarrow
\begin{bmatrix}
\pmb{\Lambda} \\ \mathbf{Q}
\end{bmatrix}, \ \
\begin{bmatrix}
\mathbf{F}_1 \\ \mathbf{F}_2
\end{bmatrix}
\leftrightarrow
\begin{bmatrix}
\mathbf{F}_1 \\ \mathbf{B}
\end{bmatrix}.
\end{equation}

The first formulations of the HDG method \cite{Carrero2005,Cockburn2009} dealt with linear elliptic problems and imposed particular conditions, such as the requirement on the viscous numerical fluxes to be adjoint-consistent (see for example \cite{Arnold2002}).
When solving nonlinear conservation laws, additional constraints are imposed for a DG method to be hybridizable with $\pmb{\lambda} = \numflux{q}$. 
For example, the numerical flux is restricted to the form \cite{Nguyen2009,Peraire2011}
\begin{equation} \label{eq:HDGconstraint}
\numflux{f} = \blocktensor{f} (\hat{\mathbf{q}}) \cdot \vec{n} + \mat{S}(\mathbf{q}^N,\hat{\mathbf{q}}) (\mathbf{q}^N - \hat{\mathbf{q}}),
\end{equation}
where $\mathbf{q}^N$ is the solution on the element $e$, $\numflux{q}$ is the numerical trace of it, and $\mat{S}$ is a stabilizing function. Some broadly used numerical flux functions, such as the Lax-Friedrichs flux, can be expressed in this form, but others (e.g. Roe) cannot.

Without the constraint \eqref{eq:HDGconstraint}, it would not be possible to condense the system as a function of $\numflux{Q}$. 
However, we would like to point out that, in purely advective nonlinear conservation laws, it is possible to use $\pmb{\lambda} = \numflux{f}^{\ a}$ and ignore the restriction \eqref{eq:HDGconstraint}.

\end{appendices}

\end{document}